# CHAPTER 1

# A COMPARATIVE STUDY OF PORTFOLIO OPTIMIZATION METHODS FOR THE INDIAN STOCK MARKET

JAYDIP SEN, ARUP DASGUPTA, PARTHA PRATIM SENGUPTA & SAYANTANI ROY CHOUDHURY

## Introduction

The design of optimized portfolios has remained a research topic of broad and intense interest among researchers of quantitative and statistical finance for a long time. An optimum portfolio allocates the weights to a set of capital assets in a way that optimizes the return and risk of those assets. Markowitz in his seminal work proposed the *mean-variance optimization* approach which is based on the mean and covariance matrix of returns (Markowitz, 1952). The *mean-variance portfolio* (MVP) design works on an algorithm, known as the *critical line algorithm* (CLA). The CLA algorithm, despite the elegance of its theoretical framework, has some major limitations. One of the major problems is the adverse effects of the estimation errors in its expected returns and covariance matrix on the performance of the portfolio. Since it is extremely challenging to accurately estimate the expected returns of an asset from its historical prices, it is a popular practice to use either a minimum variance portfolio or an optimized risk portfolio with the maximum Sharpe ratio as better proxies for the expected returns. However, due to the inherent complexity, several factors have been used to explain the expected returns.

The *hierarchical risk parity* (HRP) algorithm attempts to address three major shortcomings of quadratic optimization methods which are particularly relevant to the CLA used in the MVP approach to portfolio design (de Prado, 2016). These problems are instability, concentration, and underperformance. Unlike the CLA, the HRP algorithm does not require the covariance matrix of return values to be invertible. Hence, the HRP portfolios can deliver good results even if the covariance matrices are ill-degenerated or singular, which is an impossibility for a quadratic optimizer like CLA. Interestingly, even though MVP's objective is to minimize the variance, HRP is proven to have a lower probability of yielding lower out-of-sample variance than MVP. Given the fact that future returns cannot be forecasted with sufficient accuracy, many researchers have proposed risk-based asset allocation using the covariance matrix of the returns. However, this approach brings in another problem of instability that arises because the quadratic programming methods require the inversion of a covariance matrix whose all eigenvalues must be positive (Baily & de Prado, 2012). HRP is a new portfolio design method that addresses the pitfalls of the quadratic optimization-based MVP approach using techniques of graph theory and machine learning (de Prado, 2016). This method exploits the features of the covariance matrix without the requirement of its invertibility.

The HERC portfolio optimization uses an integrated approach to machine learning and a top-down recursive bisection method of the HRP portfolio method (Raffinot, 2018). The proponents of the HERC method identified several shortcomings of the HRP portfolio optimization. The single linkage-based cluster trees constructed in the HRP method usually lead to deep and wide trees and suboptimal allocation of weights to the clusters. The HRP algorithm usually involves higher computations. Finally, the recursive bijection approach used in HRP bisects the cluster tree before the weight allocation instead of directly allocating the weights based on the dendrogram of clustering. This makes the computed weights inaccurate. The HERC algorithm avoids these problems using a top-down recursive bisection and a naive risk parity within the clusters.

This chapter presents a comparative study of the three portfolio optimization methods, MVP, HRP, and HERC, on the Indian stock market, particularly focusing on the stocks chosen from 15 sectors listed on the National Stock Exchange (NSE) of India. The top stocks of each cluster are identified based on their free-float market capitalization from the NSE's report published on July 1, 2022 (NSE Website). For each sector, three portfolios are designed on stock prices from July 1, 2019, to June 30, 2022, following three portfolio optimization approaches. The portfolios are tested over the period from July 1, 2022, to June 30, 2023. For evaluation of the performances of the portfolios, three metrics are used (i) cumulative returns, (ii) annual volatilities, and (iii) Sharpe ratios. based on their cumulative returns. For each sector, the portfolios that yield the highest cumulative return, the lowest volatility, and the maximum Sharpe Ratio over the training and the test periods are identified.

The work has three unique contributions. First, it presents an efficient portfolio design approach using three well-known portfolio optimization methods for the Indian stock market. Fifteen sectors each containing ten critical stocks are used in designing the portfolios. Second, the portfolios are backtested using several metrics including, cumulative returns, annual volatilities, and Sharpe ratios. The results of the evaluation identify the best-performing portfolio corresponding to each sector of stocks over the training and the test periods. Finally, the results of this study provide a deep insight into the current profitability of the sectors that will be useful for investors in the Indian stock market.

The chapter is organized as follows. The section titled *Related Work* presents some of the existing portfolio design approaches proposed in the literature. Next, the section titled *Methodology* presents the research approach followed in the current work. The section titled *Results* presents an extensive set of results and a detailed analysis of the observations. Finally, the chapter is concluded in the section titled *Conclusion*.

## Related Work

Portfolio design and optimization is a challenging problem for which numerous solutions and approaches have been proposed by researchers. Portfolio design and optimization is a challenging problem that has attracted considerable attention from researchers. Numerous approaches have been proposed to solve this complex problem involving robust stock price prediction and the formation of the optimized combination of stocks to maximize the return on investment. Machine learning models have been extensively used by researchers in predicting future stock prices (Carta et al., 2021; Chatterjee et al., 2021; Mehtab & Sen, 2021; Mehtab & Sen, 2020a; Mehtab & Sen, 2019; Mehtab et al., 2021; Sarmento & Horta, 2020; Sen, J., 2018a; Sen & Datta Chaudhuri, 2017a). The prediction accuracies of the models are found to have been improved by the use of deep learning architectures and algorithms (Chatterjee et al., 2021; Chen et al., 2018; Chong et al., 2017; Sen & Mehtab, 2021b; Mehtab & Sen, 2021; Mehtab & Sen, 2020a; Mehtab & Sen, 2020b; Mehtab & Sen, 2019; Mehtab et al., 2021; Mehtab, et al., 2020; Sen, 2018a; Sen & Mehtab, 2021a; Sen & Mehtab, 2021b; Sen et al., 2021a; Sen et al., 2021b; Sen et al., 2021i; Sen et al., 2020; Sen & Mehtab, 2022b; Thormann et al., 2021; Tran et al., 2019). Several approaches to text mining have been effectively applied on social media and the web to improve prediction accuracies even further (Li & Pan, 2022; Mehtab & Sen, 2019; Thormann et al., 2021; Zhang et al., 2021). Among the other alternative approaches for stock price prediction, time series decomposition-based statistical and econometric approaches are also quite popular (Chatterjee et al., 2021; Cheng et al., 2018; Sen, 2022a; Sen, 2018b; Sen, 2017; Sen & Datta Chaudhuri, 2018; Sen & Datta Chaudhuri, 2017b; Sen & Datta Chaudhuri, 2016a; Sen & Datta Chaudhuri, 2016b; Sen & Datta Chaudhuri, 2016c; Sen & Datta Chaudhuri, 2016d; Sen & Datta Chaudhuri, 2015). For estimating the future volatility and risk of stock portfolios the use of several variants of GARCH has been proposed in some works (Sen et al., 2021d). Over the last few years, reinforcement learning has been extensively used in robust and accurate prediction of stock prices and portfolio design (Brim, 2020; Fengqian & Chao, 2020; Kim et al., 2022; Kim & Kim, 2019; Lei et al., 2020; Li et al., 2019; Lu et al., 2021; Park & Lee, 2021; Sen, 2023; Sen, 2022d).

The classical mean-variance optimization approach is the most well-known method for portfolio optimization (Sen & Mehtab, 2022a; Sen et al., 2021e; Sen et al., 2021g; Sen et al., 2021h; Sen & Sen, 2023). Several alternatives to the mean-variance approach to portfolio optimization have also been proposed by some researchers. Notable among these methods are multiobjective optimization (Wang et al, 2022; Zheng & Zheng, 2022), eigen portfolios using principal component analysis (Sen & Dutta, 2022b; Sen & Mehtab, 2022a), risk parity-based methods (Sen & Dutta, 2022a; Sen & Dutta, 2022c; Sen

& Dutta, 2021; Sen et al., 2021c; Sen et al., 2021f), and swarm intelligence-based approaches (Corazza et al., 2021; Thakkar & Chaudhuri, 2021). The use of genetic algorithms (Kaucic et al., 2019), fuzzy sets (Karimi et al., 2022), prospect theory (Li et al., 2021), and quantum evolutionary algorithms (Chou et al., 2021) are also proposed in the literature.

As an alternative to portfolios with multiple stocks, pair-trading portfolios involving two stocks have also been proposed by researchers in the literature (Flori & Regoli, 2021; Gupta & Chatterjee, 2020; Ramos-Requena et al., 2021; Sen, 2022b; Sen, 2022c).

The current work presents a comprehensive study of the performances of three different approaches to portfolio design, MVP, HRP, and HERC, on 15 important sectors of stocks listed on the NSE of India. To the best of the knowledge and belief of the authors, no such studies have been done so far in this direction. Hence, the results of this work are expected to be useful to financial analysts and investors interested in the Indian stock market.

## Methodology

The details of the data used and the methodology followed in this work are presented in this section. This section discusses the methodology followed in this work especially focusing on the steps involved in designing the three portfolios for each of the sectors. The methodology involves in following seven steps.

**(i)** ***Choice of the sectors for analysis:*** Fifteen important sectors are first selected from those listed in the NSE so that they exhibit diversity in the Indian stock market. The chosen fifteen sectors are (i) *auto*, (ii) *banking*, (iii) *consumer durables*, (iv) *financial services*, (v) *fast-moving consumer goods* (FMCG), (vi) *information technology* (IT), (vii) *media*, (viii) *metal*, (ix) *mid-small IT and telecom*, (x) *oil and gas*, (xi) *pharma*, (xii) *private banks*, (xiii) *PSU banks*, (xiv) *realty*, and (xv) *NIFTY 50*. The monthly reports of the NSE identify the ten stocks with the maximum free-float capitalization from each sector. In this work, the report published on June 30, 2022, is used for identifying the ten stocks from each of the fourteen sectors, and the 50 stocks from NIFTY 50 (NSE Website).

**(ii)** ***Extraction of historical stock prices from the web:*** From the Yahoo Finance website, the historical daily prices of the stocks are extracted from July 1, 2019, to June 30, 2023, using the *DataReader* function of the *pandas_datareader* module of Python. The portfolios are built on the records from July 1, 2019, to June 30, 2022. The testing is done on the records from July 1, 2022, to June 30, 2023. The *close* values of the stocks are used in designing the portfolios.

**(iii)** ***Derivation of the return and volatility of portfolios:*** This step involves the computation of the daily return values for the stocks in the fourteen sectors (including the NIFTY 50 stocks). The daily return values reflect the percentage change in the *close* values for successive days. For computing the daily returns, the *pct_change* function of Python is used. Using the daily returns, the daily volatility values are obtained by computing the square root of the variance of the daily return values. Assuming that there are 250 operational days in a calendar year, the annual volatility values for the stocks are derived by multiplying the daily volatilities by a square root of 250. Next, the covariance matrix of the daily returns of the stocks for a sector is computed. Based on the covariance matrix of the returns of the stocks, the portfolio annual return and annual risk for a sector are computed. If a portfolio involves $n$ stocks and if $w_i$ represents the weight assigned to the stock $i$ which has an annual return of $R_i$, then the annual return ($R$) of the portfolio is given by (1).

$$R = \sum_{i=1}^{n} w_i * R_i \quad (1)$$

The variance ($V$) of a portfolio is given by (2).

$$V = \sum_{i=1}^{n} w_i s_i^2 + 2 * \sum_{i,j} w_i * w_j * cov(i,j) \quad (2)$$

In (2) $s_i$ represents the annual standard deviation of the stock $i$, and $cov\ (i,j)$ is the covariance between the returns of stocks $i$ and $j$. The square root of $V$, i.e., the standard deviation of the annual return indicates the annual risk associated with the portfolio.

**(iv) *Designing the MVP portfolios:*** At this step, the *mean-variance portfolio* (MVP) for each sector is designed. The design of this portfolio involves maximization of the risk-adjusted return. For this purpose, first, two concepts need to be understood, (i) Sharpe ratio, and (ii) efficient frontiers of portfolios.

The *Sharpe ratio* (SR) of a portfolio is given by (3)

$$SR = \frac{R_c - R_f}{\sigma_c} \tag{3}$$

In (3), $R_c$, $R_f$, and $\sigma_c$ denote the return of the current portfolio, the risk-free portfolio, and the standard deviation of the current portfolio, respectively. The risk-free portfolio is a portfolio with a volatility value of 1%. The mean-variance portfolio optimization involves the maximization of the Sharpe Ratios for a set of portfolios.

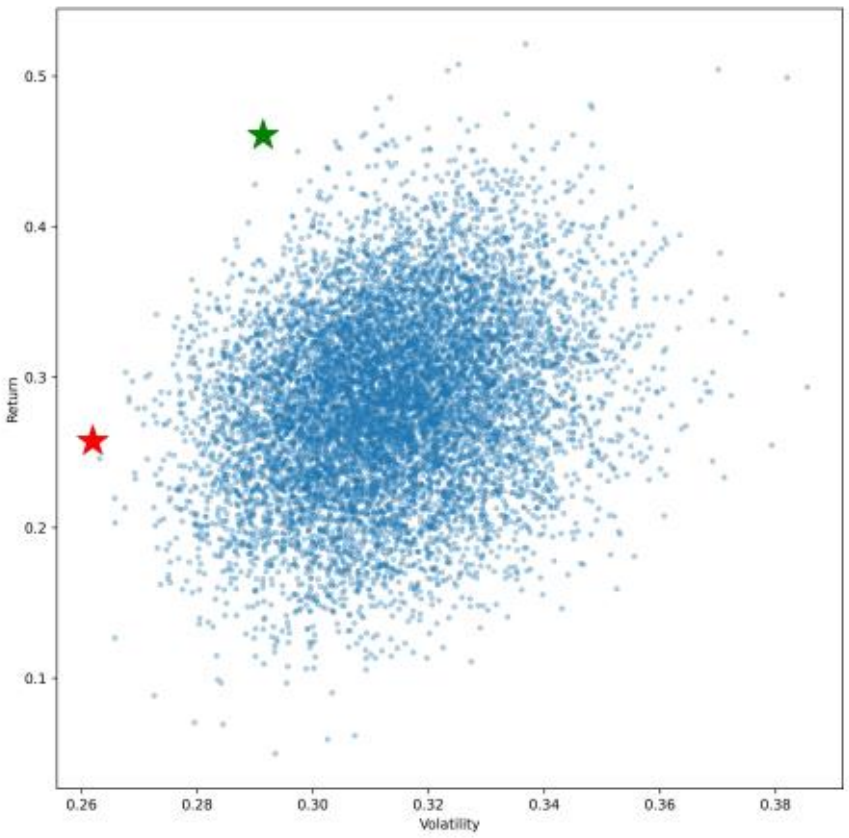

**Figure 1.1.** The efficient frontier of 10,000 candidate portfolios. The minimum-risk portfolio is indicated by the red star, while the green star represents the portfolio with the maximum Sharpe ratio (i.e., the mean-variance optimized portfolio)

For identifying the portfolio with the maximum Sharpe ratio, the *efficient frontiers* of a set of candidate portfolios need to be constructed. For a given portfolio of stocks, the efficient frontier is that plots the return along the *y*-axis and the volatility (i.e., risk) on the *x*-axis. The points on the contour of an efficient frontier indicate the portfolios with the maximum return for a given value of volatility or those with the minimum value of volatility for a given return. Since the volatility is plotted along the *x*-axis, the minimum risk portfolio is identified by the leftmost point on the efficient frontier. On the other hand, the optimum portfolio that maximizes the Sharpe ratio is identified by the point with the highest return/risk ratio. Figure 1.1 depicts the efficient frontier for several candidate portfolios, with the minimum-risk portfolio and the portfolio with the maximum Sharpe ratio identified.

The MVP portfolios are built for the 15 sectors of stocks. This involved creating 10000 candidate portfolios choosing the weights randomly and then finding out the portfolio with the maximum Sharpe ratio for each sector. The portfolios are built on the historical stock prices from July 1, 2019, to June 30, 2022.

**(v)** ***Designing the HRP portfolios:*** At this step, the HRP portfolios are designed for each sector. The HRP portfolio design involves three phases: (a) *tree clustering*, (b) *quasi-diagonaliz*ation, and (c) *recursive bisection*. These steps are described in the following.

***Tree Clustering***: The tree clustering used in the HRP algorithm is an agglomerative clustering algorithm. To design the agglomerative clustering algorithm, a *hierarchy* class is first created in Python. The hierarchy class contains a *dendrogram* method that receives the value returned by a method called *linkage* defined in the same class. The linkage method receives the stock price data after pre-processing and transformation and computes the minimum distances between stocks based on their return values. There are several options for computing the distance. However, the *ward distance* is a good choice since it minimizes the variances in the distance between two clusters in the presence of high volatility in the stock return values. In this work, the ward distance has been used as a method to compute the distance between two clusters. The linkage method performs the clustering and returns a list of the clusters formed. The computation of linkages is followed by the visualization of the clusters through a dendrogram. In the dendrogram, the leaves represent the individual stocks, while the root depicts the cluster containing all the stocks. The distance between each cluster formed is represented along the *y*-axis, longer arms indicate less correlated clusters. Figure 1.2 exhibits a typical dendrogram of the agglomerative clustering done by the HRP portfolio optimization method.

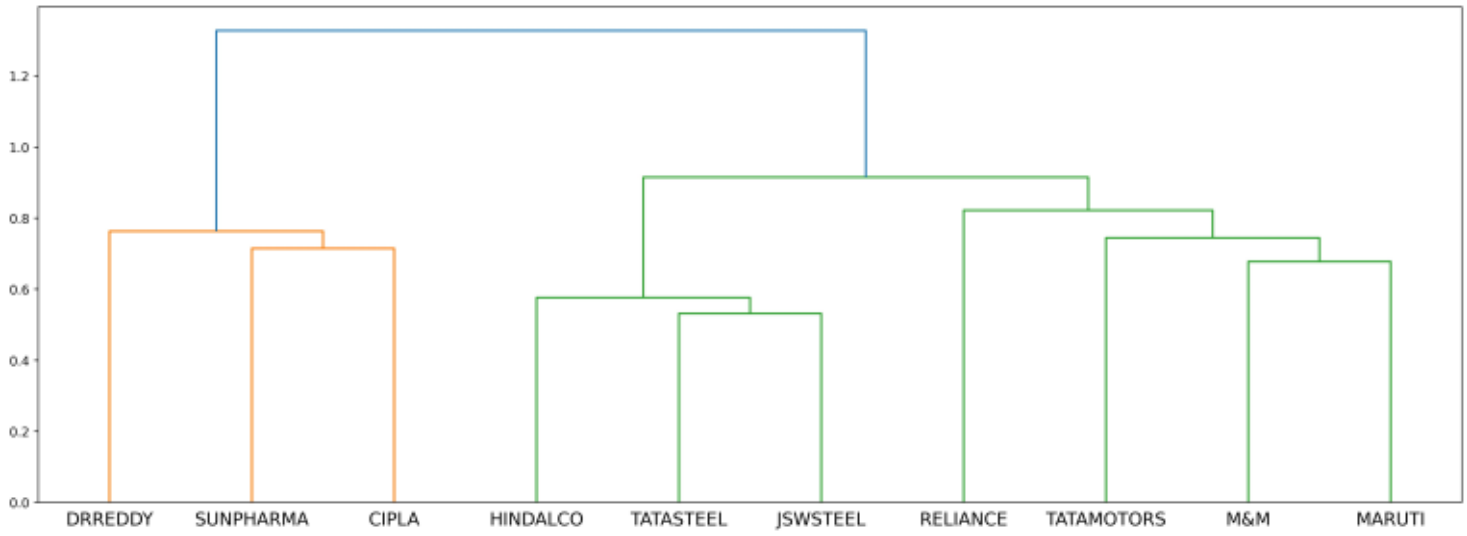

**Figure 1.2.** The dendrogram of the agglomerative clustering is created by the hierarchical risk parity method. The *x*-axis shows the ten stocks on which the clustering has been done, the *y*-axis depicts the ward distance.

***Quasi-Diagonalization***: In this step, the rows, and the columns of the covariance matrix of the return values of the stocks are reorganized in such a way that the largest values lie along the diagonal. Without requiring a change in the basis of the covariance matrix, the quasi-diagonalization yields a very important property of the matrix – the assets (i.e., stocks) with similar return values are placed closer, while disparate assets are put at a far distance. The working principles of the algorithm are as follows. Since each row of the linkage matrix merges two branches into one, the clusters ($C_{N-1}$, 1) and ($C_{N-2}$, 2) are replaced with their constituents recursively, until there are no more clusters to merge. This recursive merging of clusters preserves the original order of the clusters (Baily & de Prado, 2012). The output of the algorithm is a sorted list of the original stocks.

***Recursive Bisection:*** The quasi-diagonalization step transforms the covariance matrix into a quasi-diagonal form. It is proven mathematically that the allocation of weights to the assets in an inverse ratio to their variance is an optimal allocation for a *quasi-diagonal matrix* (Baily & de Prado, 2012). This allocation may be done in two different ways. In the *bottom-up* approach, the variance of a contiguous subset of stocks is computed as the variance of an inverse-variance allocation of the composite cluster. In the alternative top-down approach, the allocation among two adjacent subsets of stocks is done in inverse proportion to their aggregated variances. In the current implementation, the *top-down* approach is followed. A python function *computeIVP* computes the inverse-variance portfolio based on the computed variances of two clusters as its given input. The variance of a cluster is computed using another Python function called *clusterVar*. The output of the *clusterVar* function is used as the input to another Python function called *recBisect* which computes the final weights allocated to the individual stocks based on the recursive bisection algorithm.

The HRP method performs the weight allocation to $n$ assets in the best case and in the average in time $T(n) = O(\log_2 n)$, while its worst-case complexity is given by $T(n) = O(n)$. The complexity of the algorithm is directly proportional to the height of the cluster tree. Unlike the MVP approach, which is an approximate algorithm, the HRP is an exact and deterministic algorithm.

The HRP portfolios for the 15 sectors are built on the historical stock prices from July 1, 2019, to June 30, 2022.

**(vi)** ***Designing the HERC portfolios:*** At this step, the HERC portfolios are designed for each sector. The HERC portfolio is a risk-based portfolio optimization method that integrates machine learning and the top-down recursive bisection approach of HRP for portfolio optimization (Raffinot, 2018).

The proponents of the HERC method identify several shortcomings of the HRP portfolio optimization. The first problem is the linkage metric the HRP method uses to combine the clusters. The use of *single linkage clustering* constructs the tree based on the distance between the two closest points in the clusters, which results in a chaining effect making the tree very deep and wide. This makes dense clustering difficult and the weight allocation suboptimal. Second, in the HRP portfolio algorithm, there is a high possibility of large trees getting formed from a large dataset of stocks which may result in a very high computational task and possible overfitting of the model. Third, the recursive bisection step of the HRP method bisects the tree before allocating the weights instead of allocating the weights based on the constructed dendrogram. This results in inaccuracies in the allocated weights. Finally, the HRP method computes the weights based on the variances of the clusters. Accordingly, the assets in clusters with minimum variance receive higher weights. Since the estimates of risk computed on the past variances of the stocks are very unreliable and unstable, the weight allocation by HRP may not be very accurate for the out-of-sample data.

The HERC portfolio optimization involves the following four steps: (a) *hierarchical tree clustering*, (b) *selecting the optimal number of clusters*, (c) *top-down recursive bisection*, and (d) *naive risk parity* within the clusters.

Step (a) of the HERC method that calculates the distance matrix from the correlation matrix for cluster formation remains identical to that of the HRP method.

Step (b) of the HERC differs from the HRP approach. In this step, the optimal number of clusters is identified. While HRP does not involve any computation toward finding the optimal number of clusters, HERC uses the gap index method for determining the number of clusters to be used (Tibshirani et al., 2001). After the optimal number of clusters is determined, the top-down recursive bisection step computes the weight for each cluster.

In Step (c), the clustering tree bisects the cluster at a given level into two sub-clusters. The weights assigned to the sub-clusters are in the ratio of their contributions to the overall risk of the original cluster. Suppose, for a cluster *C*, the clustering algorithm has formed to sub-clusters $C_1$ and $C_2$. The weights assigned to the sub-clusters, $W_1$, and $W_2$, are given by (4) and (5), respectively.

$$W_1 = \frac{R_1}{R_1 v + R_2} \tag{4}$$

$$W_2 = 1 - W_1 \tag{5}$$

In (4) and (5), $R_1$ and $R_2$ represent the risk contributions of the sub-clusters $C_1$ and $C_2$, respectively, to cluster *C*. Several alternative metrics exist for computing risk such as variance, standard deviation, conditional value at risk, conditional drawdown as risk, etc. The risk involved in each cluster is the additive risk contribution of all the individual members in that cluster as expressed in (5). The weight allocation to the clusters is done through the entire tree until all the clusters (i.e., the assets at the leaf level) are assigned weights.

Finally, in step (d), the weights are assigned to the assets within the clusters using a naive risk parity approach based on the inverse of the assets' risks. This is illustrated using the above example in which cluster *C* is divided into two sub-clusters, $C_1$ and $C_2$. Let us assume that $C_1$ contains two stocks $S_1$ and $S_2$. The weights for $S_1$ and $S_2$ are to be determined. The naïve risk parity weights for $S_1$ and $S_2$ are given by (6) and (7), respectively.

$$W_{nrp}^{S1} = \frac{\frac{1}{R_1}}{\frac{1}{R_1}+\frac{1}{R_2}} \quad (6)$$

$$W_{nrp}^{S2} = \frac{\frac{1}{R_2}}{\frac{1}{R_1}+\frac{1}{R_2}} \quad (7)$$

The final weights for the stocks $S_1$ and $S_2$ are obtained by multiplying their respective naïve parity weights by the weights of the cluster to which they belong. i.e., cluster $C_1$. The final weights for stocks $S_1$ and $S_2$ are computed using (8) and (9), respectively.

$$W_{final}^{S1} = W_{nrp}^{S1} * W_1 \quad (8)$$

$$W_{final}^{S2} = W_{nrp}^{S2} * W_1 \quad (9)$$

In (8) and (9), $W_{final}^{S1}$ and $W_{final}^{S2}$ represent the final weights assigned to the assets $S_1$ and $S_2$, respectively, $W_1$ denotes the weight allocated to the cluster to which they belong, i.e., the cluster $C_1$ (derived in (1)), and $W_{nrp}^{S1}$ and $W_{nrp}^{S2}$ denote the naïve parity weights for stocks $S_1$ and $S_2$, respectively. This method is repeated till the weights for all the assets in all clusters are computed.

The HERC algorithm has the same time complexity for computation as the HRP algorithm.

The HERC portfolios for the 15 sectors are built on the historical stock prices from July 1, 2019, to June 30, 2022.

**(vii)** ***Visual presentation of the portfolios:*** The portfolios are now represented in the form of pie diagrams, in which the weights allocated to the stocks are shown as percentage figures. The weights assigned to the stocks are also listed in tabular format for every sector for a better understanding and readability purposes of their magnitudes and relative importance. The tables and charts are created using the properties of the *pandas* data frames and various functions defined in the *matplotlib* library.

**(viii)** ***Computation of the portfolio cumulative returns:*** In the final step, based on the weights allocated by the three portfolios (i.e., MVP, HRP, and HERC) to each stock in a given sector, the daily cumulative daily returns, the annual volatilities, and the Sharpe ratios for the portfolios are computed over the training and the test periods. The weighted sum of the daily return values of the stocks in a given portfolio is used to compute the portfolio returns. The cumulative returns for the three portfolios for each of the 15 sectors are then plotted for the training and the test data points. The numeric values of the cumulative returns, annual volatilities, and the Sharpe ratios of the three portfolios are also listed in the tables. To compute the annual volatilities of the stocks, first, the annual variances of the stock returns are computed by multiplying the variances of their daily returns by a factor of 252, assuming that there are 252 working days in a calendar year. The square root of the weighted variances of the stocks of a portfolio yields the annual volatility of the portfolio. The Sharpe ratio of a portfolio is computed as the ratio of its annual return to its annual volatility. For a given sector, the portfolios yielding the maximum cumulative return the lowest annual volatility, and the highest Sharpe ratio are identified as these portfolios have performed better in comparison to their counterparts on the chosen three metrics.

## Experimental Results

This section presents the detailed results and analysis of the portfolios. The fifteen sectors which are studied in this work are the following (i) *auto*, (ii) *banking*, (iii) *consumer durables*, (iv) financial services, (v) *FMCG*, (vi) *IT*, (vii) *media*, (viii) *metal*, (ix) *mid-small IT and telecom*, (x) *oil and gas*, (xi) *pharma*, (xii) *private banks*, (xiii) *PSU banks*, (xiv) *realty*, and (xv) *NIFTY 50*. The MVP, HRP, and HERC portfolios are implemented using Python 3.9.8 and its associated libraries *numpy*, *pandas*, *matplotlib*, *statsmodels*, *seaborn*, and *riskfolio-lib*. The portfolio models are trained and tested on the GPU environment of Google Colab (Google Colab).

***Auto sector:*** As per the report published by the NSE on July 30, 2022, the ten stocks of the *auto* sector with the largest free-float market capitalization and their contributions (in percent) to the overall sectoral index are the following: Mahindra & Mahindra (M&M): 17.92%, Maruti Suzuki India (MARUTI): 17.71, Tata Motors (TATAMOTORS): 15.39, Bajaj Auto (BAJAJ-AUTO): 7.57, Eicher Motors (EICHERMOT): 6.25%, Hero MotoCorp (HEROMOTOCO): 5.65%, Tube Investment of India (TIINDIA): 4.36%, TVS Motor Company (TVSMOTOR): 4.35, Ashok Leyland (ASHOKLEY): 3.59%, and Bharat Forge (BHARATFORG): 3.24% (NSE Website). The ticker names of the stocks are mentioned in parentheses. The ticker name of a stock is its unique identifier for a given stock exchange.

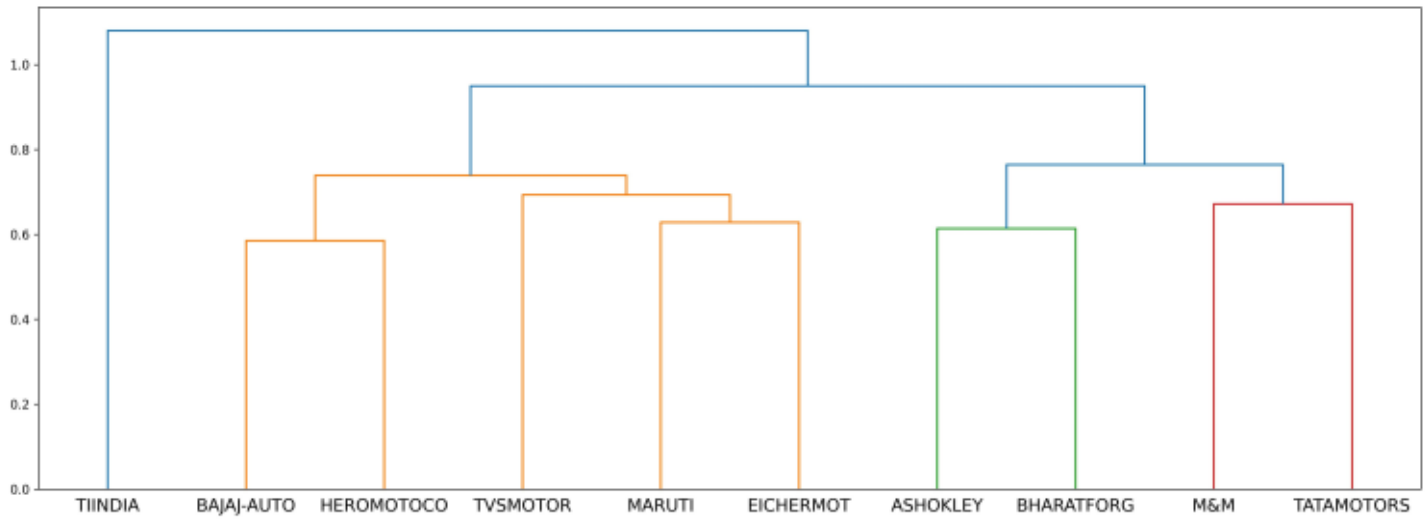

**Figure 1.3.** The dendrogram of the agglomerative clustering of the stocks from the *auto* sector is created based on the historical stock prices from July 1, 2019, to June 30, 2022.

The dendrogram of the clustering of the stocks of the auto sector is shown in Figure 1.3. The *y*-axis of the dendrogram depicts the ward linkage values, where a longer length of the arms signifies a higher distance and hence a cluster with less compactness. For example, the cluster containing the stocks BAJAJ-AUTO and HEROMOTO is the most compact, while the one containing M&M and TATAMOTORS is the least homogeneous. Figure 1.4 depicts the weight allocation done by the MVP and the HRP portfolios to the auto sector stocks.

Table 1.1 shows the weight allocations for the three portfolios in tabular format.

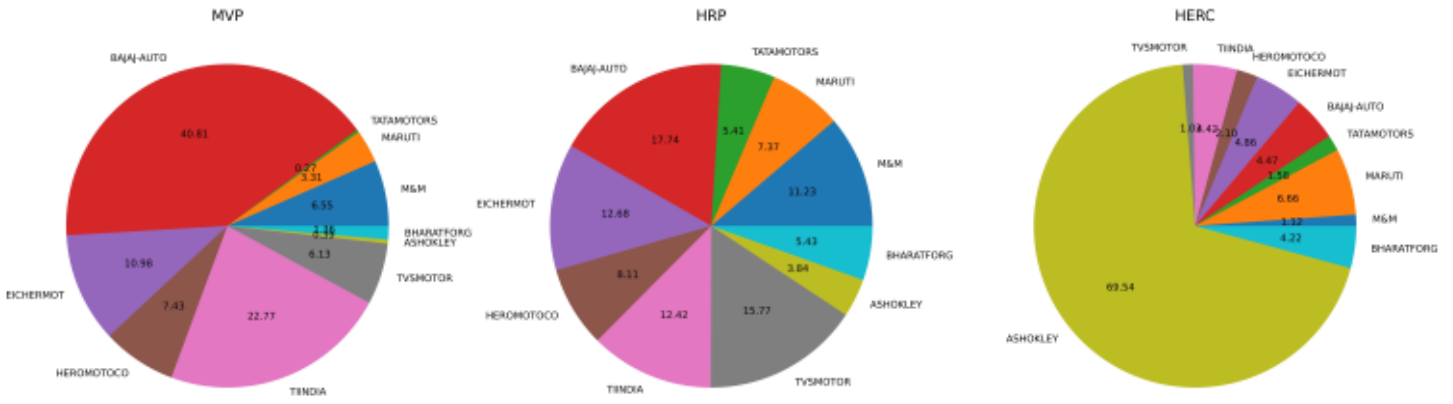

**Figure 1.4.** The allocation of weights done by the MVP, HRP, and HERC algorithms for the stocks of the *auto* sector based on the historical stock prices from July 1, 2019, to June 30, 2022.

**TABLE 1.1.** THE PORTFOLIO COMPOSITIONS OF THE AUTO SECTOR (PERIOD: JULY 1, 2019 – JUNE 30, 2022)

| Stock | MVP Portfolio | HRP Portfolio | HERC Portfolio |
|---|---|---|---|
| M&M | 0.0655 | 0.1123 | 0.0112 |
| MARUTI | 0.0331 | 0.0737 | 0.0666 |
| TATAMOTORS | 0.0027 | 0.0541 | 0.0158 |
| BAJAJ-AUTO | 0.4081 | 0.1774 | 0.0447 |
| EICHERMOT | 0.1098 | 0.1268 | 0.0486 |
| HEEROMOTOCO | 0.0743 | 0.0811 | 0.0210 |
| TIINIDA | 0.2277 | 0.1242 | 0.0442 |
| TVSMOTOR | 0.0613 | 0.1577 | 0.0103 |
| ASHOKLEY | 0.0039 | 0.0384 | 0.6954 |
| BHARATFORG | 0.0136 | 0.0543 | 0.0422 |

The cumulative returns yielded by the portfolios over the training and the test periods are depicted in Figure 1.5 and Figure 1.6, respectively. Table 1.2 presents the cumulative returns, annual volatilities, and the Sharpe ratios of the three portfolios of the auto sector stocks over the training and the test periods. The highest cumulative return, the lowest volatility, and the maximum Sharpe ratio over the portfolio training and test periods are indicated in red.

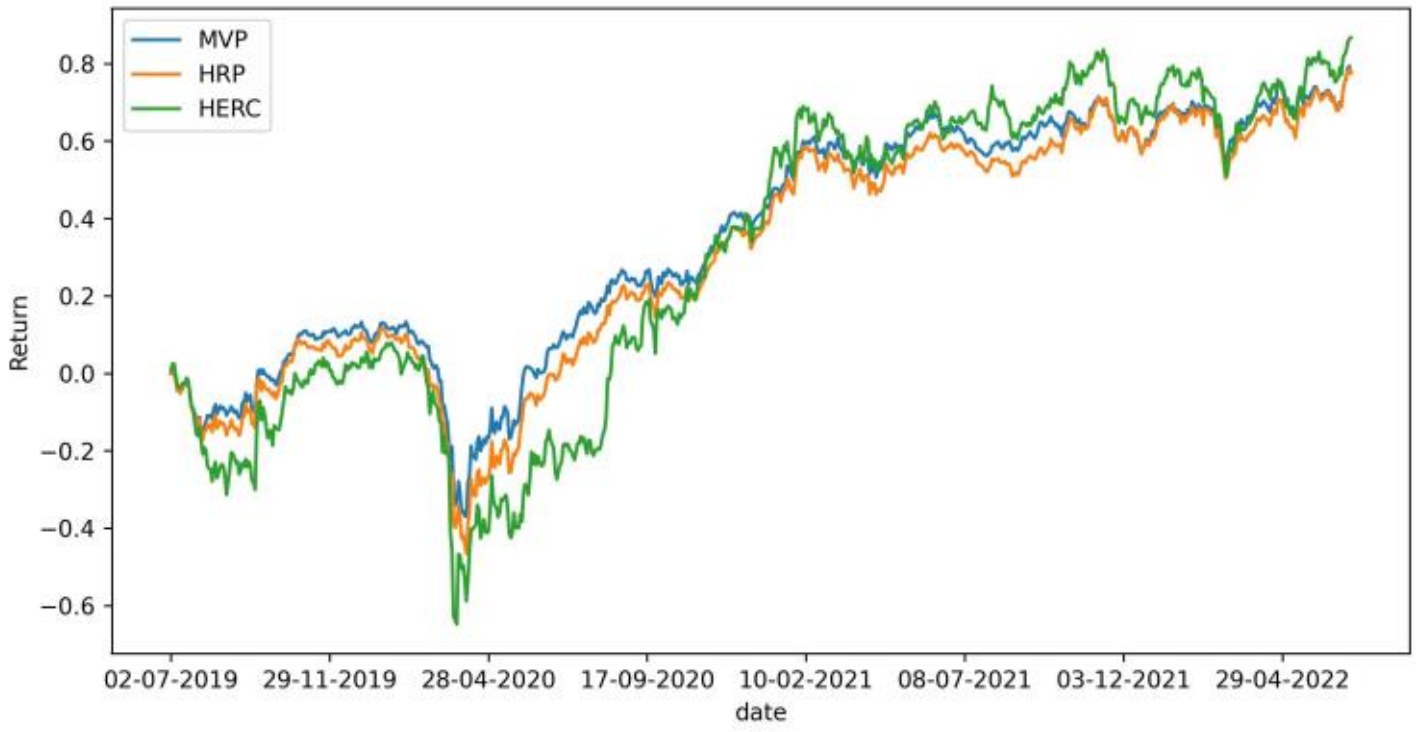

**Figure 1.5.** The cumulative returns yielded by the MVP, HRP, and HERC portfolios of the stocks from the *auto* sector on the training data from July 1, 2019, to June 30, 2022.

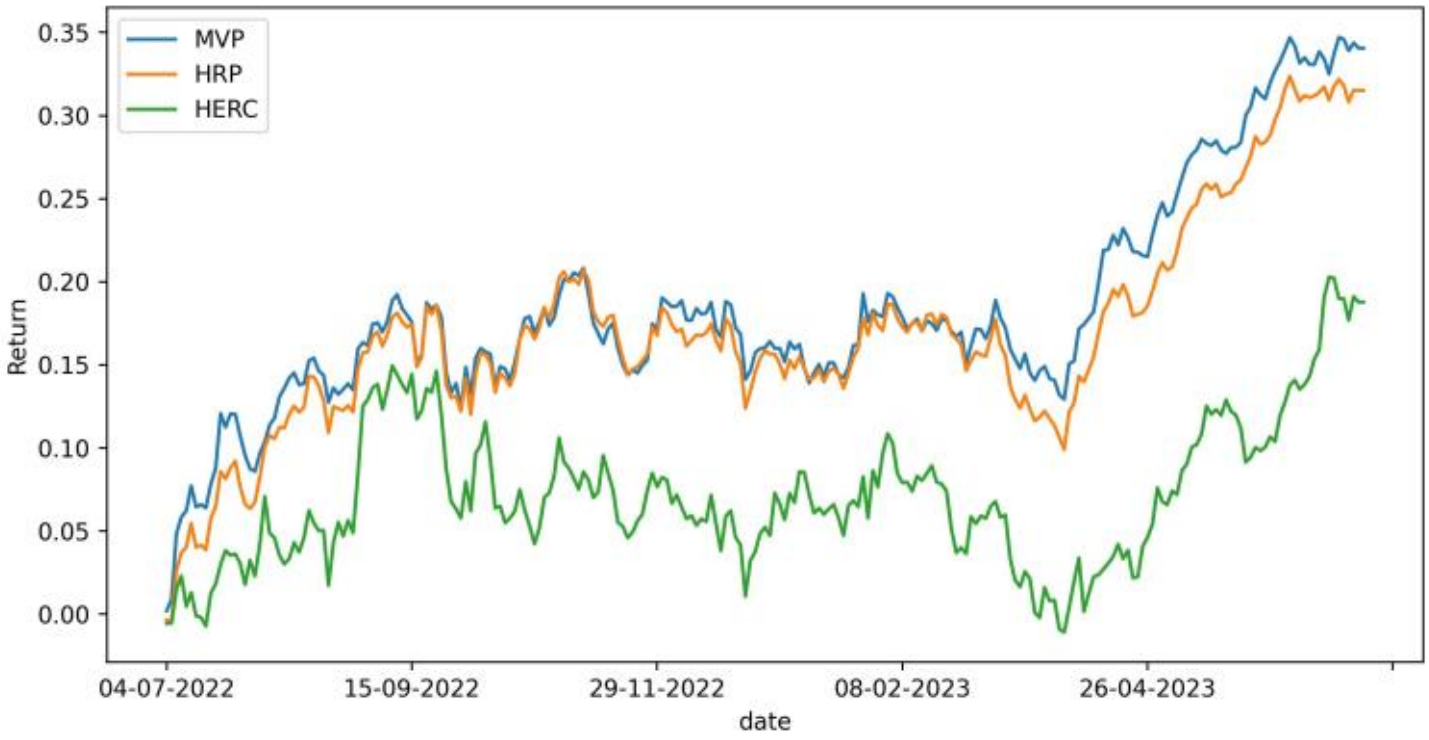

**Figure 1.6.** The cumulative returns yielded by the MVP, HRP, and HERC portfolios of the stocks from the *auto* sector on the test data from July 1, 2022, to June 30, 2023.

**TABLE 1.2.** THE PERFORMANCE RESULTS OF THE PORTFOLIOS OF THE AUTO SECTOR STOCKS

| Period | MVP Portfolio | | | HRP Portfolio | | | HERC Portfolio | | |
|---|---|---|---|---|---|---|---|---|---|
| | Ret (%) | Vol (%) | SR | Ret (%) | Vol (%) | SR | Ret (%) | Vol (%) | SR |
| Training | 26.32 | **25.35** | **1.0385** | 26.28 | 27.18 | 0.9670 | **29.38** | 41.92 | 0.7009 |
| Test | **35.03** | 16.07 | **2.1795** | 32.41 | **15.39** | 2.1061 | 19.32 | 20.52 | 0.9416 |

***Banking sector:*** As per the report published by the NSE on June 30, 2022, the ten stocks with the largest free-float market capitalization in the *banking* sector and their contributions (in percent) to the overall index of the sector are as follows: (i) HDFC Bank (HDFCBANK): 28.42%, (ii) ICICI Bank (ICICIBANK): 24.04%, (iii) State Bank of India (SBIN): 9.89%, (iv) Kotak Mahindra Bank (KOTAKBANK): 9.40%, (v) Axis Bank (AXISBANK): 9.35%, (vi) IndusInd Bank (INDUSINDBK): 6.74%, (vii) Bank of Baroda (BANKBARODA): 2.75%, (viii) AU Small Finance Bank (AUBANK): 2.56%, (ix) Federal Bank (FEDERALBNK): 2.33%, and (x) IDFC First Bank (IDFCFIRSTB): 1.98% (NSE Website). The ticker names of the stocks are mentioned in parentheses.

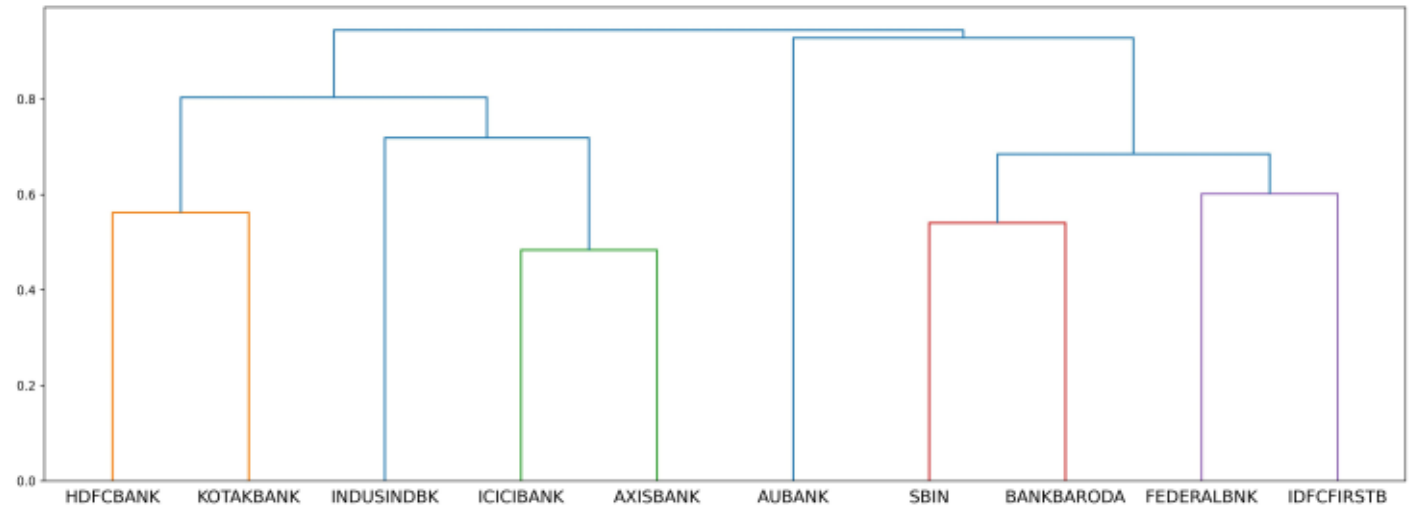

**Figure 1.7.** The dendrogram of the agglomerative clustering of the stocks from the *banking* sector is created based on the historical stock prices from July 1, 2019, to June 30, 2022.

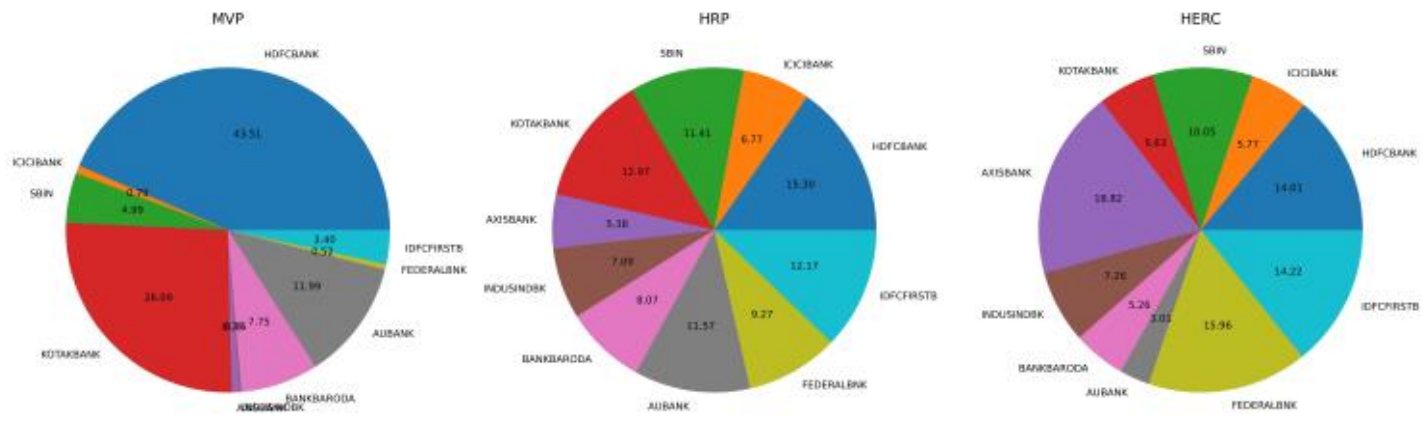

**Figure 1.8.** The allocation of weights done by the MVP, HRP, and HERC algorithms for the stocks of the *banking* sector based on the historical stock prices from July 1, 2019, to June 30, 2022.

The dendrogram of the clustering of the stocks of the banking sector is shown in Figure 1.7. The cluster containing the stocks ICICIBANK and AXISBANK is the most compact, while the one containing FEDERALBNK and IDFCFIRSTB is the least homogeneous.

**TABLE 1.3.** THE PORTFOLIO COMPOSITIONS OF THE BANKING SECTOR
(PERIOD: JULY 1, 2019 – JUNE 30, 2022)

| Stock | MVP Portfolio | HRP Portfolio | HERC Portfolio |
|---|---|---|---|
| HDFCBANK | 0.4351 | 0.1530 | 0.1401 |
| ICICIBANK | 0.0079 | 0.0677 | 0.0577 |
| SBIN | 0.0499 | 0.1141 | 0.1005 |
| KOTAKBANK | 0.2609 | 0.1297 | 0.0563 |
| AXISBANK | 0.0079 | 0.0538 | 0.1882 |
| INDUSINDBK | 0.0015 | 0.0709 | 0.0726 |
| BANKBARODA | 0.0775 | 0.0807 | 0.0526 |
| AUBANK | 0.1199 | 0.1157 | 0.0301 |
| FEDERALBNK | 0.0052 | 0.0927 | 0.1596 |
| IDFCFIRSTB | 0.0340 | 0.1217 | 0.1422 |

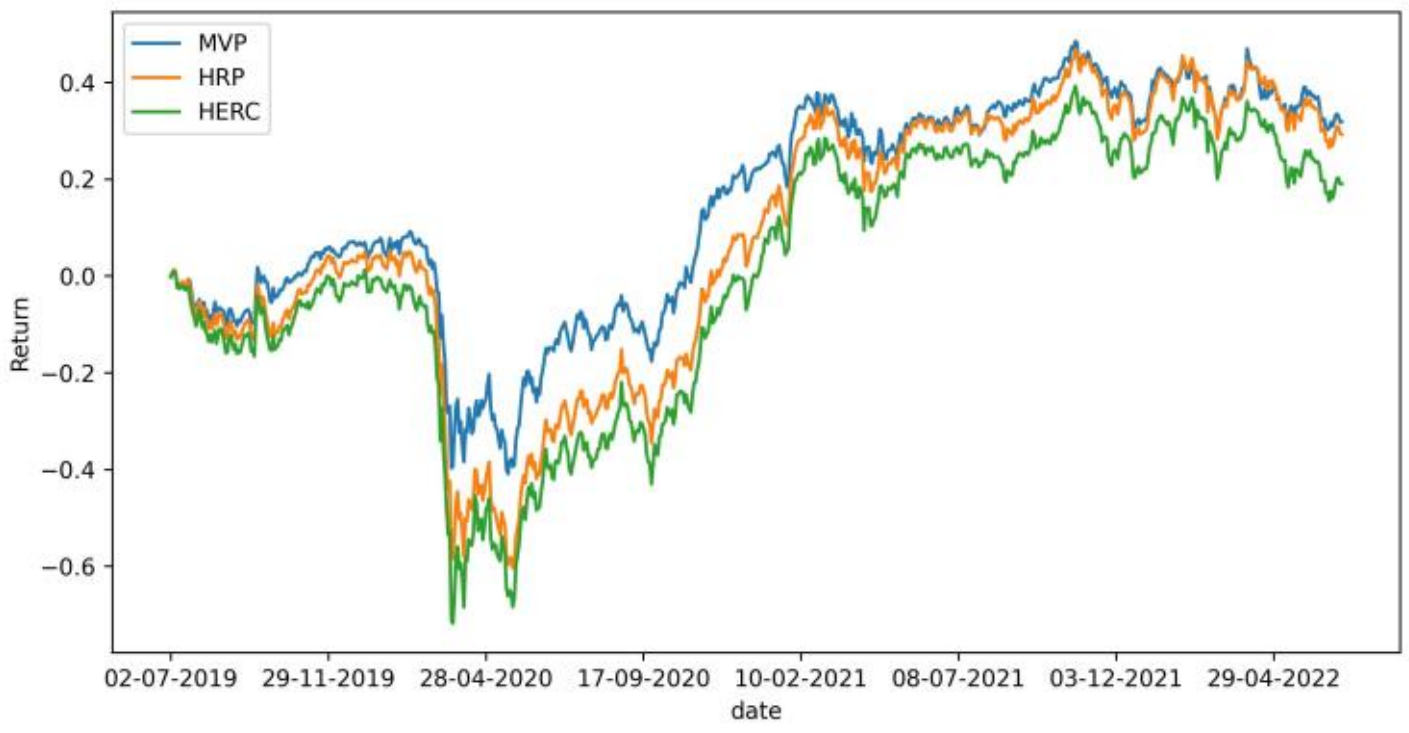

**Figure 1.9.** The cumulative returns yielded by the MVP, HRP, and HERC portfolios of the stocks from the *banking* sector on the training data from July 1, 2019, to June 30, 2022.

Figure 1.8 depicts the weight allocation done by the three portfolios to the banking sector stocks. Table 1.3 shows the weight allocations for the three portfolios in tabular format.

The cumulative returns yielded by the portfolios over the training and the test periods are depicted in Figure 1.9 and Figure 1.10, respectively. Table 1.4 presents the cumulative returns, annual volatilities, and the Sharpe ratios of the three portfolios of the banking sector stocks over the training and the test periods. The highest cumulative return, the lowest volatility, and the maximum Sharpe ratio over the portfolio training and test periods are indicated in red.

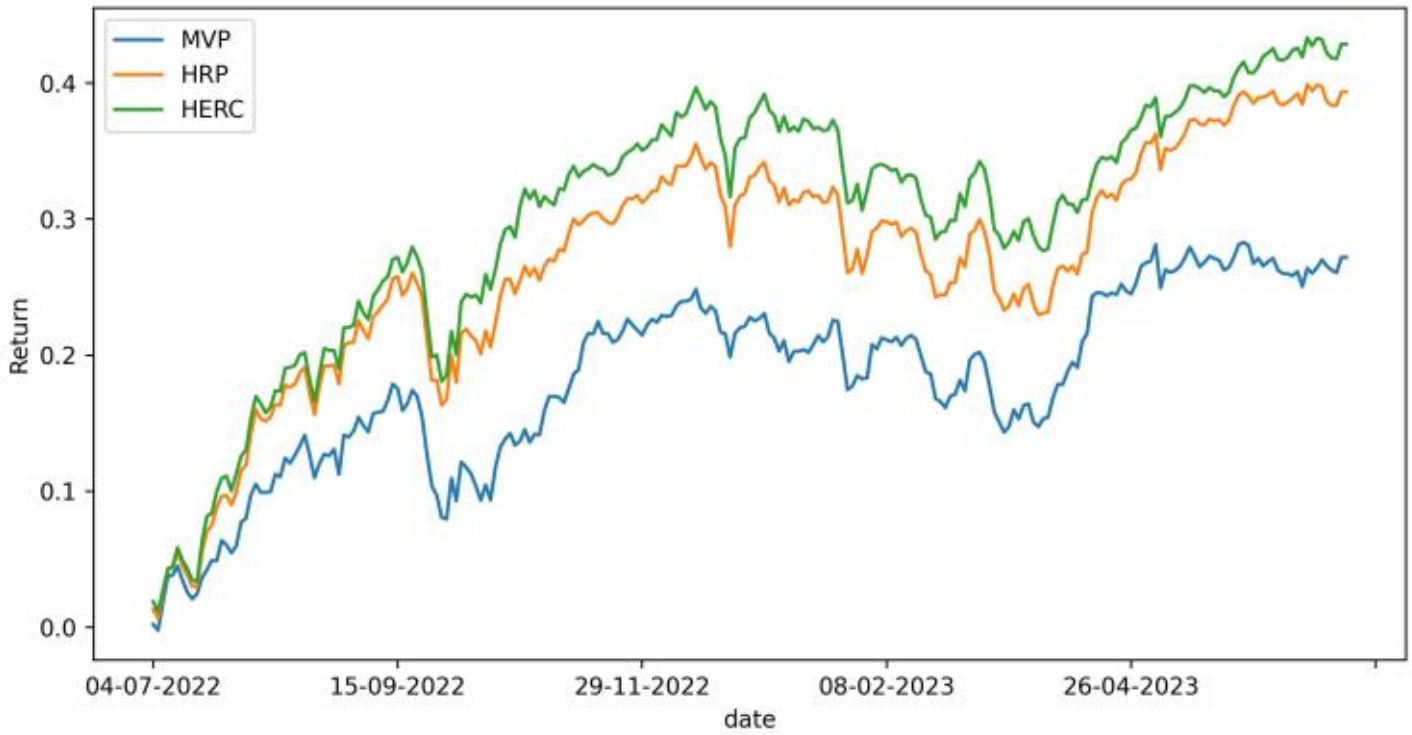

**Figure 1.10.** The cumulative returns yielded by the MVP, HRP, and HERC portfolios of the stocks from the *banking* sector on the test data from July 1, 2022, to June 30, 2023.

**TABLE 1.4.** THE PERFORMANCE RESULTS OF THE PORTFOLIOS OF THE BANKING SECTOR STOCKS

| **Period** | **MVP Portfolio** | | | **HRP Portfolio** | | | **HERC Portfolio** | | |
|---|---|---|---|---|---|---|---|---|---|
| | **Ret (%)** | **Vol (%)** | **SR** | **Ret (%)** | **Vol (%)** | **SR** | **Ret (%)** | **Vol (%)** | **SR** |
| Training | **10.81** | **28.79** | **0.3755** | 9.91 | 31.92 | 0.3104 | 6.45 | 34.12 | 0.1890 |
| Test | 27.98 | **15.88** | 1.6436 | 40.49 | 17.73 | 2.0356 | **44.10** | 18.82 | **2.1931** |

***Financial Services sector:*** Based on the NSE's report published on June 30, 2022, the ten stocks that have the maximum free-float market capitalization and their respective contributions to the overall index to the *financial services* sector are as follows: (i) HDFC Bank (HDFCBANK): 32.08%, (ii) ICICI Bank (ICICIBANK): 21.53%, (iii) Kotak Mahindra Bank (KOTAKBANK): 8.41%, (iv) Axis Bank (AXISBANK): 8.29%, (v) State Bank of India (SBIN): 7.54%, (vi) Bajaj Finance (BAJFINANCE): 6.17%, (vii) Bajaj Finserv (BAJAJFINSV): 2.74%, (viii) HDFC Life Insurance Company (HDFCLIFE): 2.20%, (ix) SBI Life Insurance Company (SBILIFE): 1.83%, and (x) Shriram Finance (SHRIRAMFIN): 1.60% (NSE Website). The ticker names, the unique identifier for the stocks, are mentioned in parentheses.

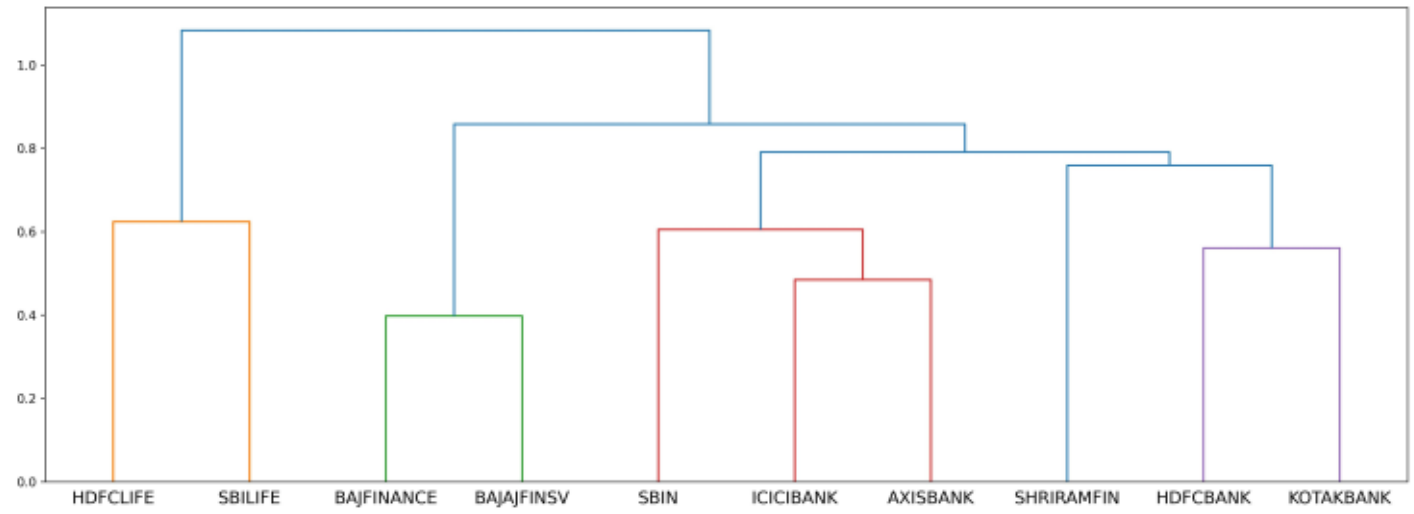

**Figure 1.11.** The dendrogram of the agglomerative clustering of the stocks from the *financial services* sector is created based on the historical stock prices from July 1, 2019, to June 30, 2022.

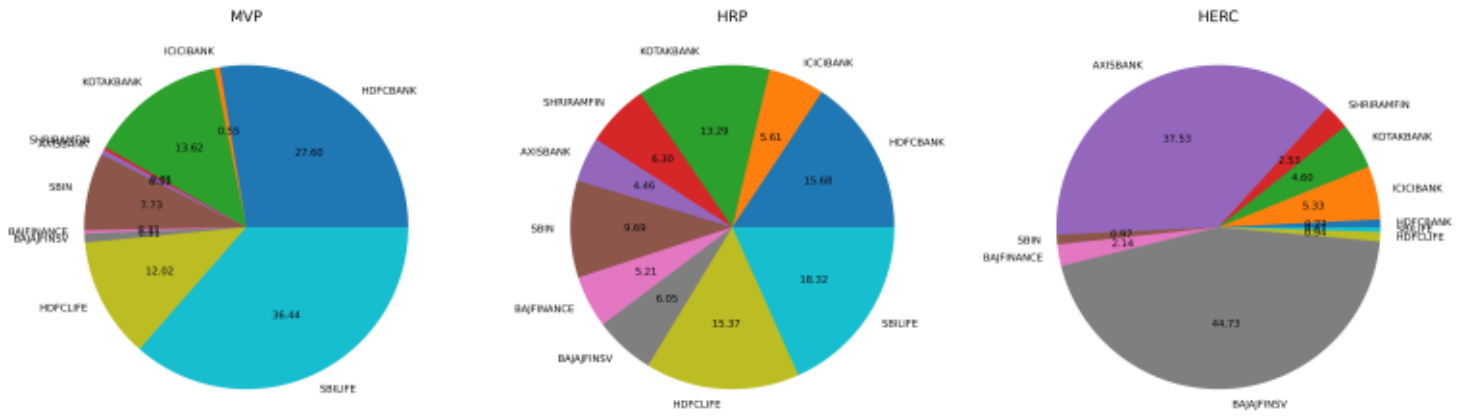

**Figure 1.12.** The allocation of weights done by the MVP, HRP, and HERC algorithms for the stocks from the *financial services* sector based on the historical stock prices from July 1, 2019, to June 30, 2022.

The dendrogram of the clustering of the stocks of the financial services sector is shown in Figure 1.11. The cluster containing the stocks BAJFINANCE and BAJAJFINSV is the most compact, while the one containing HDFCLIFE and SBILIFE is the least homogeneous. Figure 1.12 depicts the weight allocation done by the three portfolios to the financial services sector stocks. Table 1.5 shows the weight allocations for the three portfolios in tabular format.

**TABLE 1.5.** THE PORTFOLIO COMPOSITIONS OF THE FINANCIAL SERVICES SECTOR
(PERIOD: JULY 1, 2019 – JUNE 30, 2022)

| Stock | MVP Portfolio | HRP Portfolio | HERC Portfolio |
|---|---|---|---|
| HDFCBANK | 0.2760 | 0.1568 | 0.0077 |
| ICICIBANK | 0.0055 | 0.0561 | 0.0533 |
| KOTAKBANK | 0.1362 | 0.1329 | 0.0460 |
| AXISBANK | 0.0051 | 0.0446 | 0.3753 |
| SBIN | 0.0773 | 0.0969 | 0.0097 |
| BAJFINANCE | 0.0031 | 0.0521 | 0.0214 |
| BAJAJFINSV | 0.0091 | 0.0605 | 0.4473 |
| HDFCLIFE | 0.1202 | 0.1537 | 0.0094 |
| SBILIFE | 0.3644 | 0.1832 | 0.0047 |
| SHRIRAMFIN | 0.0031 | 0.0630 | 0.0253 |

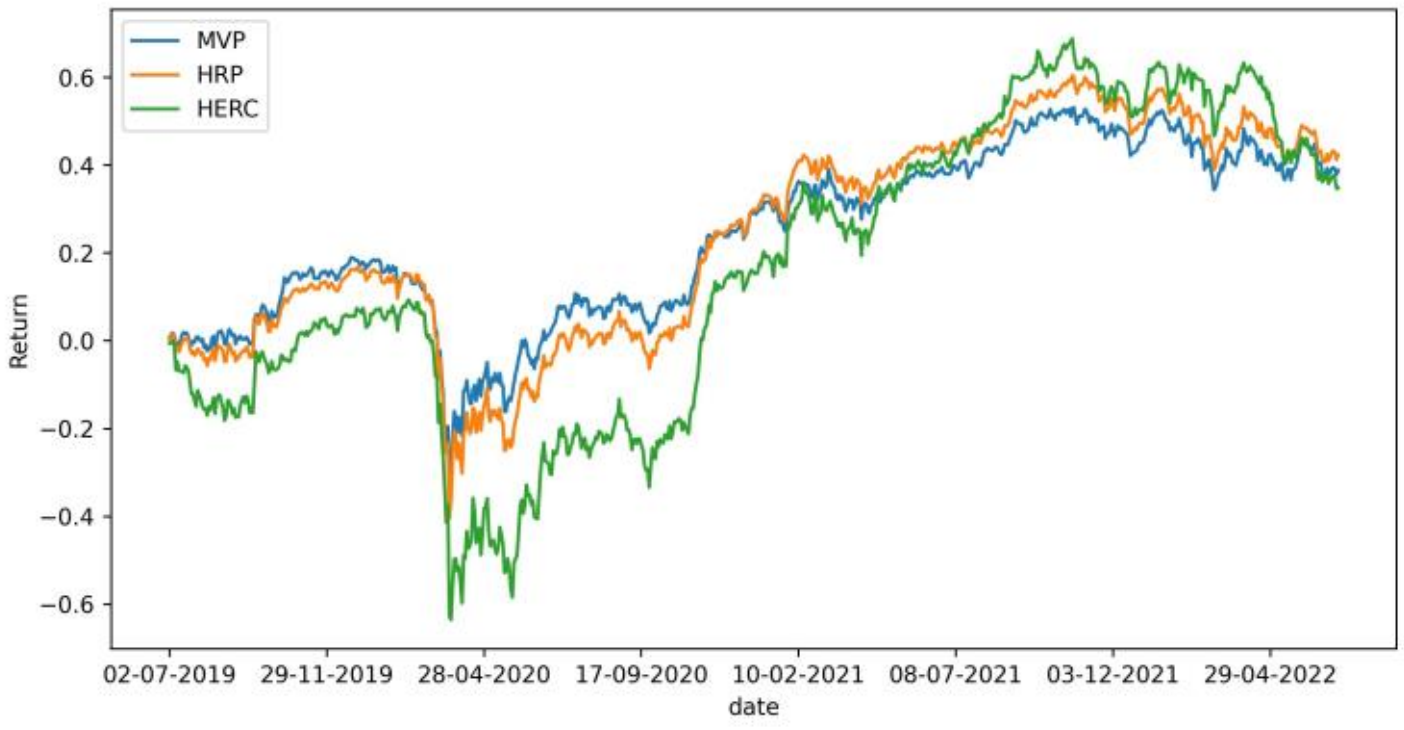

**Figure 1.13.** The cumulative returns yielded by the MVP, HRP, and HERC portfolios of the stocks from the *financial services* sector on the training data from July 1, 2019, to June 30, 2022.

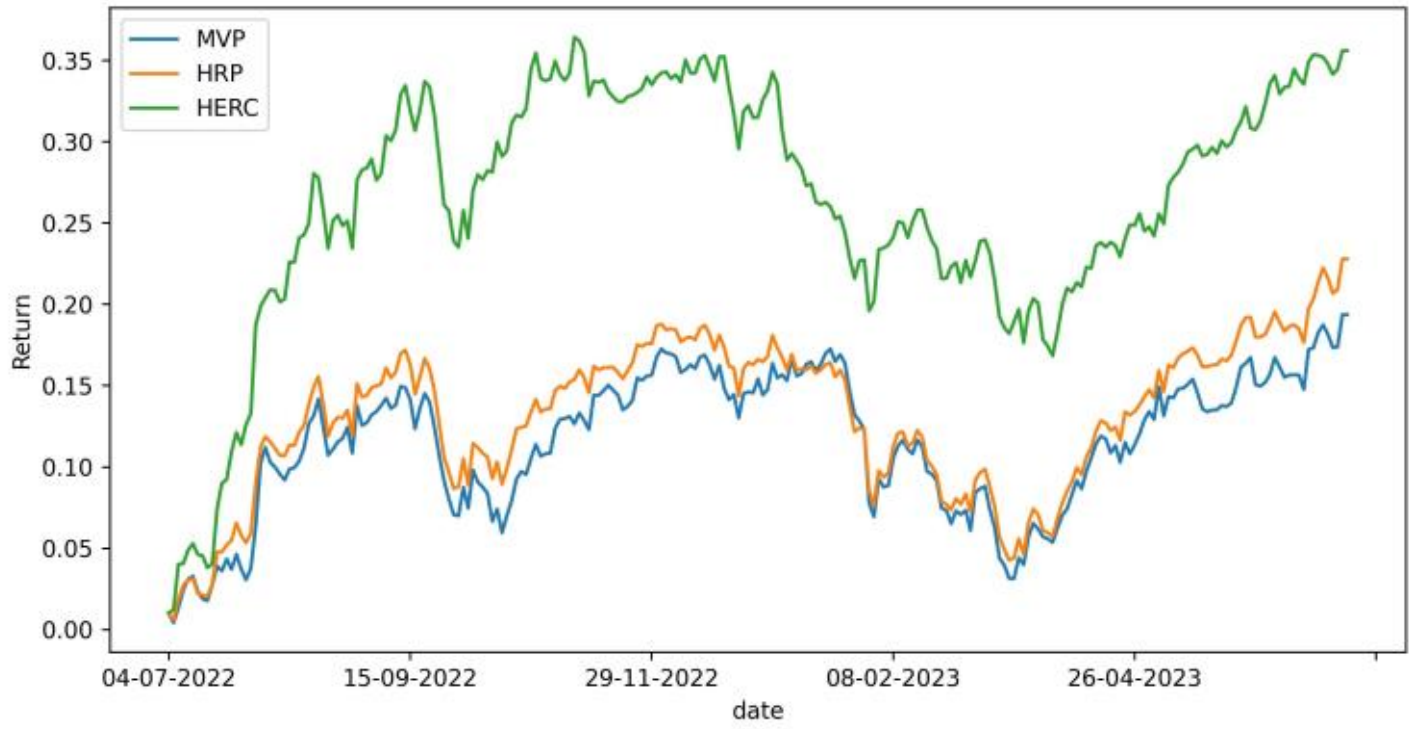

**Figure 1.14.** The cumulative returns yielded by the MVP, HRP, and HERC portfolios of the stocks from the *financial services* on the test data from July 1, 2022, to June 30, 2023.

The cumulative returns yielded by the portfolios over the training and the test periods are depicted in Figure 1.13 and Figure 1.14, respectively. Table 1.6 presents the cumulative returns, annual volatilities, and the Sharpe ratios of the three portfolios of the financial services sector stocks over the training and the test periods. The highest cumulative return, the lowest volatility, and the maximum Sharpe ratio over the portfolio training and test periods are indicated in red.

**TABLE 1.6.** THE PERFORMANCE RESULTS OF THE PORTFOLIOS OF THE FINANCIAL SERVICES SECTOR STOCKS

| Period | MVP Portfolio | | | HRP Portfolio | | | HERC Portfolio | | |
|---|---|---|---|---|---|---|---|---|---|
| | Ret (%) | Vol (%) | SR | Ret (%) | Vol (%) | SR | Ret (%) | Vol (%) | SR |
| Training | 6.81 | **24.02** | 0.2834 | **10.48** | 25.88 | **0.4051** | 9.56 | 32.25 | 0.2956 |
| Test | **19.79** | **13.39** | 1.4775 | 22.19 | 13.97 | 1.5889 | 30.71 | 18.72 | **1.6407** |

***Consumer Durables sector:*** As NSE's report published on June 30, 2022, the ten stocks from the *consumer durables* sector that have the largest free-float market capitalization and their contributions (in percent) to the overall index of the sector are as follows: (i) Titan Company (TITAN): 32.75%, (ii) Havells India (HAVELLS): 14.95%, (iii) Crompton Greaves Consumer Electricals (CROMPTON): 8.40%,

(iv) Voltas (VOLTAS): 7.96%, (v) Dixon Technologies (DIXON): 6.80%, (vi) Kajaria Ceramics (KAJARIACER): 5.32%, (vii) Bata India (BATAINDIA): 4.99%, (viii) Blue Star (BLUESTARCO): 3.98%, (ix) Rajesh Exports (RAJESHEXPO): 3.08%, and (x) Relaxo Footwears (RELAXO): 3.05% (NSE Website). The ticker names of the stocks are mentioned in parentheses.

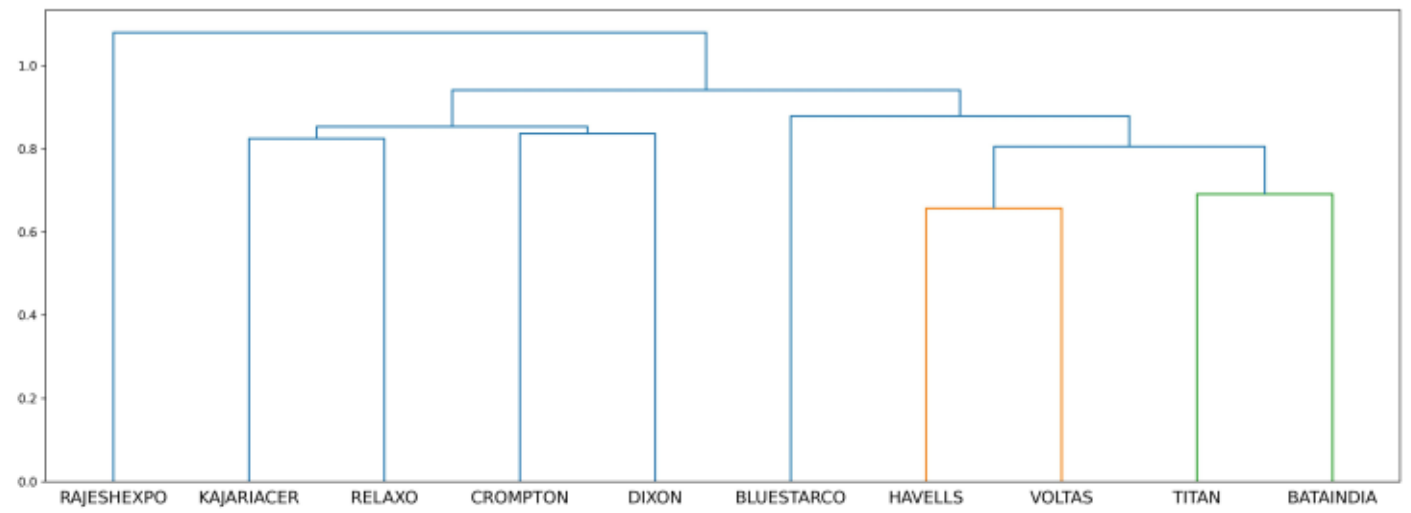

**Figure 1.15.** The dendrogram of the agglomerative clustering of the stocks from the *consumer durables* sector is created based on the historical stock prices from July 1, 2019, to June 30, 2022.

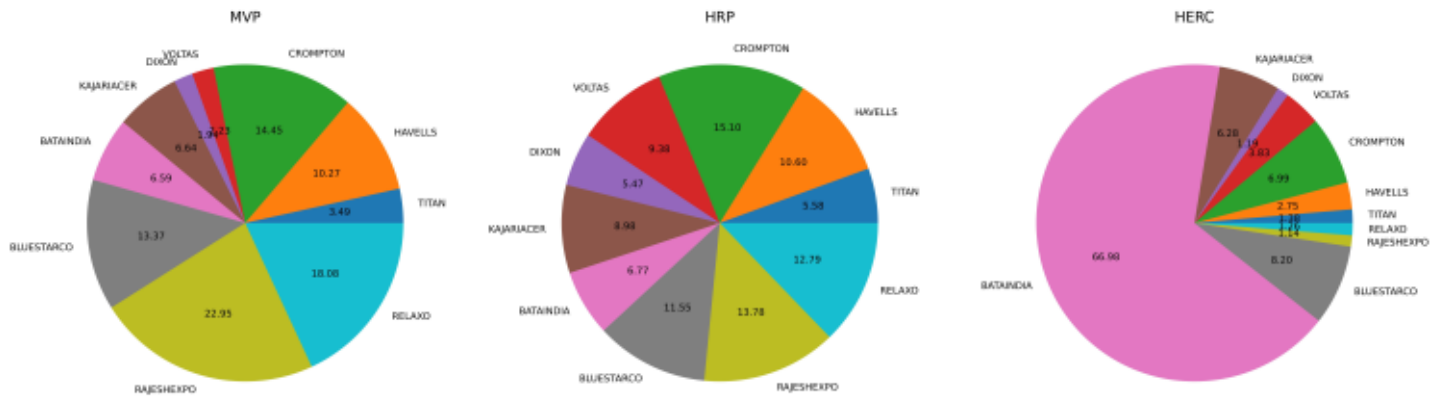

**Figure 1.16.** The allocation of weights done by the MVP, HRP, and HERC algorithms for the stocks of the *consumer durables* sector based on the historical stock prices from July 1, 2019, to June 30, 2022.

The dendrogram of the clustering of the stocks of the consumer durables sector is shown in Figure 1.15. The cluster containing the stocks HAVELLS and VOLTAS is the most compact, while the one containing CROMPTON and DIXON is the least homogeneous. Figure 1.16 depicts the weight allocation done by the three portfolios to the consumer durables sector stocks. Table 1.7 shows the weight allocations for the three portfolios in tabular format.

**TABLE 1.7.** THE PORTFOLIO COMPOSITIONS OF THE CONSUMER DURABLES SECTOR (PERIOD: JULY 1, 2019 – JUNE 30, 2022)

| Stock | MVP Portfolio | HRP Portfolio | HERC Portfolio |
|---|---|---|---|
| TITAN | 0.0349 | 0.0558 | 0.0138 |
| HAVELLS | 0.1027 | 0.1060 | 0.0275 |
| CROMPTON | 0.1445 | 0.1510 | 0.0699 |
| VOLTAS | 0.0223 | 0.0938 | 0.0383 |
| DIXON | 0.0194 | 0.0547 | 0.0119 |
| KAJARIACER | 0.0664 | 0.0898 | 0.0628 |
| BATAINDIA | 0.0659 | 0.0677 | 0.6698 |
| BLUESTARCO | 0.1337 | 0.1155 | 0.0820 |
| RAJESHEXPO | 0.2295 | 0.1378 | 0.0114 |
| RELAXO | 0.1808 | 0.1279 | 0.0126 |

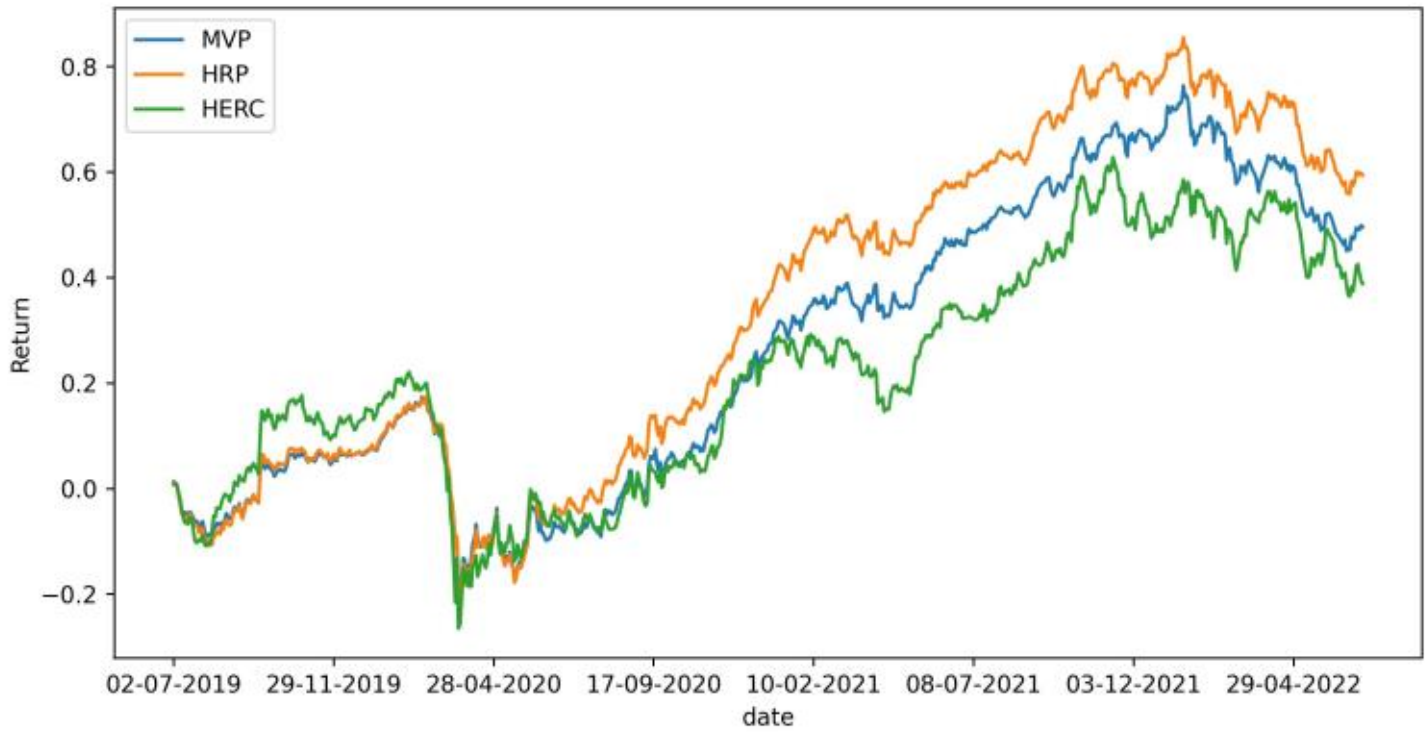

**Figure 1.17.** The cumulative returns yielded by the MVP, HRP, and HERC portfolios of the stocks from the *consumer durables* sector on the training data from July 1, 2019, to June 30, 2022.

The cumulative returns yielded by the portfolios over the training and the test periods are depicted in Figure 1.17 and Figure 1.18,

respectively. Table 1.8 presents the cumulative returns, annual volatilities, and the Sharpe ratios of the three portfolios of the consumer durables sector stocks over the training and the test periods. The highest cumulative return, the lowest volatility, and the maximum Sharpe ratio over the portfolio training and test periods are indicated in red.

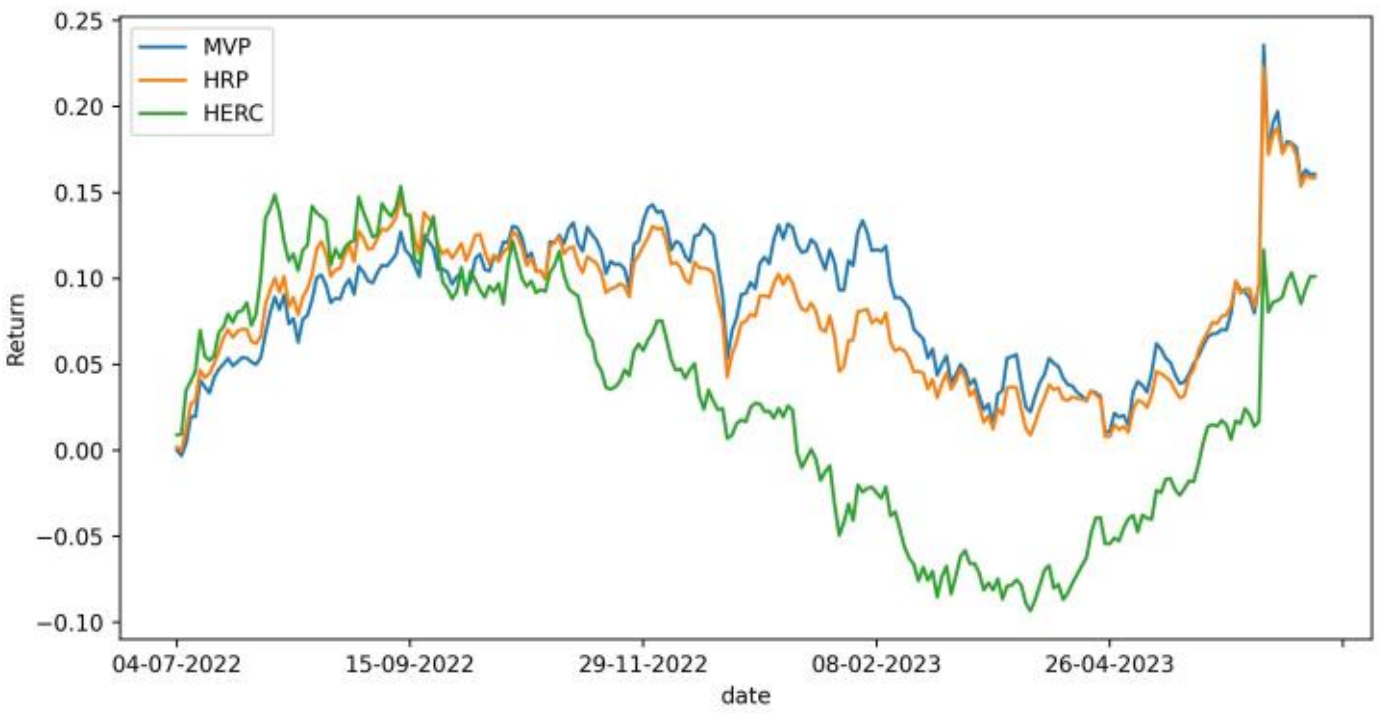

**Figure 1.18.** The cumulative returns yielded by the MVP, HRP, and HERC portfolios of the stocks from the *consumer durables* sector on the test data from July 1, 2022, to June 30, 2023.

**TABLE 1.8.** THE PERFORMANCE RESULTS OF THE PORTFOLIOS OF THE CONSUMER DURABLES SECTOR STOCKS

| Period | MVP Portfolio | | | HRP Portfolio | | | HERC Portfolio | | |
|---|---|---|---|---|---|---|---|---|---|
| | Ret (%) | Vol (%) | SR | Ret (%) | Vol (%) | SR | Ret (%) | Vol (%) | SR |
| Training | 16.80 | **20.05** | 0.8380 | **20.13** | 20.63 | **0.9757** | 13.16 | 26.33 | 0.4998 |
| Test | **16.52** | 21.18 | 0.7802 | 16.30 | 18.92 | **0.8615** | 10.41 | **18.56** | 0.5610 |

***FMCG sector:*** Based on the NSE's report published on June 30, 2022, the ten stocks that have the maximum free-float market capitalization in the FMCG sector, and their contributions to the overall index of the sector are as follows: (i) ITC (ITC): 32.04%, (ii) Hindustan Unilever (HINDUNILVR): 21.71%, (iii) Nestle India (NESTLEIND): 7.64%, (iv) Britannia Industries (BRITANNIA): 6.24%, (v) Tata Consumer Products (TATACONSUM): 5.63%, (vi) Godrej Consumer Products (GODREJCP): 4.32%, (vii) Varun Beverages (VBL): 4.15%, (viii) Dabur India (DABUR): 3.71%, (ix)

United Spirits (MCDOWELL-N): 3.26%, and (x) Marico (MARICO): 3.20% (NSE Website). The ticker names of the stocks are mentioned in parentheses. The ticker names serve as the unique identifiers for the stocks on a given stock exchange.

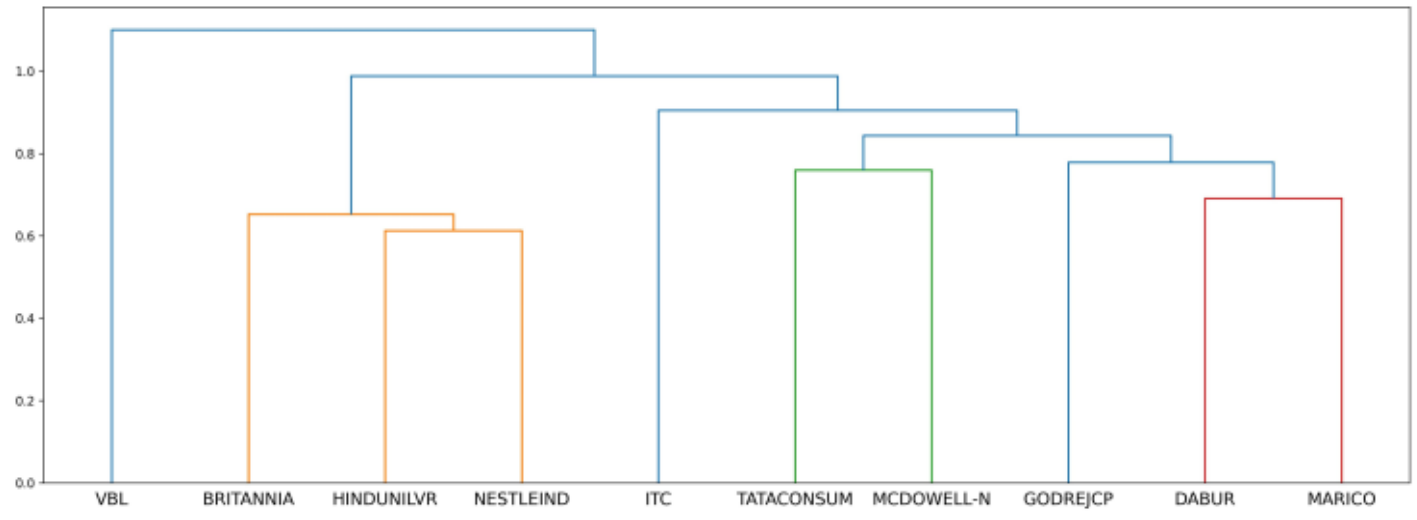

**Figure 1.19.** The dendrogram of the agglomerative clustering of the stocks from the FMCG sector is created based on the historical stock prices from July 1, 2019, to June 30, 2022.

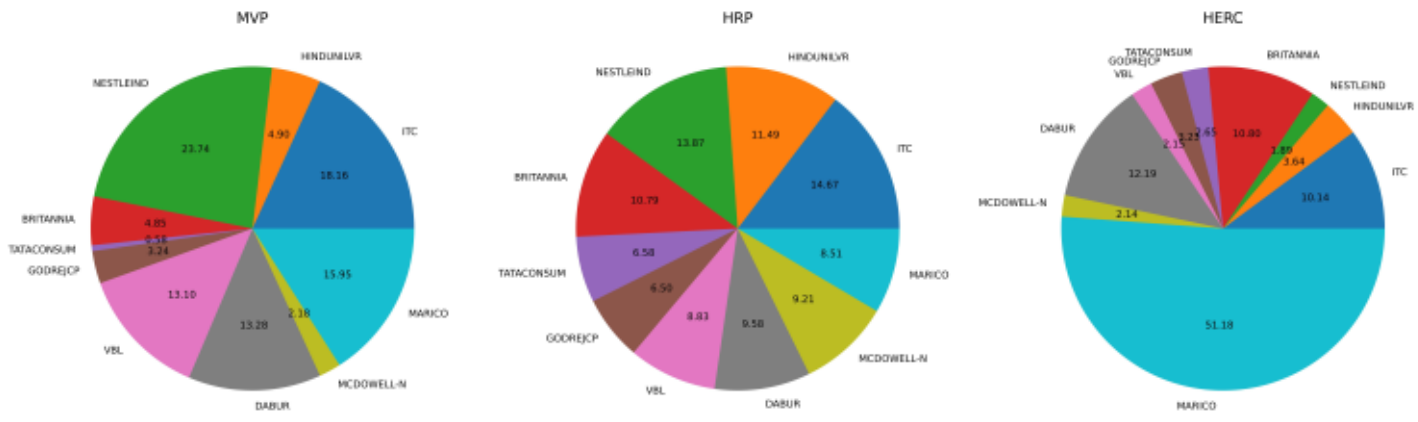

**Figure 1.20.** The allocation of weights done by the MVP, HRP, and HERC algorithms for the stocks of the FMCG sector based on the historical stock prices from July 1, 2019, to June 30, 2022.

The dendrogram of the clustering of the stocks of the FMCG sector is shown in Figure 1.19. The cluster containing the stocks HINDUNILVR and NESTLEIND is the most compact, while the one containing TATACONSUM and MCDOWELL-N is the least homogeneous. Figure 1.20 depicts the weight allocation done by the three portfolios to the FMCG sector stocks. Table 1.9 shows the weight allocations for the three portfolios in tabular format.

**TABLE 1.9.** THE PORTFOLIO COMPOSITIONS OF THE FMCG (PERIOD: JULY 1, 2019 – JUNE 30, 2022)

| Stock | MVP Portfolio | HRP Portfolio | HERC Portfolio |
|---|---|---|---|
| ITC | 0.1816 | 0.14673 | 0.1014 |
| HINDUNILVR | 0.0490 | 0.1149 | 0.0364 |
| NESTLEIND | 0.2374 | 0.1387 | 0.0189 |
| BRITANNIA | 0.0485 | 0.1079 | 0.1080 |
| TATACONSUM | 0.0058 | 0.0658 | 0.0265 |
| GODREJCP | 0.0324 | 0.0650 | 0.0323 |
| VBL | 0.1310 | 0.0883 | 0.0215 |
| DABUR | 0.1328 | 0.0958 | 0.1219 |
| MCDOWELL-N | 0.0218 | 0.0921 | 0.0214 |
| MARICO | 0.1595 | 0.0851 | 0.5118 |

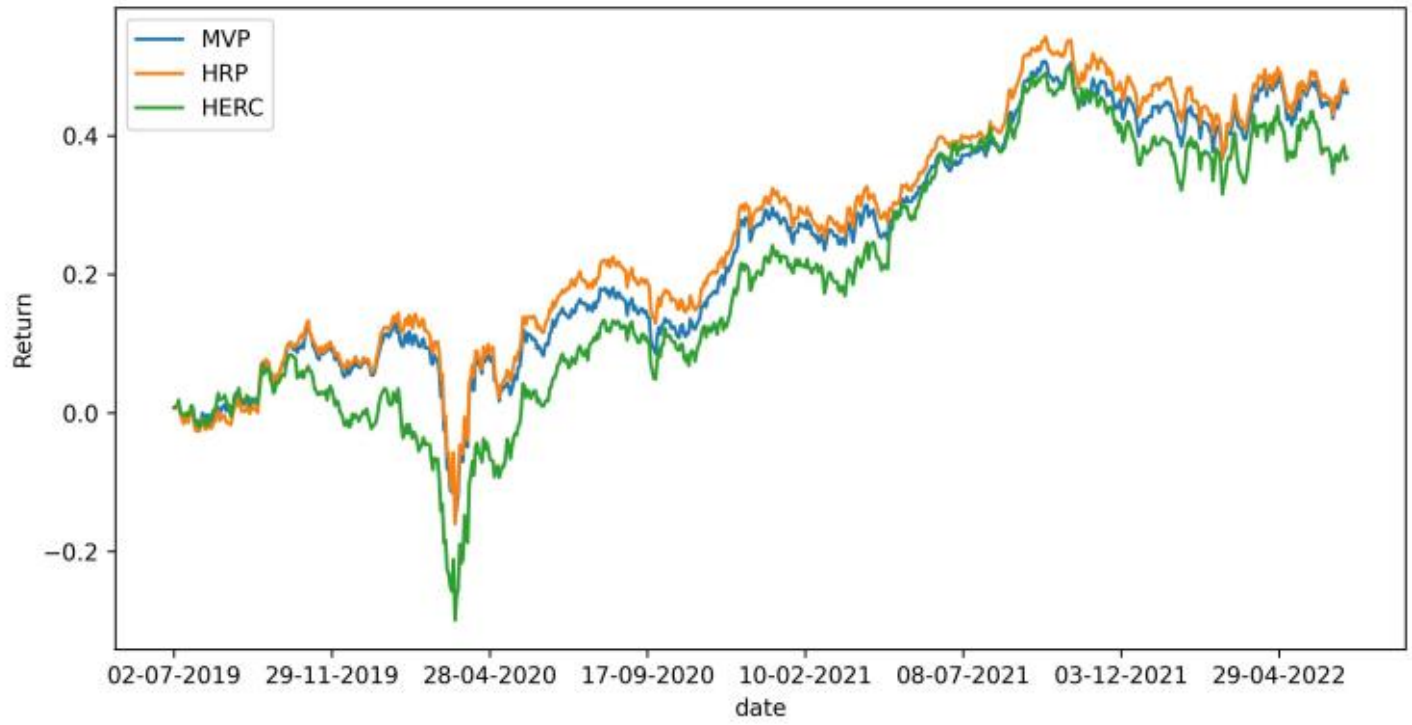

**Figure 1.21.** The cumulative returns yielded by the MVP, HRP, and HERC portfolios of the stocks from the FMCG sector on the training data from July 1, 2019, to June 30, 2022.

The cumulative returns yielded by the portfolios over the training and the test periods are depicted in Figure 1.21 and Figure 1.22, respectively. Table 1.10 presents the cumulative returns, annual volatilities, and the Sharpe ratios of the three portfolios of the FMCG sector stocks over the training and the test periods. The highest cumulative return, the lowest volatility, and the maximum Sharpe ratio over the portfolio training and test periods are indicated in red.

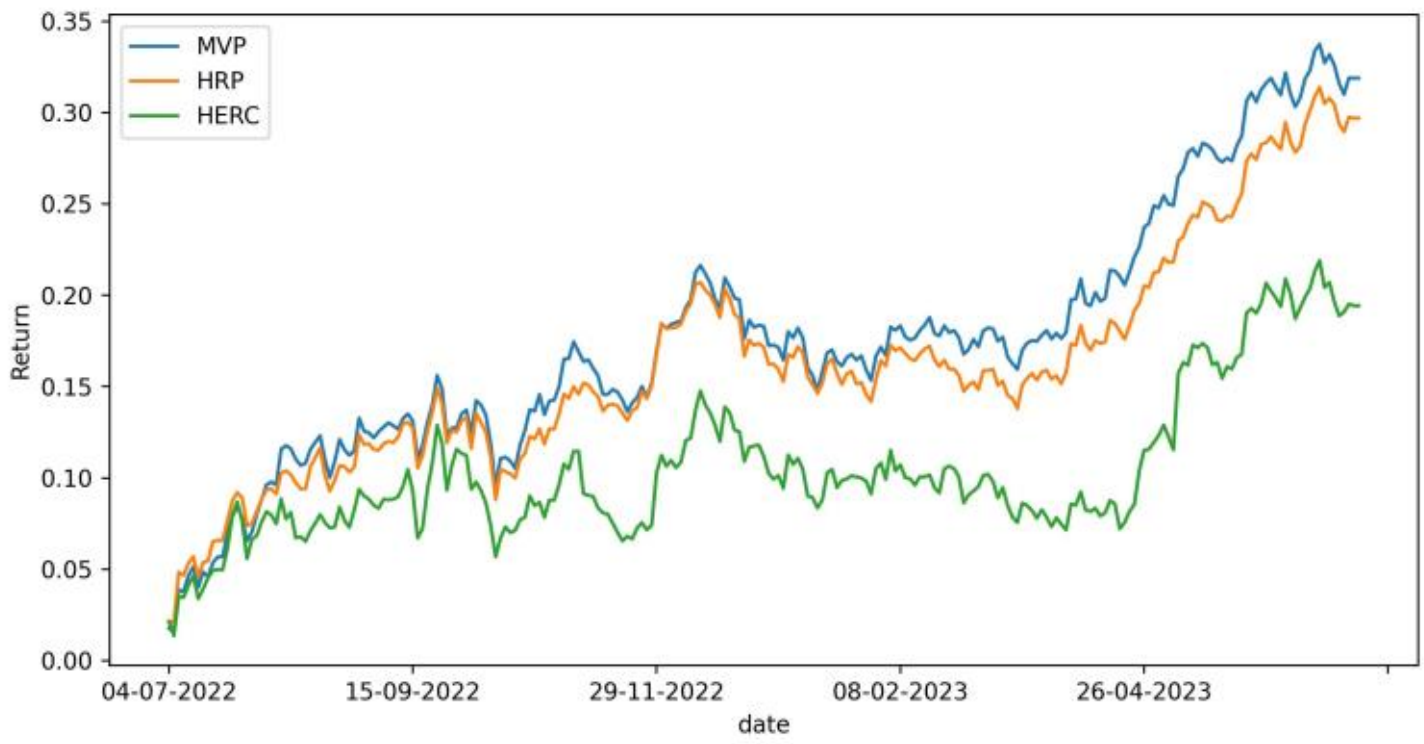

**Figure 1.22.** The cumulative returns yielded by the MVP, HRP, and HERC portfolios of the stocks from the FMCG sector on the test data from July 1, 2022, to June 30, 2023.

**TABLE 1.10.** THE PERFORMANCE RESULTS OF THE PORTFOLIOS OF THE FMCG SECTOR STOCKS

| **Period** | **MVP Portfolio** | | | **HRP Portfolio** | | | **HERC Portfolio** | | |
|---|---|---|---|---|---|---|---|---|---|
| | **Ret (%)** | **Vol (%)** | **SR** | **Ret (%)** | **Vol (%)** | **SR** | **Ret (%)** | **Vol (%)** | **SR** |
| Training | 15.65 | **18.33** | **0.8539** | **15.83** | 18.97 | 0.8347 | 12.47 | 20.53 | 0.6071 |
| Test | **32.80** | 12.75 | **2.5717** | 30.54 | **12.13** | 2.5170 | 19.98 | 14.30 | 1.3977 |

***Information Technology (IT) sector:*** As per the report published by the NSE on June 30, 2022, the ten stocks with the largest free-float market capitalization and their respective contributions (in percent) to the overall index of the IT sector are as follows: (i) Infosys (INFY): 27.32%, (ii) Tata Consultancy Services (TCS): 26.32%, (iii) Wipro (WIPRO): 9.44%, (iv) Tech Mahindra (TECHM): 9.31, (v) HCL Technologies (HCLTECH): 8.87%, (vi) LTIMindtree (LTIM): 7.05%, (vii) Persistent Systems (PERSISTENT): 3.84%, (viii) Coforge (COFORGE): 3.12%, (ix) MphasiS (MPHASIS): 2.99%, and (x) L&T Technology Services (LTTS): 1.74% (NSE Website). The ticker names of the stocks are mentioned in parentheses against their names. The ticker names are the unique identifiers of the stocks in a given stock exchange.

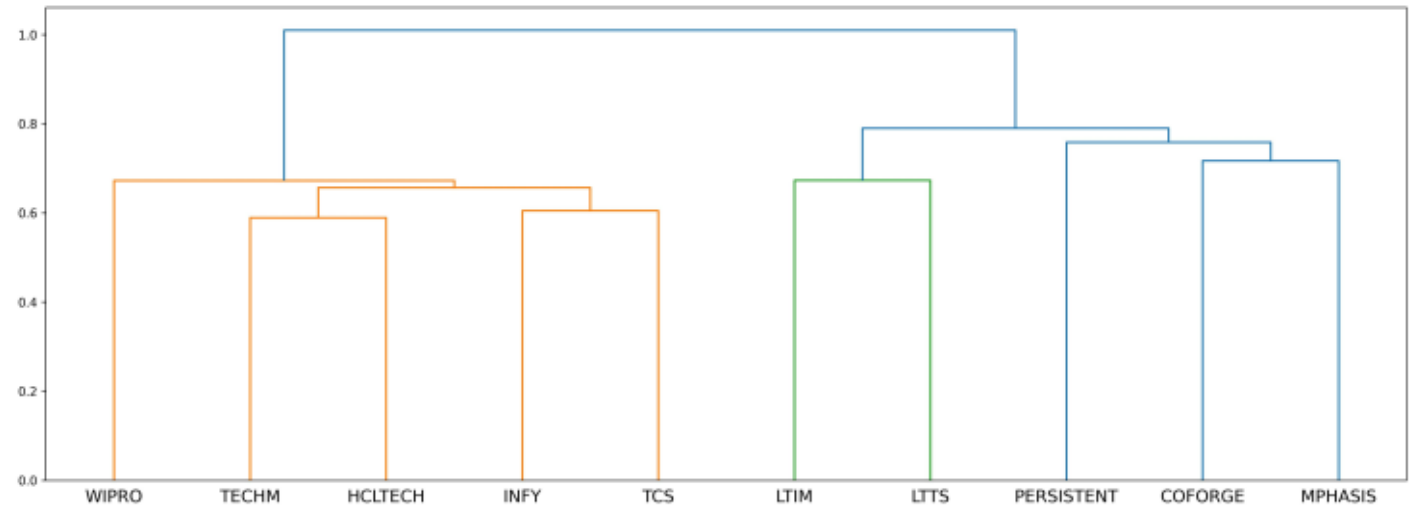

**Figure 1.23.** The dendrogram of the agglomerative clustering of the stocks from the IT sector is created based on the historical stock prices from July 1, 2019, to June 30, 2022.

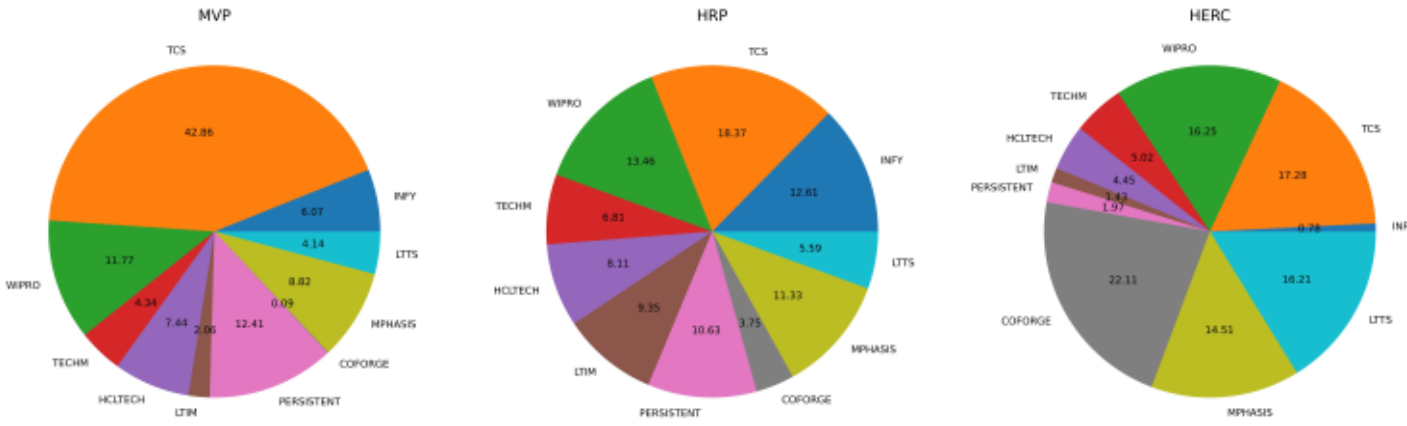

**Figure 1.24.** The allocation of weights done by the MVP, HRP, and HERC algorithms for the stocks of the IT sector based on the historical stock prices from July 1, 2019, to June 30, 2022.

The dendrogram of the clustering of the stocks of the IT sector is shown in Figure 1.23. The cluster containing the stocks TECHM and HCLTECH is the most compact, while the one containing COFORGE and MPHASIS is the least homogeneous. Figure 1.24 depicts the weight allocation done by the three portfolios to the IT sector stocks. Table 1.11 shows the weight allocations for the three portfolios in tabular format.

**TABLE 1.11.** THE PORTFOLIO COMPOSITIONS OF THE IT (PERIOD: JULY 1, 2019 – JUNE 30, 2022)

| Stock | MVP Portfolio | HRP Portfolio | HERC Portfolio |
|---|---|---|---|
| INFY | 0.0607 | 0.1261 | 0.0078 |
| TCS | 0.4286 | 0.1837 | 0.1728 |
| WIPRO | 0.1177 | 0.1346 | 0.1625 |
| TECHM | 0.0434 | 0.0681 | 0.0502 |
| HCLTECH | 0.0744 | 0.0811 | 0.0445 |
| LTIM | 0.0206 | 0.0935 | 0.0143 |
| PERSISTENT | 0.1241 | 0.1063 | 0.0197 |
| COFORGE | 0.0009 | 0.0375 | 0.2211 |
| MPHASIS | 0.0882 | 0.1133 | 0.1451 |
| LTTS | 0.0414 | 0.0559 | 0.1621 |

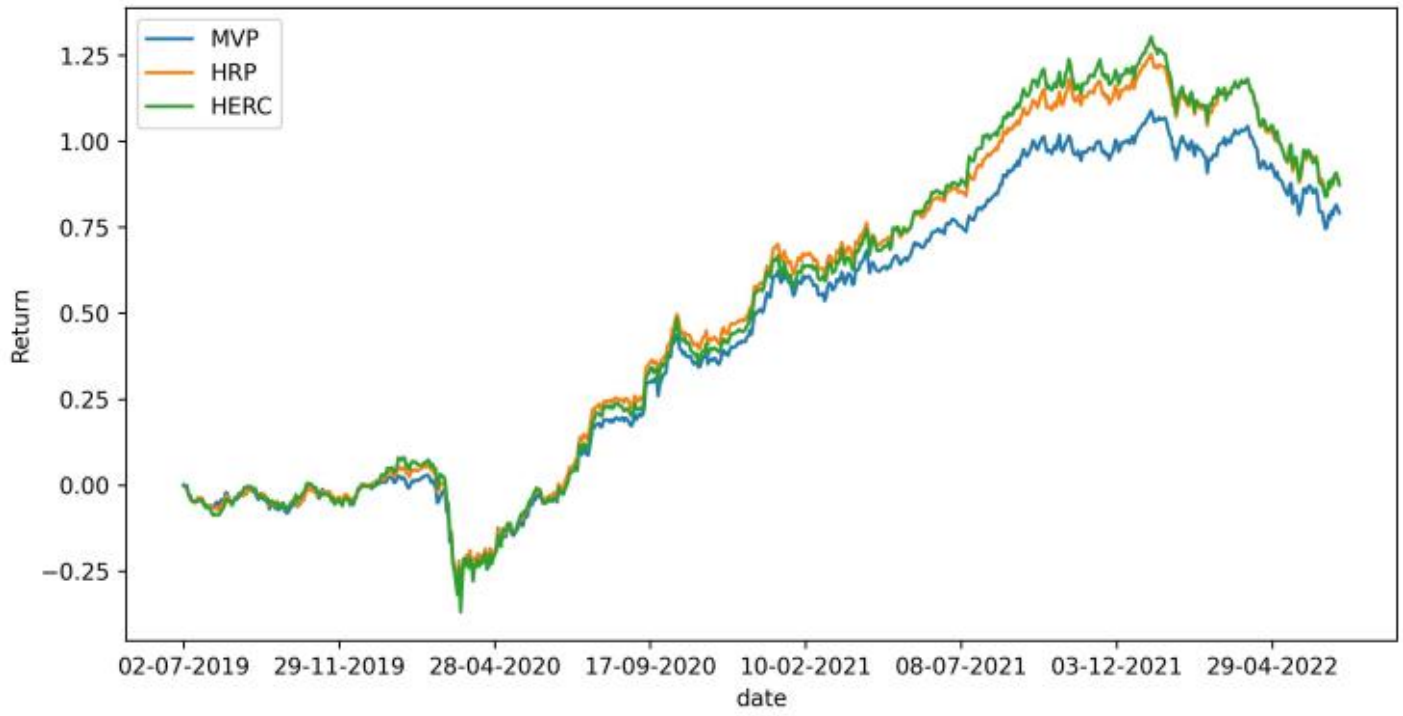

**Figure 1.25.** The cumulative returns yielded by the MVP, HRP, and HERC portfolios of the stocks from the IT sector on the training data from July 1, 2019, to June 30, 2022.

The cumulative returns yielded by the portfolios over the training and the test periods are depicted in Figure 1.25 and Figure 1.26, respectively. Table 1.12 presents the cumulative returns, annual volatilities, and the Sharpe ratios of the three portfolios of the IT sector stocks over the training and the test periods. The highest cumulative return, the lowest volatility, and the maximum Sharpe ratio over the portfolio training and test periods are indicated in red.

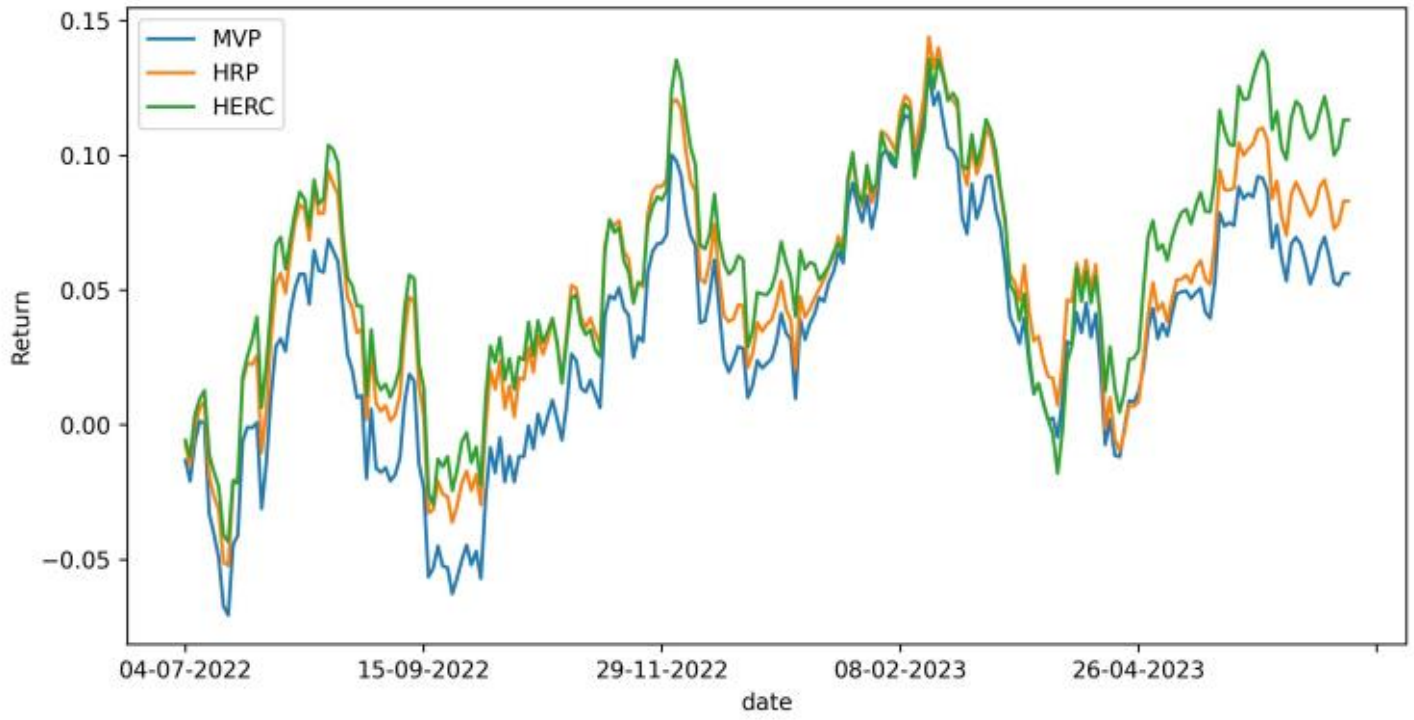

**Figure 1.26.** The cumulative returns yielded by the MVP, HRP, and HERC portfolios of the stocks from the IT sector on the test data from July 1, 2022, to June 30, 2023.

**TABLE 1.12.** THE PERFORMANCE RESULTS OF THE PORTFOLIOS OF THE IT SECTOR STOCKS

| Period | MVP Portfolio | | | HRP Portfolio | | | HERC Portfolio | | |
|---|---|---|---|---|---|---|---|---|---|
| | Ret (%) | Vol (%) | SR | Ret (%) | Vol (%) | SR | Ret (%) | Vol (%) | SR |
| Training | 26.83 | **24.24** | 1.1023 | **29.82** | 25.42 | **1.1730** | 29.60 | 28.87 | 1.0253 |
| Test | 5.78 | **19.25** | 0.3006 | 8.54% | 20.28 | 0.4214 | **11.65** | 21.00 | **0.5546** |

***Media sector:*** As per the report published by the NSE on June 30, 2022, the ten stocks that have the largest free-float market capitalization and their respective contributions to the overall index of the *media* sector are as follows: (i) Zee Entertainment Enterprises (ZEEL): 33.67%, (ii) PVR (PVRINOX): 22.86%, (iii) Sun TV Network (SUNTV): 9.46%, (iv) TV18 Broadcast (TV18BRDCST): 8.00%, (v) Nazara Technologies (NAZARA): 7.30, (vi) Dish TV India (DISHTV): 7.17%, (vii) Network18 Media & Investments (NETWORK18): 4.47%, (viii) Navneet Education (NAVNETEDUL): 3.56%, (ix) Hathway Cable & Datacom (HATHWAY): 2.15%, and (x) NDTV (NDTV): 1.35% (NSE Website). The stocks of PVRINOX and NAZARA could not be considered in the media sector portfolio as these two stocks were listed on NSE later than the starting date of the portfolio formation i.e., July 1, 2019. The listing dates for PVRINOX and NAZARA on NSE were November 11, 2022, and March 30, 2021, respectively. In place of these two stocks, the stocks of TV Today (TVTODAY) and Saregama India (SAREGAMA) are considered in the media sector portfolio as the market capitalizations of these stocks are higher among the remaining stocks in the media sector.

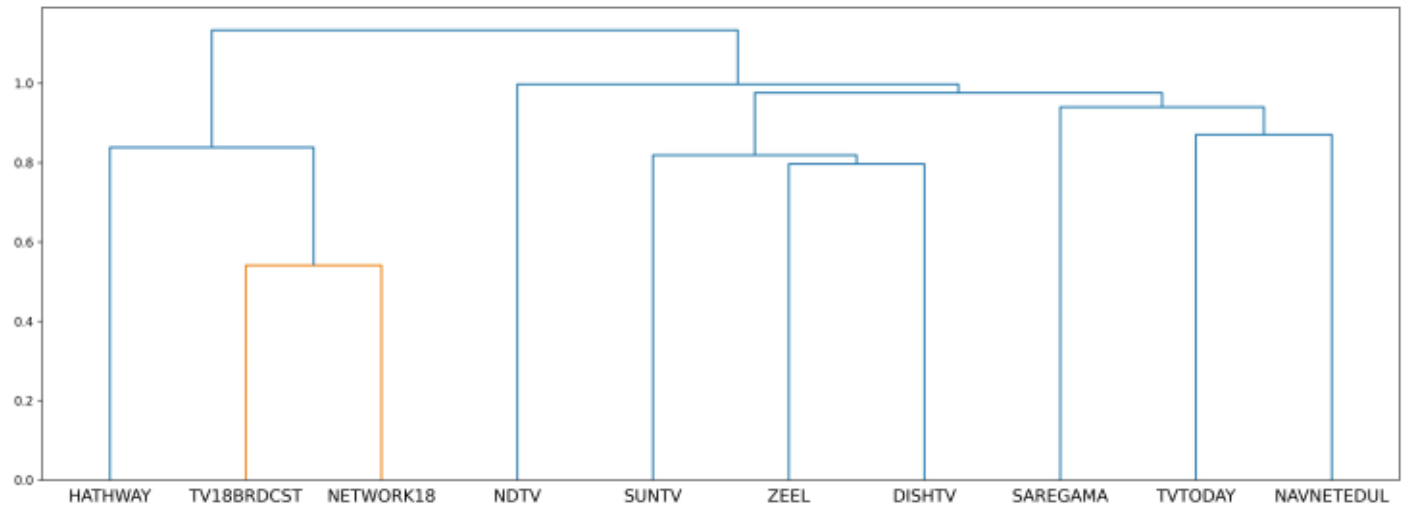

**Figure 1.27.** The dendrogram of the agglomerative clustering of the stocks from the *media* sector is created based on the historical stock prices from July 1, 2019, to June 30, 2022.

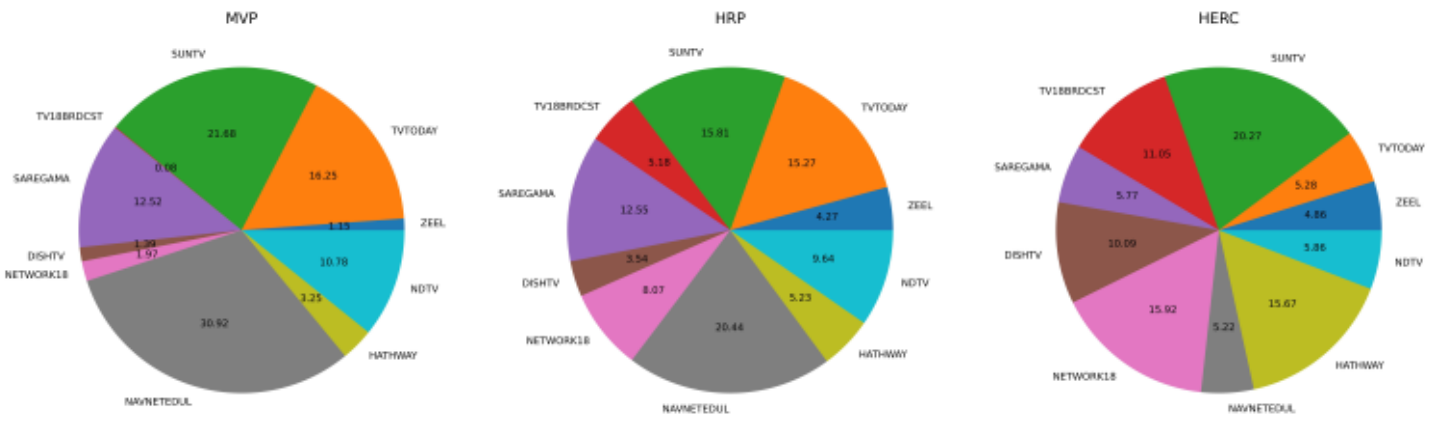

**Figure 1.28.** The allocation of weights done by the MVP, HRP, and HERC algorithms for the stocks of the *media* sector based on the historical stock prices from July 1, 2019, to June 30, 2022.

The dendrogram of the clustering of the stocks of the media sector is shown in Figure 1.27. The cluster containing the stocks TV18BRDCST and NETWORK18 is the most compact, while the one containing TVTODAY and NAVNETEDUL is the least homogeneous. Figure 1.28 depicts the weight allocation done by the three portfolios to the media sector stocks. Table 1.13 shows the weight allocations for the three portfolios in tabular format.

**TABLE 1.13.** THE PORTFOLIO COMPOSITIONS OF THE MEDIA SECTOR (PERIOD: JULY 1, 2019 – JUNE 30, 2022)

| Stock | MVP Portfolio | HRP Portfolio | HERC Portfolio |
|---|---|---|---|
| ZEEL | 0.0115 | 0.0427 | 0.0486 |
| SUNTV | 0.2168 | 0.1581 | 0.2027 |
| TV18BRDCST | 0.0008 | 0.0518 | 0.1105 |
| DISHTV | 0.0139 | 0.0354 | 0.1009 |
| NETWORK18 | 0.0197 | 0.0807 | 0.1592 |
| NAVNETEDUL | 0.3092 | 0.2044 | 0.0522 |
| HATHWAY | 0.0325 | 0.0523 | 0.1567 |
| NDTV | 0.1078 | 0.0964 | 0.0586 |
| TVTODAY | 0.1625 | 0.1527 | 0.0528 |
| SAREGAMA | 0.1252 | 0.1255 | 0.0577 |

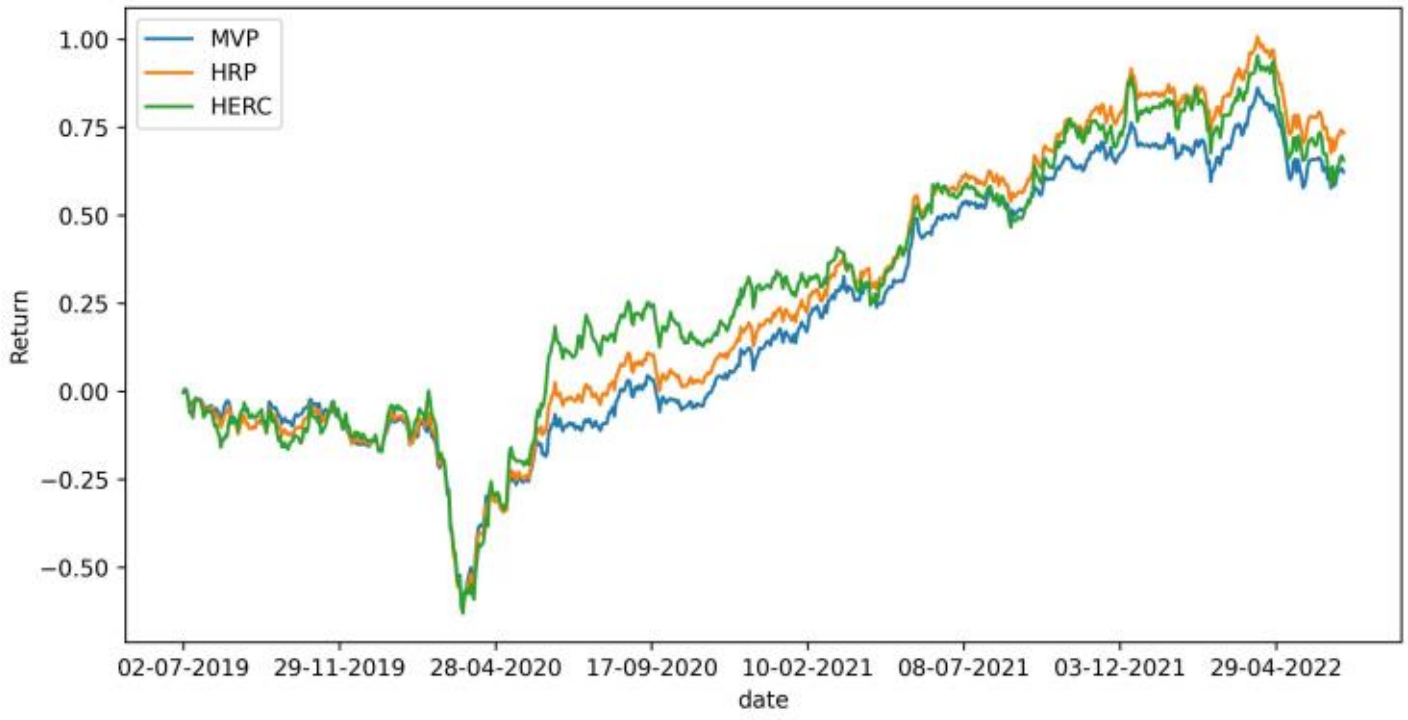

**Figure 1.29.** The cumulative returns yielded by the MVP, HRP, and HERC portfolios of the stocks from the *media* sector on the training data from July 1, 2019, to June 30, 2022.

The cumulative returns yielded by the portfolios over the training and the test periods are depicted in Figure 1.29 and Figure 1.30, respectively. Table 1.14 presents the cumulative returns, annual volatilities, and the Sharpe ratios of the three portfolios of the media sector stocks over the training and the test periods. The highest cumulative return, the lowest volatility, and the maximum Sharpe ratio over the portfolio training and test periods are indicated in red.

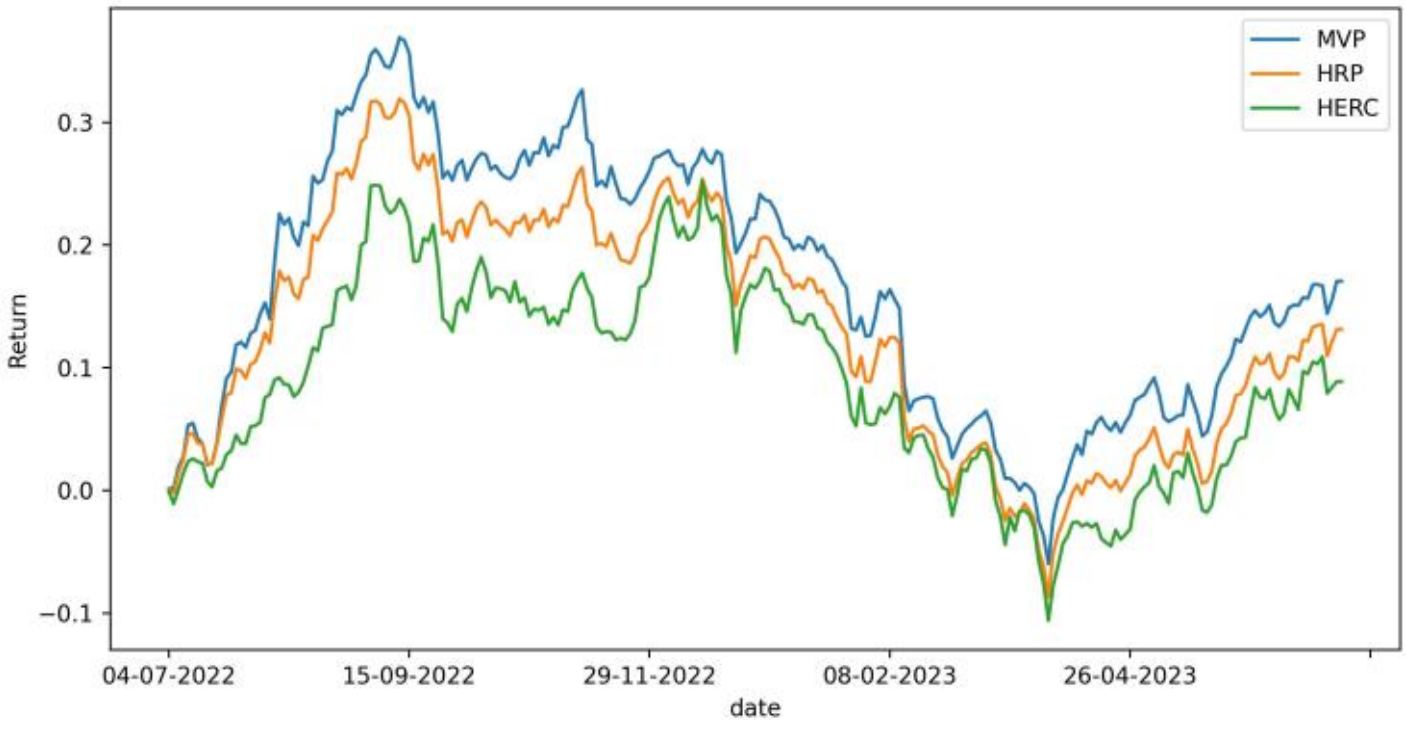

**Figure 1.30.** The cumulative returns yielded by the MVP, HRP, and HERC portfolios of the stocks from the *media* sector on the test data from July 1, 2022, to June 30, 2023.

**TABLE 1.14.** THE PERFORMANCE RESULTS OF THE PORTFOLIOS OF THE MEDIA SECTOR STOCKS

| **Period** | **MVP Portfolio** | | | **HRP Portfolio** | | | **HERC Portfolio** | | |
|---|---|---|---|---|---|---|---|---|---|
| | **Ret (%)** | **Vol (%)** | **SR** | **Ret (%)** | **Vol (%)** | **SR** | **Ret (%)** | **Vol (%)** | **SR** |
| Training | 21.11 | **26.42** | 0.7991 | **24.85** | 27.90 | **0.8906** | 22.26 | 34.02 | 0.6543 |
| Test | **17.54** | 21.66 | **0.8100** | 13.50 | **21.00** | 0.6431 | 9.13 | 22.75 | 0.4013 |

***Metal sector:*** As per the report published by the NSE on June 30, 2022, the ten stocks of the *metal* sector that have the largest free-float market capitalization and their respective contributions (in percent) to the overall index of the *metal* sector are as follows: (i) Tata Steel (TATASTEEL): 20.84%, (ii) Adani Enterprises (ADANIENT): 16.09%, (iii) JSW Steel (JSWSTEEL): 15.74%, (iv) Hindalco

Industries (HINDALCO): 14.17%, (v) Vedanta (VEDL): 6.89%, (vi) APL Apollo Tubes (APLAPOLLO): 5.75%, (vii) Jindal Steel & Power (JINDALSTEL): 5.15%, (viii) Jindal Stainless (JSL): 2.91%, (ix) Steel Authority of India (SAIL): 2.87%, and (x) NMDC (NMDC): 2.81% (NSE Website).

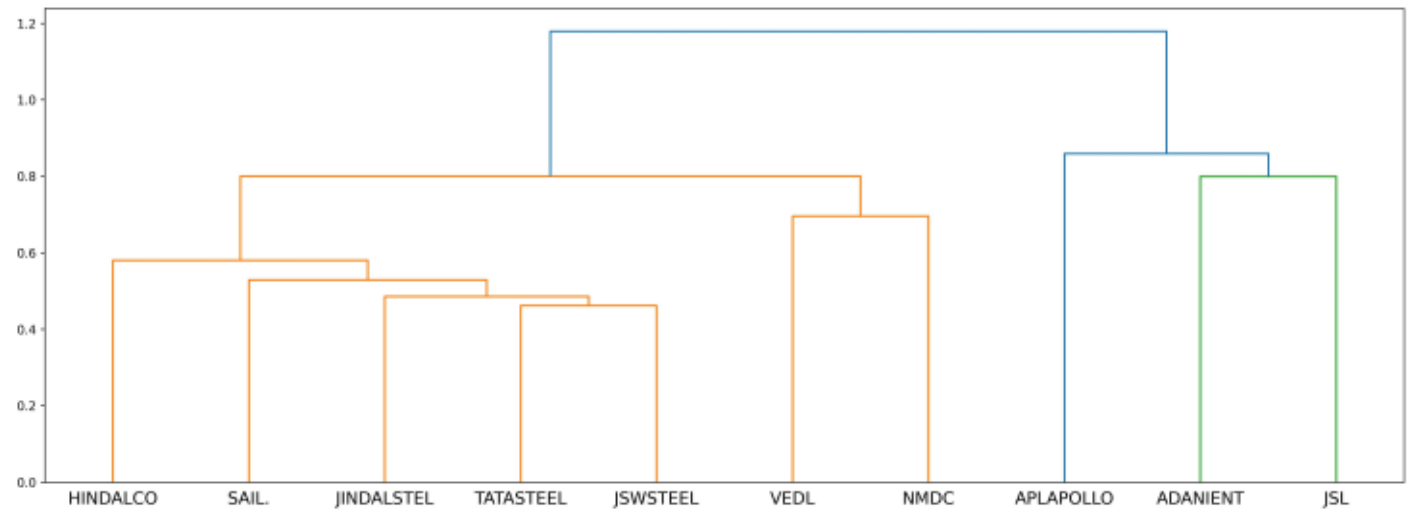

**Figure 1.31.** The dendrogram of the agglomerative clustering of the stocks from the *metal* sector is created based on the historical stock prices from July 1, 2019, to June 30, 2022.

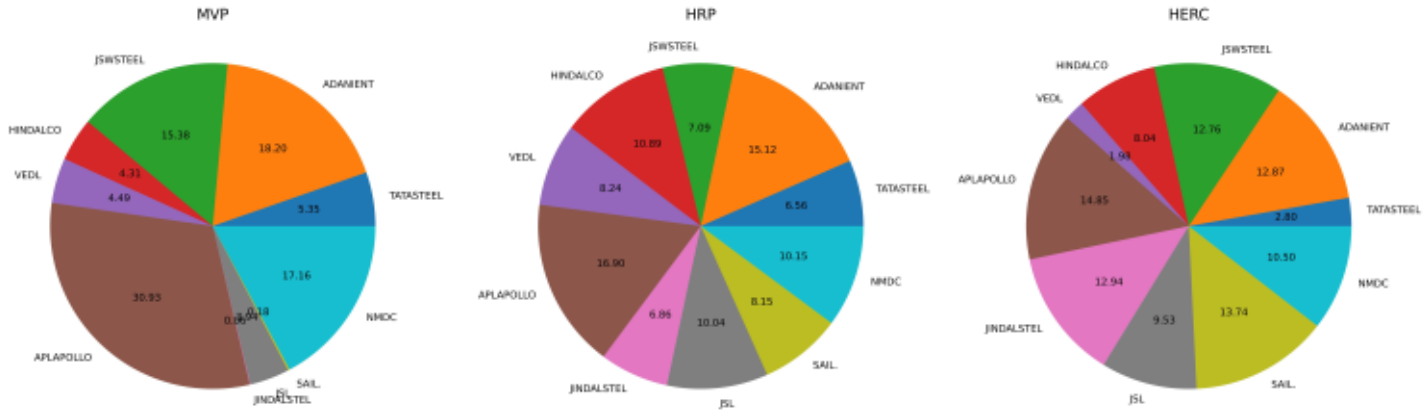

**Figure 1.32.** The allocation of weights done by the MVP, HRP, and HERC algorithms for the stocks of the *metal* sector based on the historical stock prices from July 1, 2019, to June 30, 2022.

The dendrogram of the clustering of the stocks of the metal sector is shown in Figure 1.31. The cluster containing the stocks TATASTEEL and JSWSTEEL is the most compact, while the one containing ADANIENT and JSL is the least homogeneous. Figure 1.32 depicts the weight allocation done by the three portfolios to the metal sector stocks. Table 1.15 shows the weight allocations for the three portfolios in tabular format.

**TABLE 1.15.** THE PORTFOLIO COMPOSITIONS OF THE METAL (PERIOD: JULY 1, 2019 – JUNE 30, 2022)

| Stock | MVP Portfolio | HRP Portfolio | HERC Portfolio |
|---|---|---|---|
| TATASTEEL | 0.0535 | 0.0656 | 0.0280 |
| ADANIENT | 0.1820 | 0.1512 | 0.1287 |
| JSWSTEEL | 0.1538 | 0.0709 | 0.1276 |
| HINDALCO | 0.0431 | 0.1089 | 0.0804 |
| VEDL | 0.0449 | 0.0824 | 0.0198 |
| APLAPOLLO | 0.3093 | 0.1690 | 0.1485 |
| JINDALSTEL | 0.0006 | 0.0686 | 0.1294 |
| JSL | 0.0394 | 0.1004 | 0.0953 |
| SAIL | 0.0018 | 0.0815 | 0.1374 |
| NMDC | 0.1716 | 0.1015 | 0.1050 |

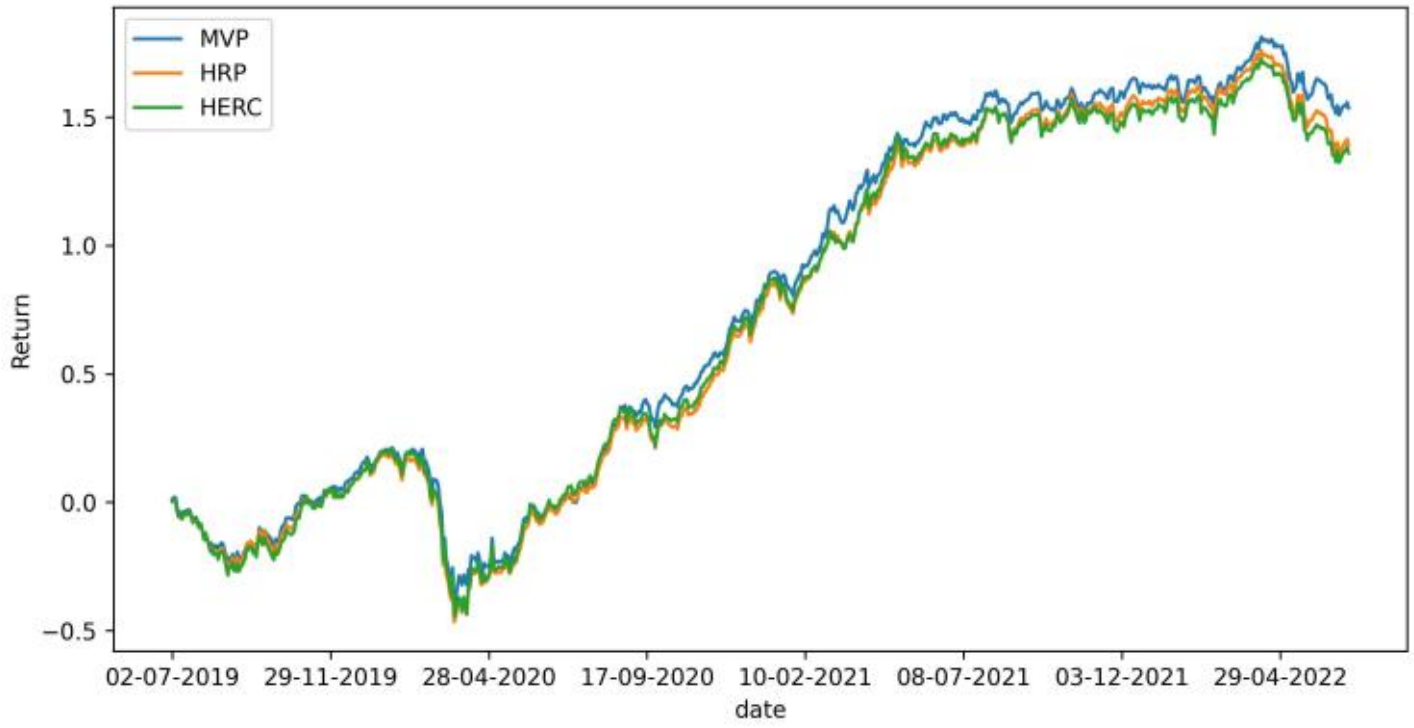

**Figure 1.33.** The cumulative returns yielded by the MVP, HRP, and HERC portfolios of the stocks from the *metal* sector on the training data from July 1, 2019, to June 30, 2022.

The cumulative returns yielded by the portfolios over the training and the test periods are depicted in Figure 1.33 and Figure 1.34,

respectively. Table 1.16 presents the cumulative returns, annual volatilities, and the Sharpe ratios of the three portfolios of the metal sector stocks over the training and the test periods. The highest cumulative return, the lowest volatility, and the maximum Sharpe ratio over the portfolio training and test periods are indicated in red.

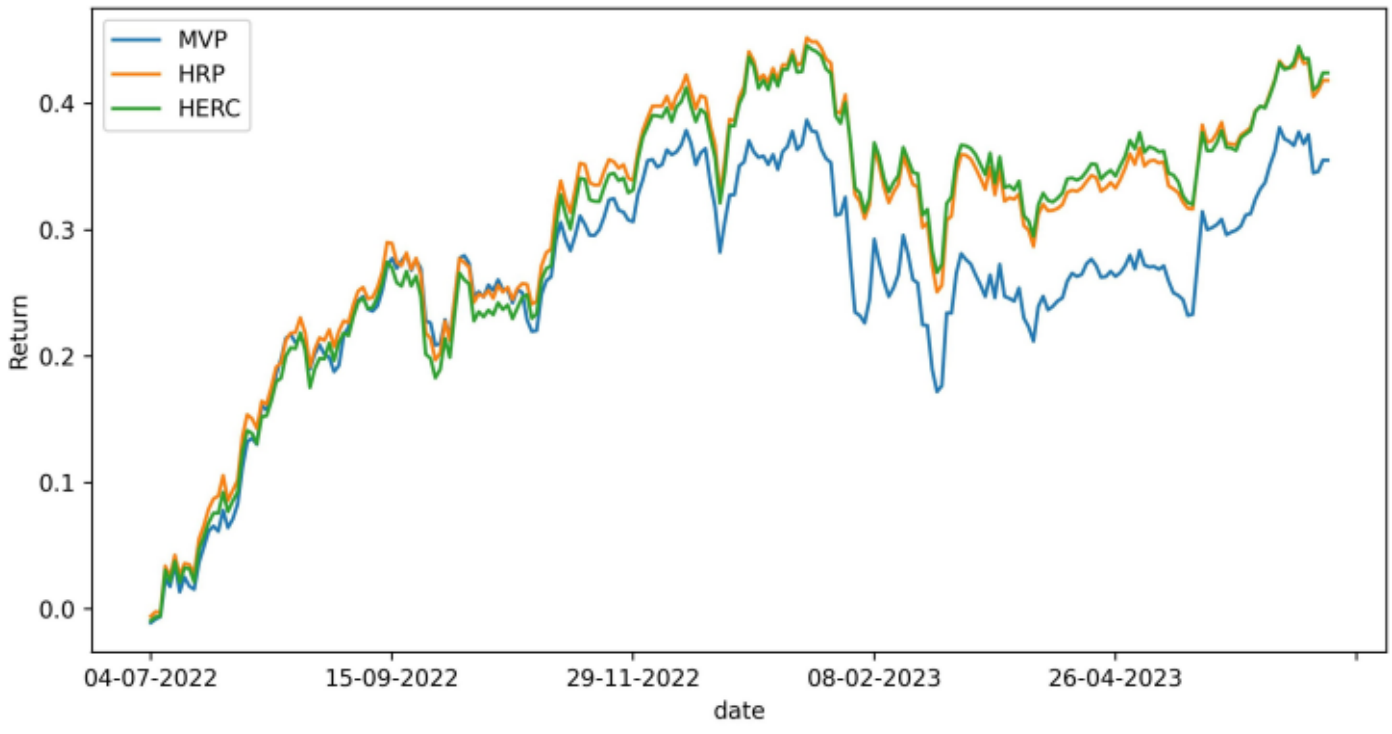

**Figure 1.34.** The cumulative returns yielded by the MVP, HRP, and HERC portfolios of the stocks from the *metal* sector on the test data from July 1, 2022, to June 30, 2023.

**TABLE 1.16.** THE PERFORMANCE RESULTS OF THE PORTFOLIOS OF THE METAL SECTOR STOCKS

| Period | MVP Portfolio | | | HRP Portfolio | | | HERC Portfolio | | |
|---|---|---|---|---|---|---|---|---|---|
| | Ret (%) | Vol (%) | SR | Ret (%) | Vol (%) | SR | Ret (%) | Vol (%) | SR |
| Training | **52.03** | **33.70** | **1.5441** | 47.14 | 35.86 | 1.3145 | 46.10 | 37.29 | 1.2363 |
| Test | **36.53** | 24.02 | 1.5210 | 43.03 | 23.70 | 1.8153 | 43.64 | **23.23** | **1.8783** |

***MidSmall IT & Telecom sector:*** This sector consists of mid-cap and small-cap stocks within the information technology and telecommunication sector. As per the report published by the NSE on June 30, 2022, the ten stocks with the largest free-float market capitalization in this sector and their respective contributions to the overall index of the sector are as follows: (i) Tata Elxsi (TATAELXSI): 10.97%, (ii) Persistent Systems (PERSISTENT): 10.74%, (iii) Tata Communications (TATACOMM): 9.22%, (iv) Coforge (COFORGE): 8.71%, (v) MphasiS (MPHASIS): 8.36%, (vi)

KPIT Technologies (KPITTECH): 7.69%, (vii) Cyient (CYIENT): 5.34%, (viii) L&T Technology Services (LTTS): 4.87%, (ix) Sonata Software (SONATSOFTW): 4.49%, and (x) Oracle Financial Services Software (OFSS): 3.99% (NSE Website). The ticker names of the stocks are mentioned in parentheses. The stocks are identified by their ticker names in a given stock exchange.

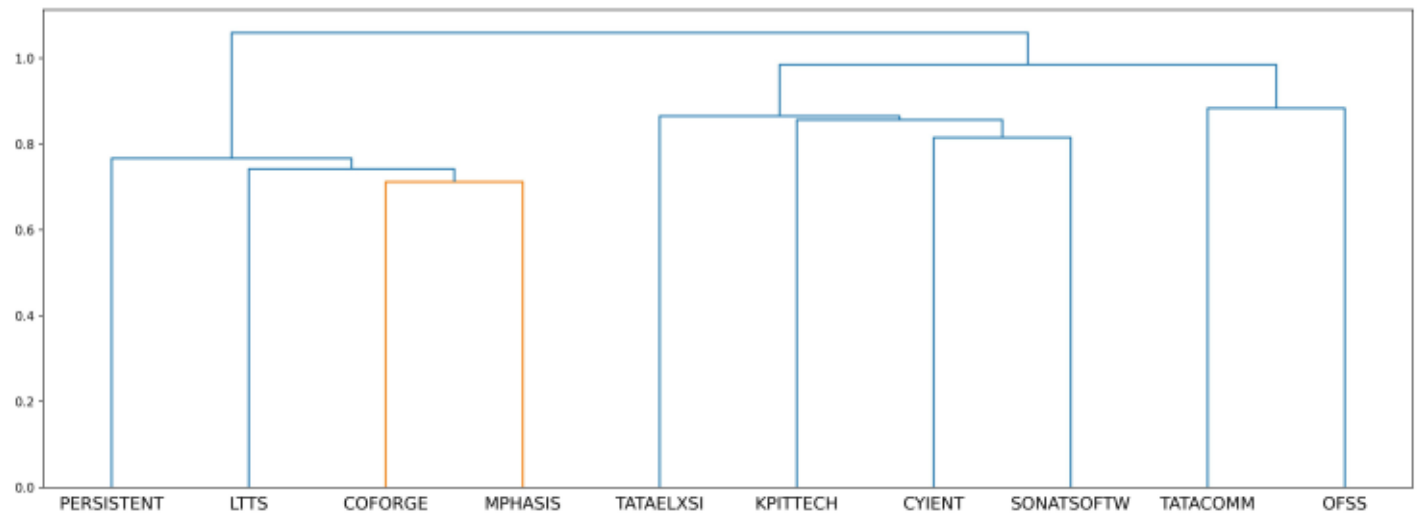

**Figure 1.35.** The dendrogram of the agglomerative clustering of the stocks from the *mid-small IT and telecom* sector based on the historical stock prices from July 1, 2019, to June 30, 2022.

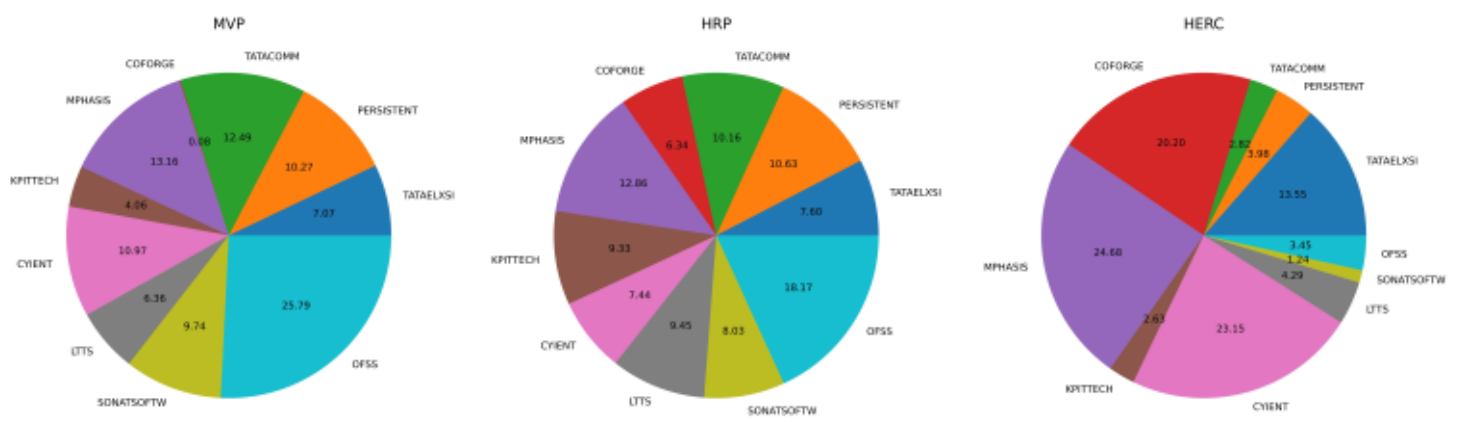

**Figure 1.36.** The allocation of weights done by the MVP, HRP, and HERC algorithms for the stocks of the *mid-small IT and telecom* sector based on the historical stock prices from July 1, 2019, to June 30, 2022.

The dendrogram of the clustering of the stocks of the *mid-small IT and telecom* sector is shown in Figure 1.35. The cluster containing the stocks COFORGE and MPHASIS is the most compact, while the one containing TATACOMM and OFSS is the least homogeneous. Figure 1.36 depicts the weight allocation done by the three portfolios to the mid-small IT and telecom sector stocks. Table 1.17 shows the weight allocations for the three portfolios in tabular format.

**TABLE 1.17.** THE PORTFOLIO COMPOSITIONS OF THE MIDSMALL IT & TELECOM SECTOR
(PERIOD: JULY 1, 2019 – JUNE 30, 2022)

| Stock | MVP Portfolio | HRP Portfolio | HERC Portfolio |
|---|---|---|---|
| TATAELXSI | 0.0707 | 0.0760 | 0.1355 |
| PERSISTENT | 0.1027 | 0.1063 | 0.0398 |
| TATACOMM | 0.1249 | 0.1016 | 0.0282 |
| COFORGE | 0.0008 | 0.0634 | 0.2020 |
| MPHASIS | 0.1316 | 0.1286 | 0.2468 |
| KPITTECH | 0.0406 | 0.0933 | 0.0263 |
| CYIENT | 0.1097 | 0.0744 | 0.2315 |
| LTTS | 0.0636 | 0.0945 | 0.0429 |
| SONATASOFTW | 0.0974 | 0.0803 | 0.0124 |
| OFSS | 0.2579 | 0.1817 | 0.0345 |

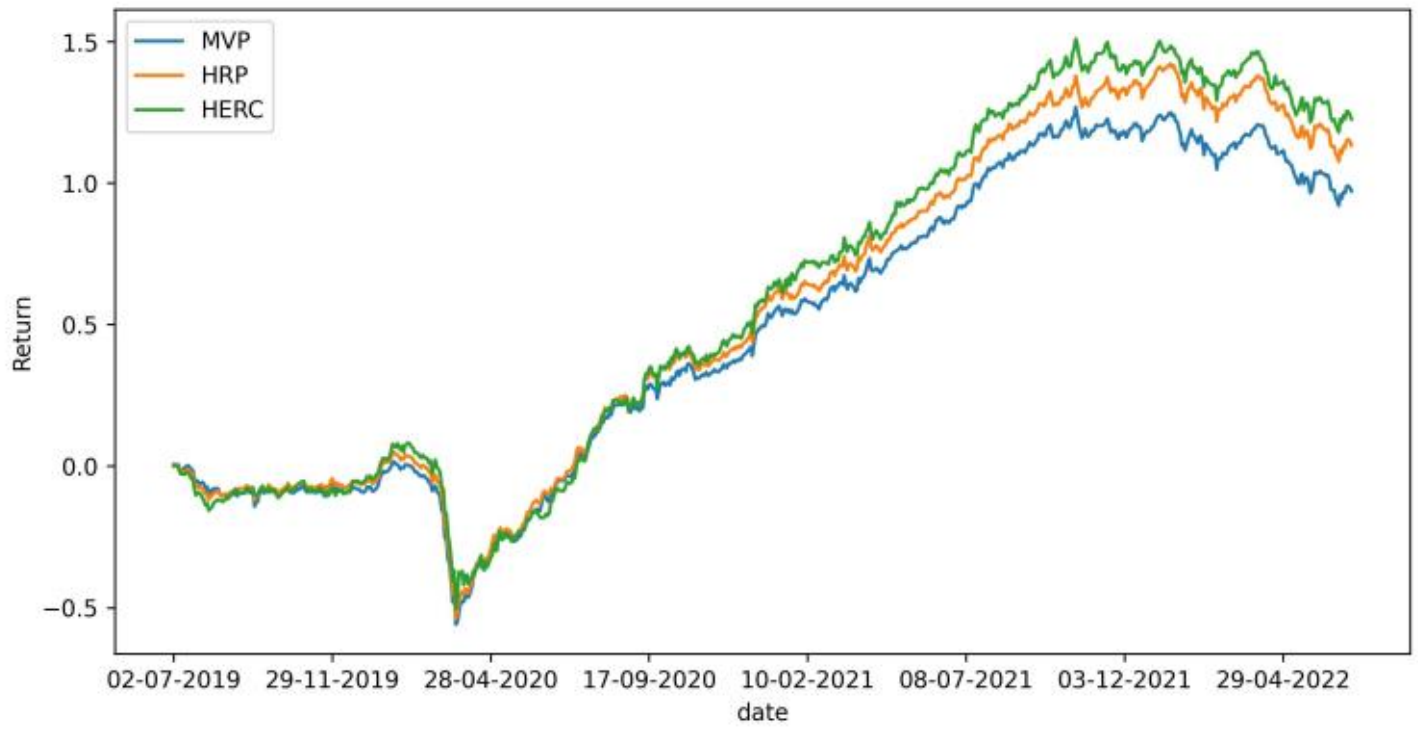

**Figure 1.37.** The cumulative returns yielded by the MVP, HRP, and HERC portfolios of the stocks from the *mid-small IT & telecom* sector on the training data from July 1, 2019, to June 30, 2022.

The cumulative returns yielded by the portfolios over the training and the test periods are depicted in Figure 1.37 and Figure 1.38,

respectively. Table 1.18 presents the cumulative returns, annual volatilities, and the Sharpe ratios of the three portfolios of the mid-small IT and telecom sector stocks over the training and test periods. The highest cumulative return, the lowest volatility, and the maximum Sharpe ratio over the portfolio training and test periods are indicated in red.

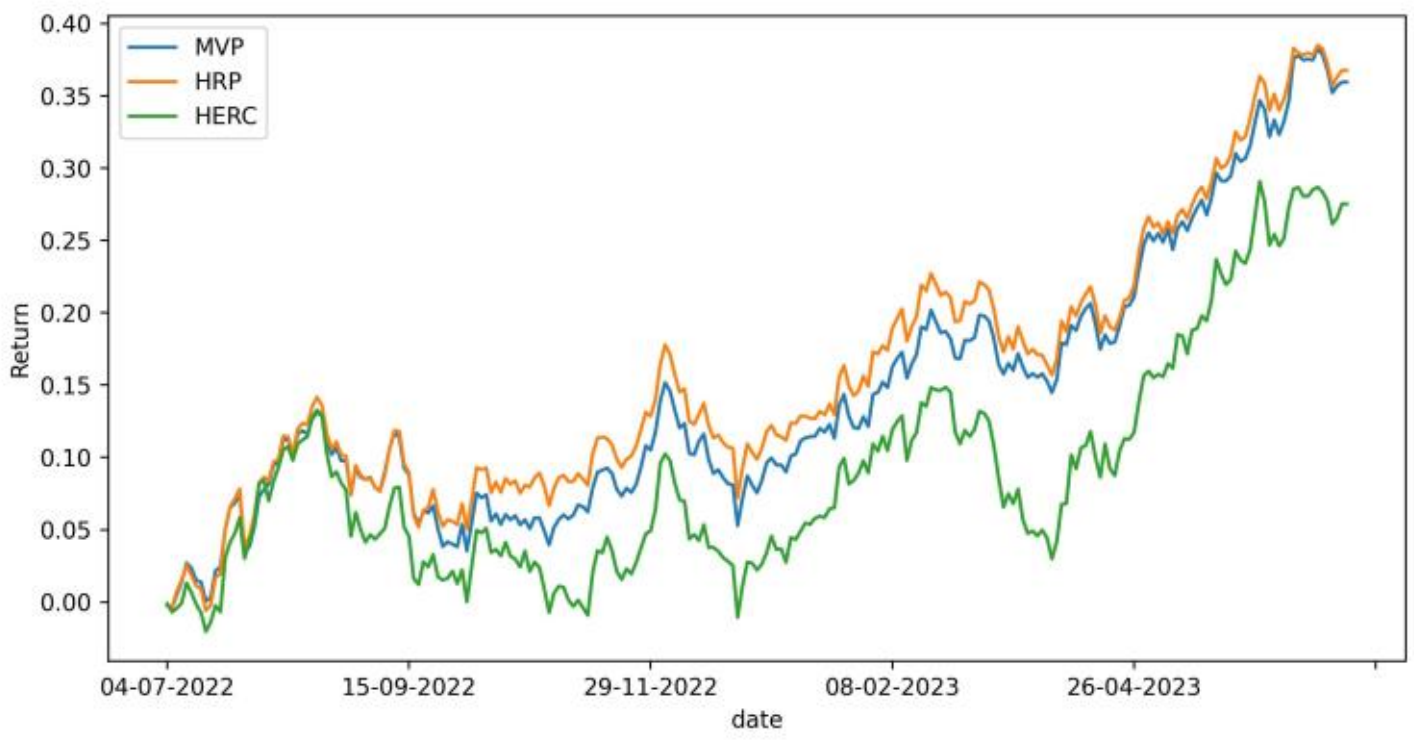

**Figure 1.38.** The cumulative returns yielded by the MVP, HRP, and HERC portfolios of the stocks from the *mid-small IT & telecom* sector on the test data from July 1, 2022, to June 30, 2023.

**TABLE 1.18.** THE PERFORMANCE RESULTS OF THE PORTFOLIOS OF THE MIDSMALL IT AND TELECOM SECTOR STOCKS

| Period | MVP Portfolio | | | HRP Portfolio | | | HERC Portfolio | | |
|---|---|---|---|---|---|---|---|---|---|
| | Ret (%) | Vol (%) | SR | Ret (%) | Vol (%) | SR | Ret (%) | Vol (%) | SR |
| Training | 32.93 | **23.96** | 1.3744 | **38.43** | 24.83 | **1.5475** | 41.57 | 28.86 | 1.4404 |
| Test | 36.98 | **16.96** | **2.1798** | **37.81** | 17.89 | 2.1131 | 28.32 | 20.33 | 1.3931 |

***Oil & Gas sector:*** The NSE's report published on June 20, 2022, the ten stocks with the largest market capitalization and their contributions to the overall index of the *oil and gas* sector are as follows: (i) Reliance Industries (RELIANCE): 30.85%, (ii) Oil & Natural Gas Corporation (ONGC): 15.92%, (iii) Bharat Petroleum Corporation (BPCL): 8.31%, (iv) Indian Oil Corporation (IOC): 7.94%, (v) GAIL India (GAIL): 7.40, (vi) Adani Total Gas (ATGL): 4.21%, (vii) Hindustan Petroleum Corporation (HINDPETRO):

4.16%, (viii) Petronet LNG (PETRONET): 4.04%, (ix) Indraprastha Gas (IGL): 3.74%, and (x) Oil India (OIL): 3.17% (NSE Website). The ticker names, which are mentioned in parentheses, are the unique identifiers of the stocks listed on a stock exchange.

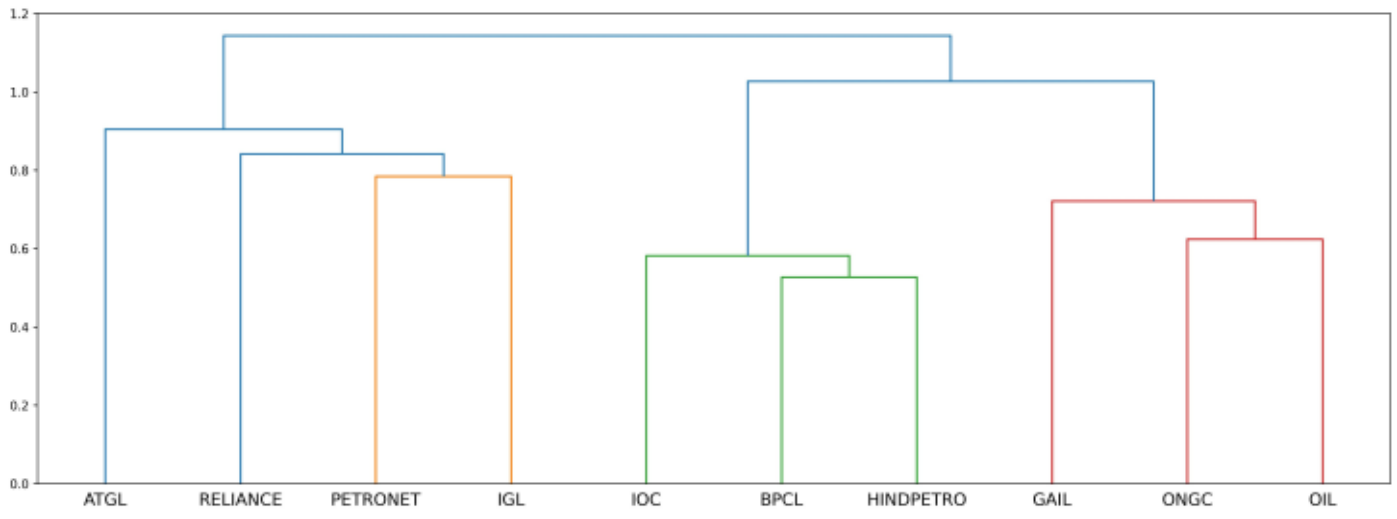

**Figure 1.39.** The dendrogram of the agglomerative clustering of the stocks from the *oil and gas* sector is created based on the historical stock prices from July 1, 2019, to June 30, 2022.

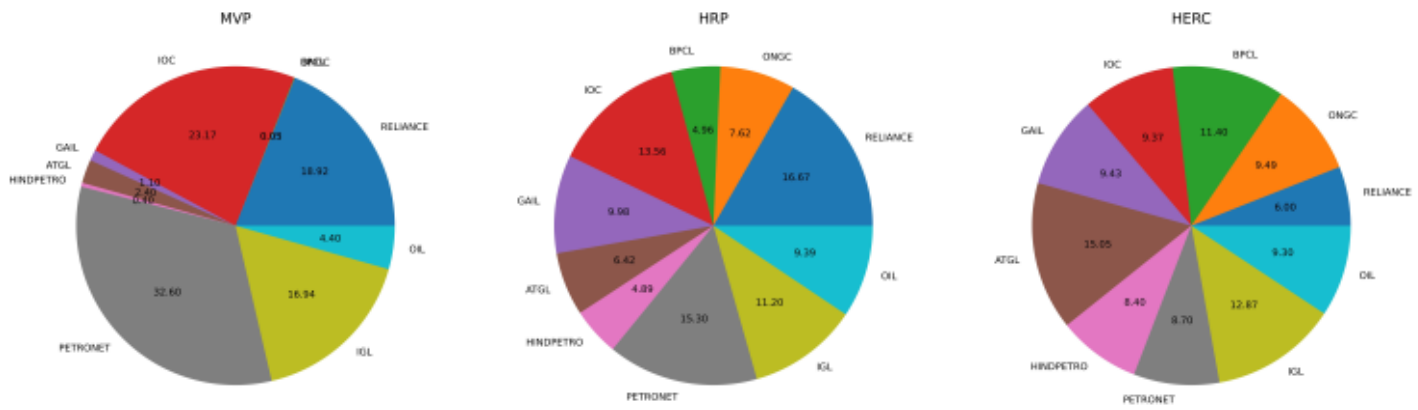

**Figure 1.40.** The allocation of weights done by the MVP, HRP, and HERC algorithms for the stocks of the *oil and gas* sector based on the historical stock prices from July 1, 2019, to June 30, 2022.

The dendrogram of the clustering of the stocks of the oil and gas sector is shown in Figure 1.39. The cluster containing the stocks BPCL and HINDPETRO is the most compact, while the one containing PETRONET and IGL is the least homogeneous. Figure 1.40 depicts the weight allocation done by the three portfolios to the oil and gas sector stocks. Table 1.19 shows the weight allocations for the three portfolios in tabular format.

**TABLE 1.19.** THE PORTFOLIO COMPOSITIONS OF THE OIL & GAS SECTOR
(PERIOD: JULY 1, 2019 – JUNE 30, 2022)

| Stock | MVP Portfolio | HRP Portfolio | HERC Portfolio |
|---|---|---|---|
| RELIANCE | 0.1892 | 0.1667 | 0.06009 |
| ONGC | 0.0003 | 0.0762 | 0.0949 |
| BPCL | 0.0005 | 0.0496 | 0.1140 |
| IOC | 0.2317 | 0.1356 | 0.0937 |
| GAIL | 0.0110 | 0.0998 | 0.0943 |
| ATGL | 0.0240 | 0.0642 | 0.1505 |
| HINDPETRO | 0.0040 | 0.0489 | 0.0840 |
| PETRONET | 0.3260 | 0.1530 | 0.0870 |
| IGL | 0.1694 | 0.1120 | 0.1287 |
| OIL | 0.0441 | 0.0939 | 0.0930 |

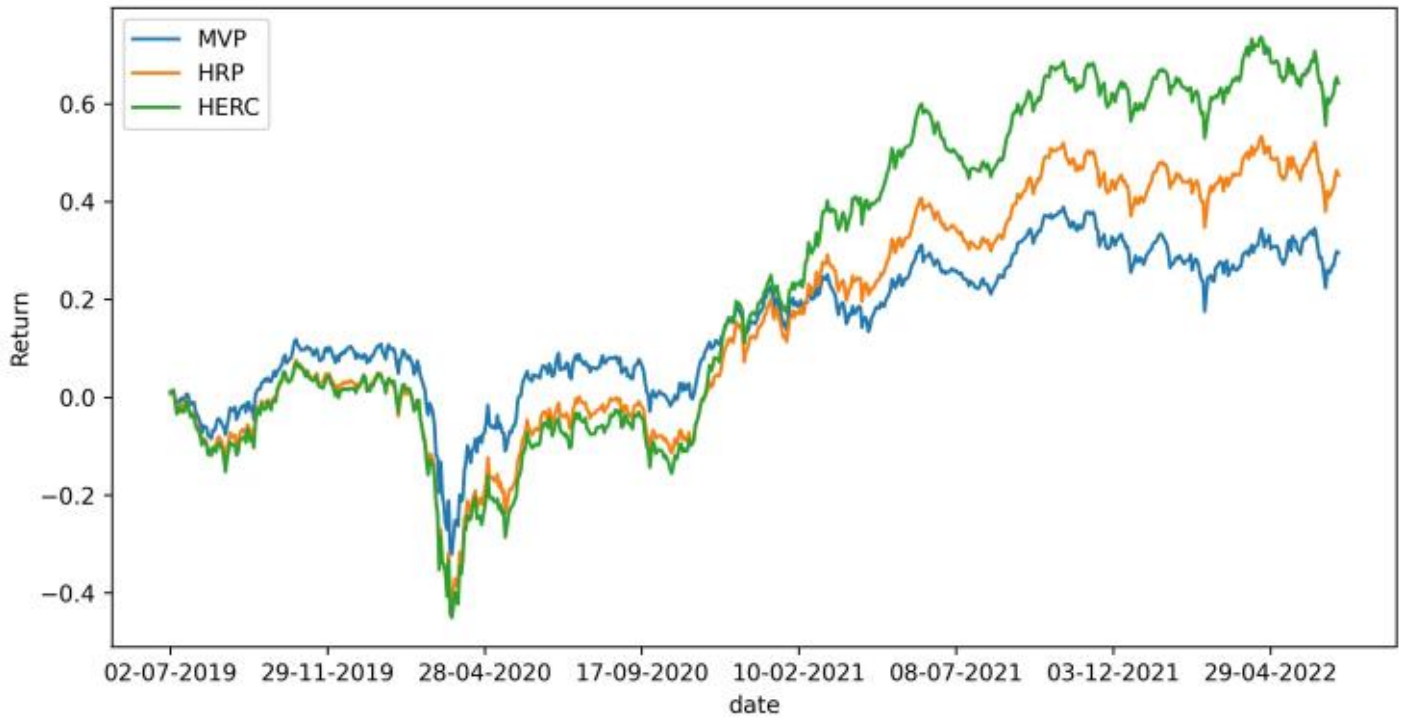

**Figure 1.41.** The cumulative returns yielded by the MVP, HRP, and HERC portfolios of the stocks from the *oil and gas* sector on the training data from July 1, 2019, to June 30, 2022.

The cumulative returns yielded by the portfolios over the training and the test periods are depicted in Figure 1.41 and Figure 1.42, respectively. Table 1.20 presents the cumulative returns, annual volatilities, and the Sharpe ratios of the three portfolios of the oil and gas sector stocks over the training and the test periods. The highest cumulative return, the lowest volatility, and the maximum Sharpe ratio over the portfolio training and test periods are indicated in red.

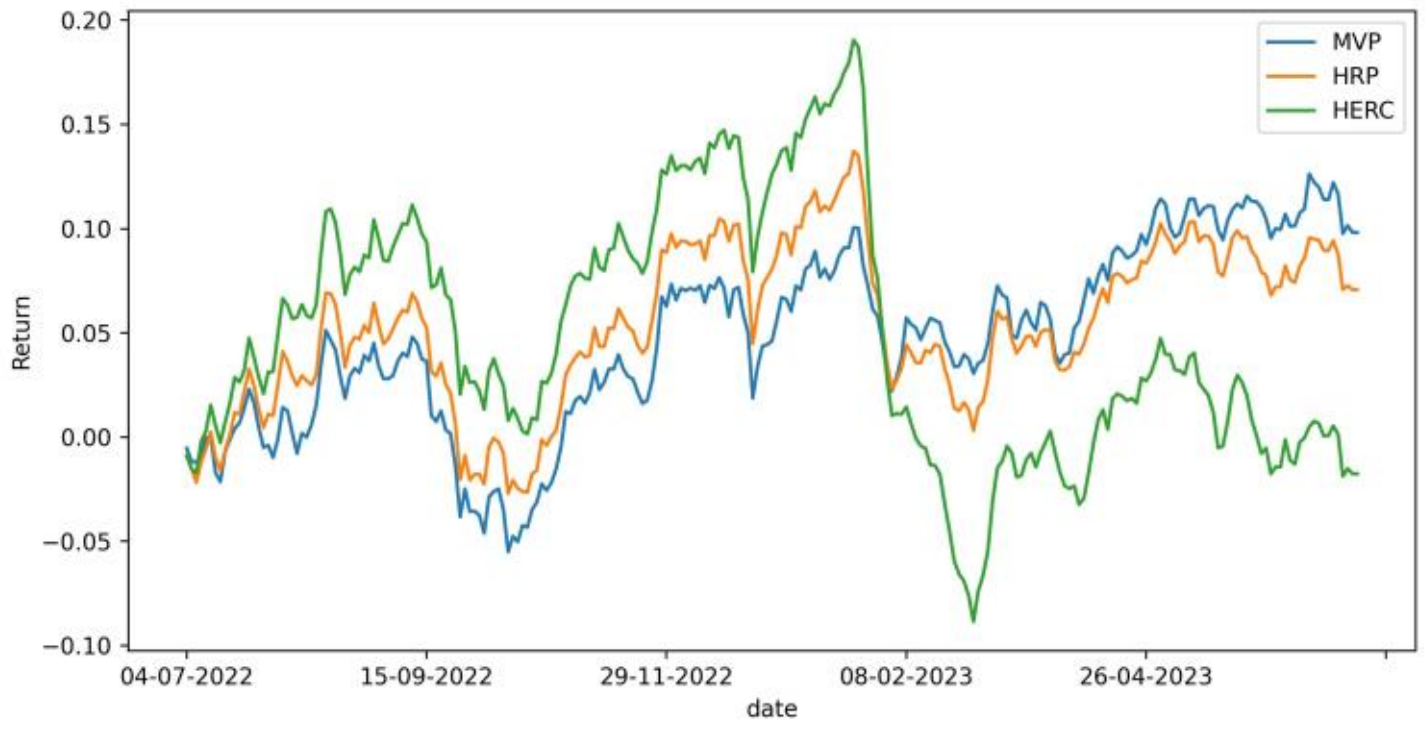

**Figure 1.42.** The cumulative returns yielded by the MVP, HRP, and HERC portfolios of the stocks from the *oil and gas* sector on the test data from July 1, 2022, to June 30, 2023.

**TABLE 1.20.** THE PERFORMANCE RESULTS OF THE PORTFOLIOS OF THE OIL AND GAS SECTOR STOCKS

| Period | MVP Portfolio | | | HRP Portfolio | | | HERC Portfolio | | |
|---|---|---|---|---|---|---|---|---|---|
| | **Ret (%)** | **Vol (%)** | **SR** | **Ret (%)** | **Vol (%)** | **SR** | **Ret (%)** | **Vol (%)** | **SR** |
| Training | 10.02 | **23.25** | 0.4310 | 15.38 | 24.86 | 0.6189 | **21.75** | 26.67 | **0.8158** |
| Test | **10.09** | **13.67** | **0.7386** | 7.27% | 13.76 | 0.5286 | -1.81 | 16.50 | -0.1097 |

***Pharma sector:*** As per the report published by the NSE on June 30, 2022, the ten stocks with the largest free-float market capitalization and their respective contributions to the overall index of the *pharma* sector are as follows: (i) Sun Pharmaceuticals Industries (SUNPHARMA): 24.59%, (ii) Dr. Reddy's Labs (DRREDDY): 13.66%, (iii) Cipla (CIPLA): 12.28%, (iv) Divi's Laboratories (DIVISLAB): 9.35%, (v) Lupin (LUPIN): 4.73%, (vi) Aurobindo Pharma (AUROPHARMA): 4.61%, (vii) Alkem Laboratories (ALKEM): 3.88%, (viii) Torrent Pharmaceuticals (TORNTPHARM): 3.65%, (ix) Zydus Lifesciences (ZYDUSLIFE): 3.19%, and (x) Laurus Labs (LAURUSLABS): 2.76%. (NSE Website). The ticker names of the stocks are mentioned in parentheses. The ticker names are the unique identifiers of the stocks listed on a stock exchange.

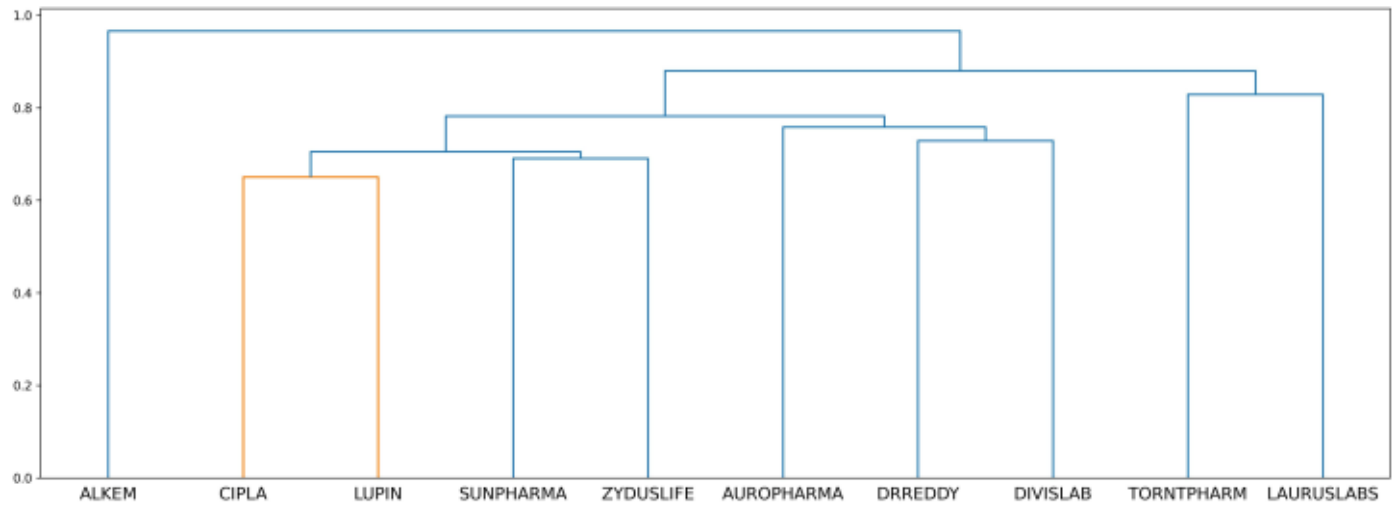

**Figure 1.43.** The dendrogram of the agglomerative clustering of the stocks from the *pharma* sector is created based on the historical stock prices from July 1, 2019, to June 30, 2022.

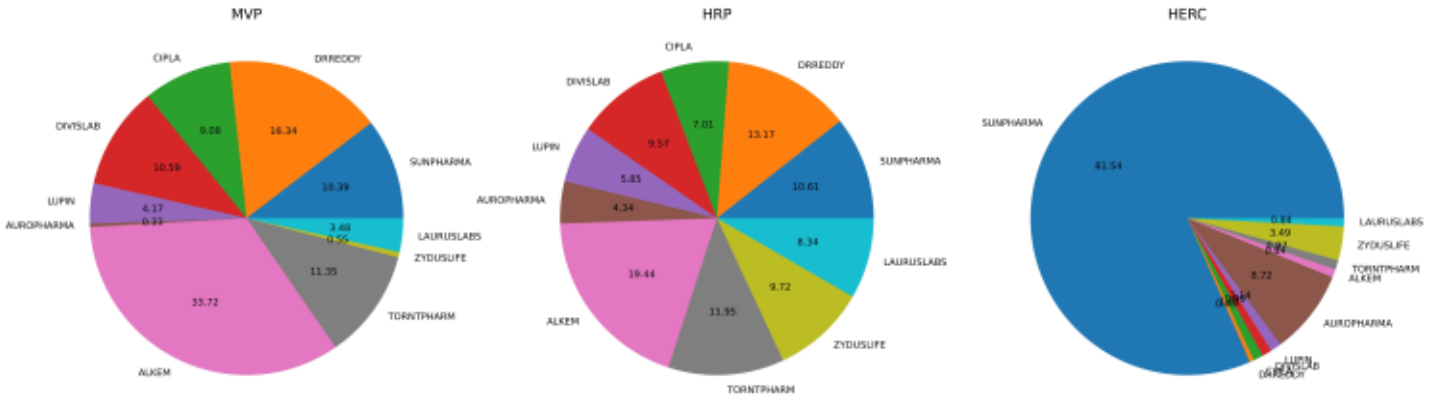

**Figure 1.44.** The allocation of weights done by the MVP, HRP, and HERC algorithms for the stocks of the *pharma* sector based on the historical stock prices from July 1, 2019, to June 30, 2022.

The dendrogram of the clustering of the stocks of the pharma sector is shown in Figure 1.43. The cluster containing the stocks CIPLA and LUPIN is the most compact, while the one containing TORNTPHARM and LAURUSLABS is the least homogeneous.

Figure 1.44 depicts the weight allocation done by the three portfolios to the pharma sector stocks. Table 1.21 shows the weight allocations for the three portfolios in tabular format.

**TABLE 1.21.** THE PORTFOLIO COMPOSITIONS OF THE PHARMA SECTOR
(PERIOD: JULY 1, 2019 – JUNE 30, 2022)

| Stock | MVP Portfolio | HRP Portfolio | HERC Portfolio |
|---|---|---|---|
| SUNPHARMA | 0.1039 | 0.1061 | 0.8154 |
| DRREDDY | 0.1634 | 0.1317 | 0.0045 |
| CIPLA | 0.0908 | 0.0701 | 0.0101 |
| DIVISLAB | 0.1059 | 0.0957 | 0.0099 |
| LUPIN | 0.0417 | 0.0585 | 0.0114 |
| AUROPHARMA | 0.0033 | 0.0434 | 0.0872 |
| ALKEM | 0.3372 | 0.1944 | 0.0084 |
| TORNTPHARM | 0.1135 | 0.1195 | 0.0097 |
| ZYDUSLIFE | 0.0055 | 0.0972 | 0.0349 |
| LAURUSLABS | 0.0348 | 0.0834 | 0.0084 |

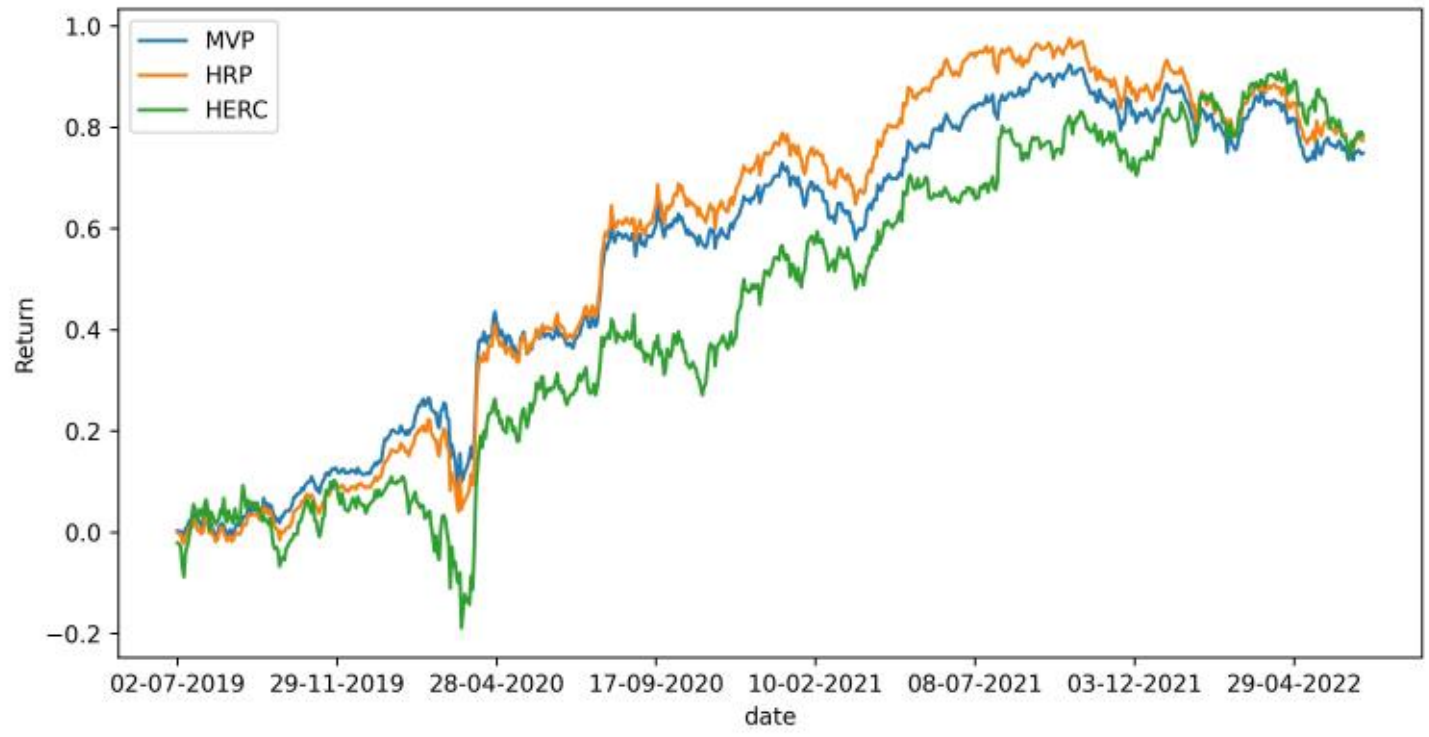

**Figure 1.45.** The cumulative returns yielded by the MVP, HRP, and HERC portfolios of the stocks from the *pharma* sector on the training data from July 1, 2019, to June 30, 2022.

The cumulative returns yielded by the portfolios over the training and the test periods are depicted in Figure 1.45 and Figure 1.46, respectively. Table 1.22 presents the cumulative returns, annual volatilities, and the Sharpe ratios of the three portfolios of the pharma sector stocks over the training and the test periods. The highest cumulative return, the lowest volatility, and the maximum Sharpe ratio over the portfolio training and test periods are indicated in red.

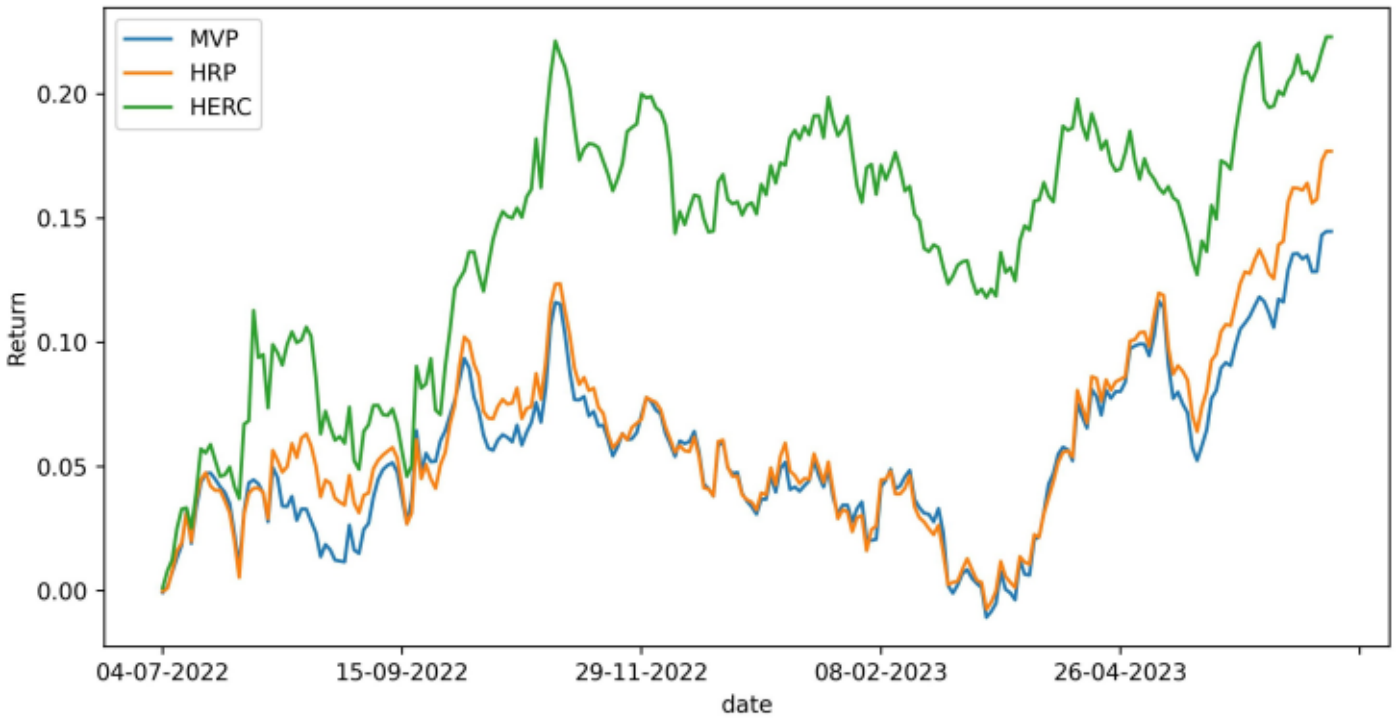

**Figure 1.46.** The cumulative returns yielded by the MVP, HRP, and HERC portfolios of the stocks from the *pharma* sector on the test data from July 1, 2022, to June 30, 2023.

**TABLE 1.22.** THE PERFORMANCE RESULTS OF THE PORTFOLIOS OF THE PHARMA SECTOR STOCKS

| **Period** | **MVP Portfolio** | | | **HRP Portfolio** | | | **HERC Portfolio** | | |
|---|---|---|---|---|---|---|---|---|---|
| | **Ret (%)** | **Vol (%)** | **SR** | **Ret (%)** | **Vol (%)** | **SR** | **Ret (%)** | **Vol (%)** | **SR** |
| Training | 25.38 | **20.71** | **1.2255** | 26.23 | 21.79 | 1.2039 | **26.53** | 29.33 | 0.9044 |
| Test | 14.88 | **12.79** | 1.1640 | 18.20 | 13.20 | 1.3788 | **22.94** | 15.89 | **1.4437** |

***Private Banks sector:*** The report published by the NSE on June 30, 2022, identified the top ten stocks in the *private banks* sector having the largest free-float market capitalization. These stocks and their respective contributions (in percent) to the overall index of the sector are as follows: (i) ICICI Bank (ICICIBANK): 25.96%, (ii) HDFC Bank (HDFCBANK): 25.17%, (iii) IndusInd Bank (INDUSINDBK): 10.58%, (iv) Kotak Mahindra Bank (KOTAKBANK): 10.15%, (v) Axis Bank (AXISBANK): 10.09%, (vi) Federal Bank (FEDERALBNK): 5.92%, (vii) IDFC First Bank (IDFCFIRSTB): 4.96%, (viii) Bandhan Bank (BANDHANBNK): 3.02%, (ix) RBL Bank (RBLBANK): 2.47%, and (x) City Union Bank (CUB): 1.75% (NSE Website). The ticker names of the stocks are mentioned in parentheses. The ticker names are the unique identifiers for the stocks listed on a stock exchange.

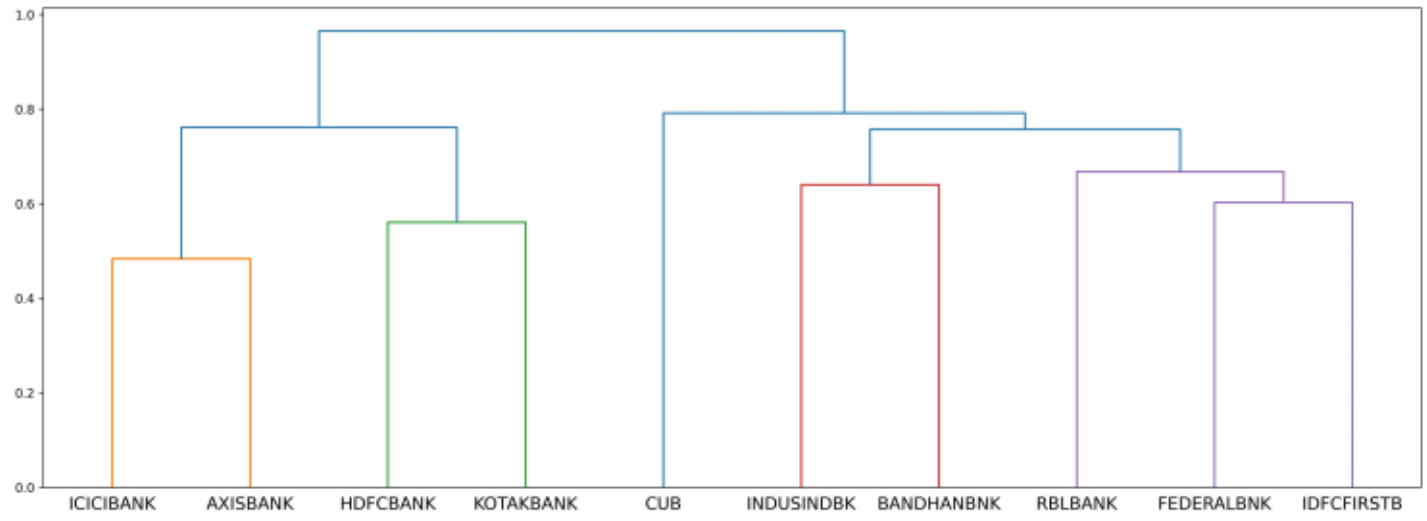

**Figure 1.47.** The dendrogram of the agglomerative clustering of the stocks from the *private banks* sector is created based on the historical stock prices from July 1, 2019, to June 30, 2022.

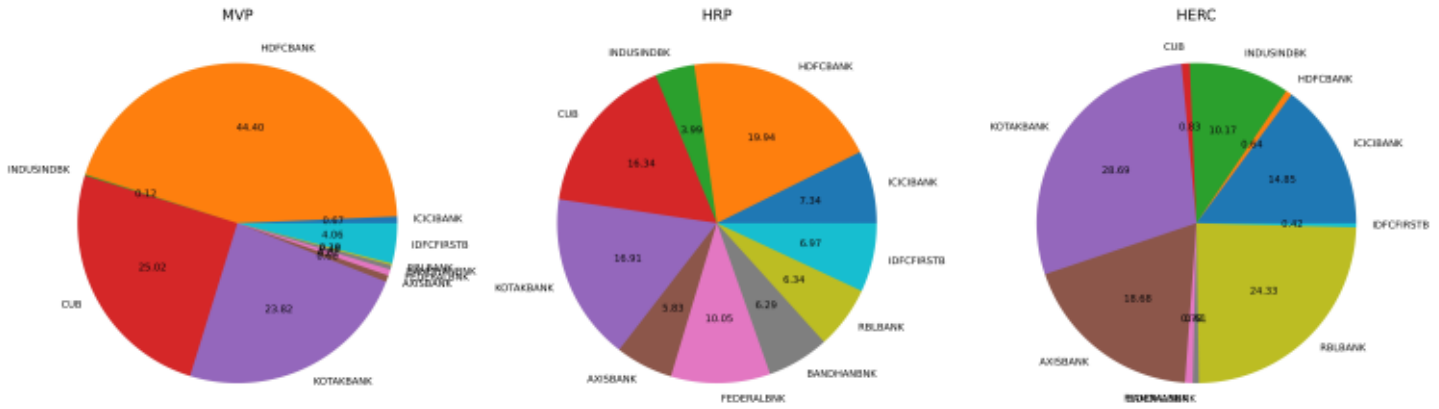

**Figure 1.48.** The allocation of weights done by the MVP, HRP, and HERC algorithms for the stocks of the *private banks* sector based on the historical stock prices from July 1, 2019, to June 30, 2022.

The dendrogram of the clustering of the stocks of the private banks sector is shown in Figure 1.47. The cluster containing the stocks ICICIBANK and AXISBANK is the most compact, while the one containing INDUSINDBK and BANDHANBNK is the least homogeneous. Figure 1.48 depicts the weight allocation done by the three portfolios to the private banks sector stocks. Table 1.23 shows the weight allocations for the three portfolios in tabular format.

**TABLE 1.23.** THE PORTFOLIO COMPOSITIONS OF THE PRIVATE BANKS SECTOR
(PERIOD: JULY 1, 2019 – JUNE 30, 2022)

| Stock | MVP Portfolio | HRP Portfolio | HERC Portfolio |
|---|---|---|---|
| ICICIBANK | 0.0067 | 0.0734 | 0.1485 |
| HDFCBANK | 0.4440 | 0.1994 | 0.0064 |
| INDUSINDBK | 0.0012 | 0.0399 | 0.1017 |
| KOTAKBANK | 0.2382 | 0.1691 | 0.2869 |
| AXISBANK | 0.0066 | 0.0583 | 0.1868 |
| FEDERALBNK | 0.0061 | 0.1005 | 0.0078 |
| IDFCFIRSTB | 0.0406 | 0.0697 | 0.0042 |
| BANDHANBNK | 0.0045 | 0.0629 | 0.0061 |
| RBLBANK | 0.0019 | 0.0634 | 0.2433 |
| CUB | 0.2502 | 0.1634 | 0.0083 |

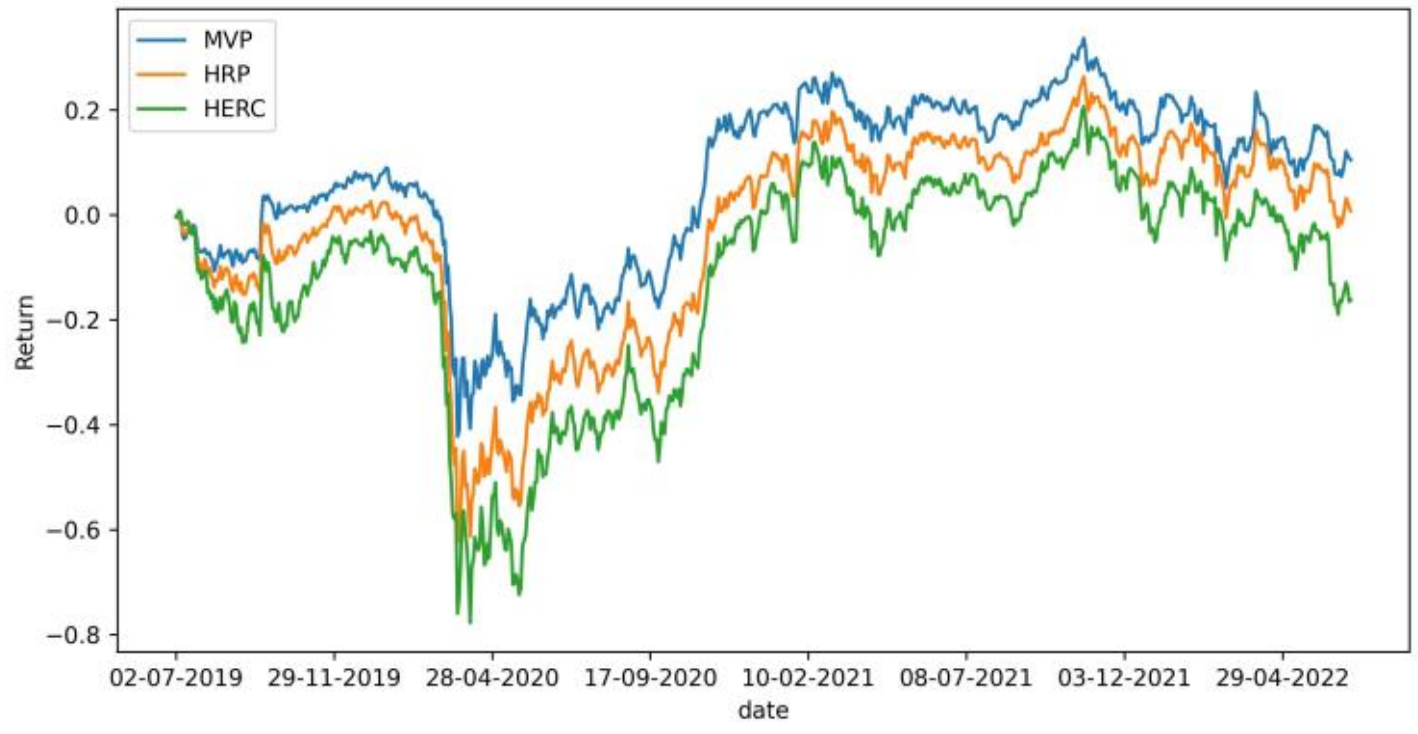

**Figure 1.49.** The cumulative returns yielded by the MVP, HRP, and HERC portfolios of the stocks from the *private banks* sector on the training data from July 1, 2019, to June 30, 2022.

The cumulative returns yielded by the portfolios over the training and the test periods are depicted in Figure 1.49 and Figure 1.50, respectively. Table 1.24 presents the cumulative returns, annual volatilities, and the Sharpe ratios of the three portfolios of the private banks sector stocks over the training and the test periods. The highest cumulative return, the lowest volatility, and the maximum Sharpe ratio over the portfolio training and test periods are indicated in red.

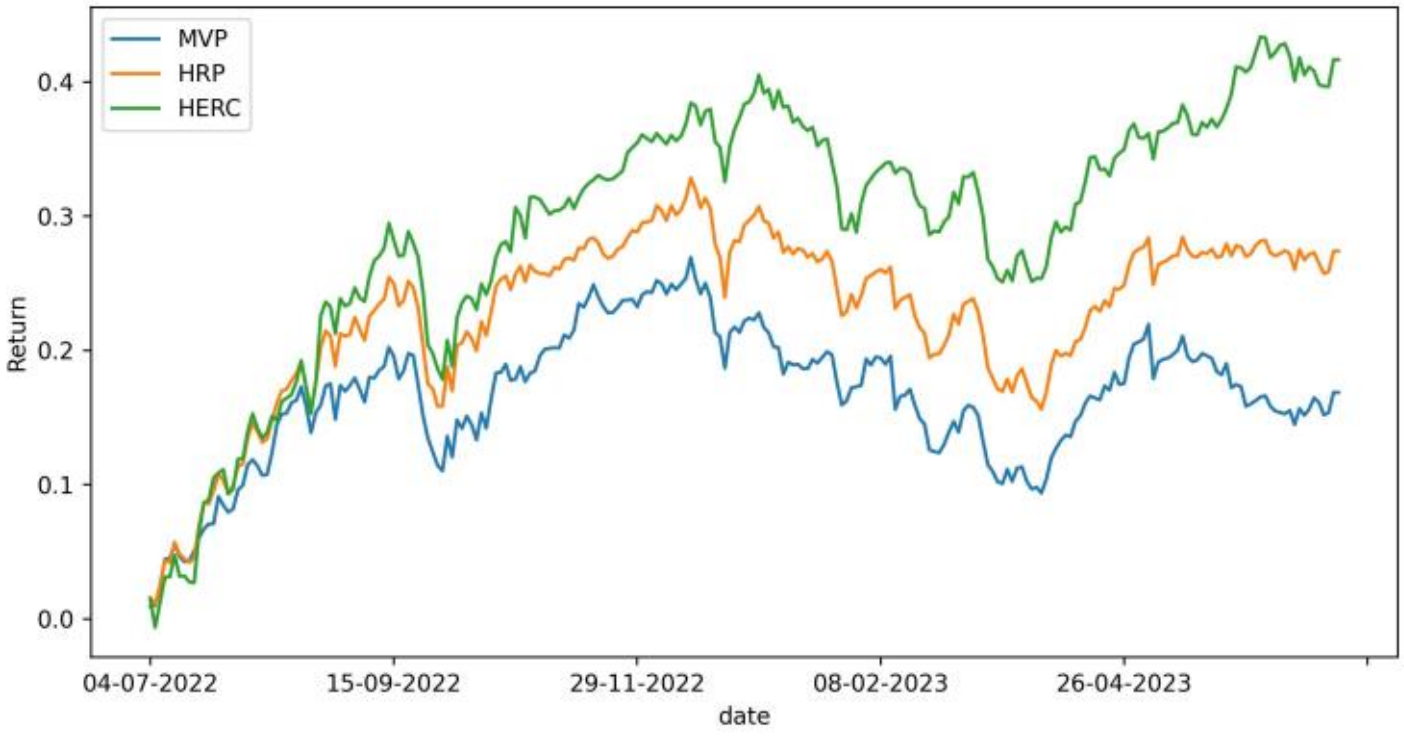

**Figure 1.50.** The cumulative returns yielded by the MVP, HRP, and HERC portfolios of the stocks from the *private banks* sector on the test data from July 1, 2022, to June 30, 2023.

**TABLE 1.24.** THE PERFORMANCE RESULTS OF THE PORTFOLIOS OF THE PRIVATE BANKS SECTOR STOCKS

| Period | MVP Portfolio | | | HRP Portfolio | | | HERC Portfolio | | |
|---|---|---|---|---|---|---|---|---|---|
| | Ret (%) | Vol (%) | SR | Ret (%) | Vol (%) | SR | Ret (%) | Vol (%) | SR |
| Training | **3.56** | **28.42** | **0.1253** | 0.27% | 31.62 | 0.0084 | -5.49 | 37.23 | -0.1474 |
| Test | 17.35 | **16.94** | 1.0245 | 28.17 | 17.65 | 1.5963 | **42.85** | 20.17 | **2.1239** |

***PSU Banks sector:*** As per the report published by the NSE on June 30, 2022, the ten stocks that have the largest free-float market capitalization in this sector, and their respective contributions (in percent) to the overall index of the sector are as follows: (i) State Bank of India (SBIN): 28.13%, (ii) Bank of Baroda (BANKBARODA): 19.53%, (iii) Punjab National Bank (PNB): 12.58%, (iv) Canara Bank (CANBK): 11.99%, (v) Union Bank of India (UNIONBANK): 8.22%, (vi) Indian Bank (INDIANB): 6.87%, (vii) Bank of India (BANKINDIA): 5.29%, (viii) Bank of Maharashtra (MAHABANK): 2.65%, (ix) Indian Overseas Bank (IOB): 1.58%, and (x) Central Bank of India (CENTRALBK): 1.46% (NSE Website). The ticker names of the stocks, which are their unique identifiers, are mentioned in parentheses.

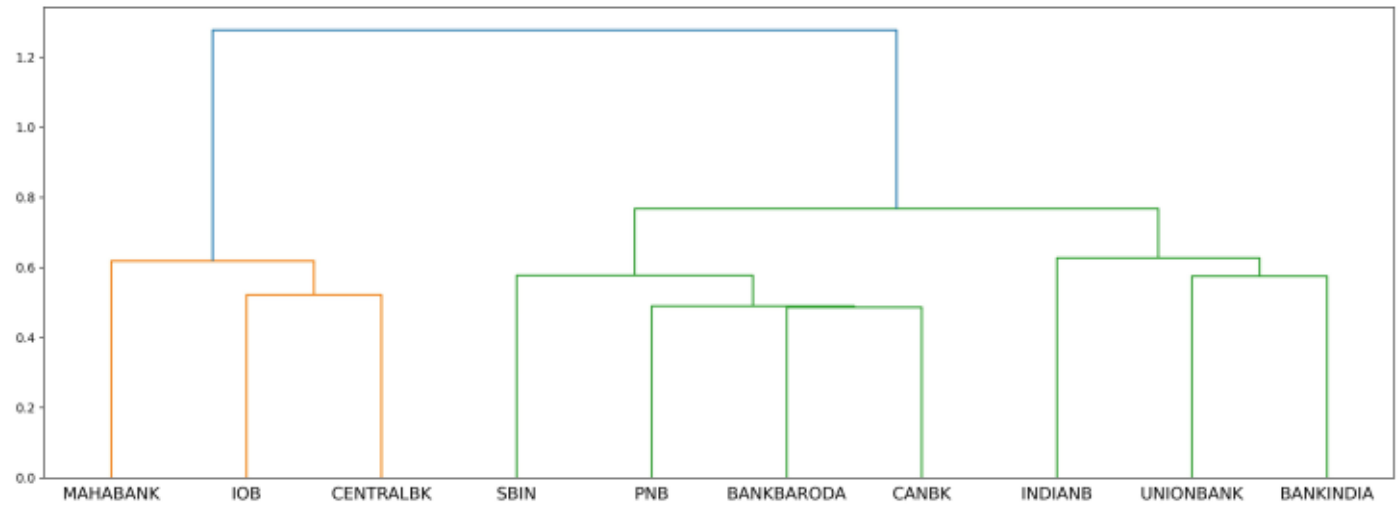

**Figure 1.51.** The dendrogram of the agglomerative clustering of the stocks from the *PSU banks* sector is created based on the historical stock prices from July 1, 2019, to June 30, 2022.

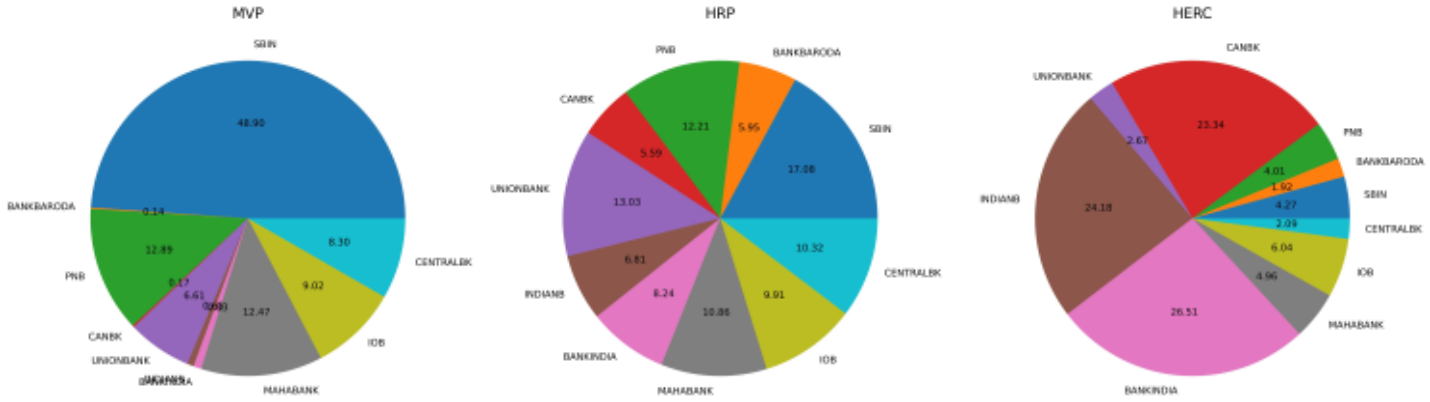

**Figure 1.52.** The allocation of weights done by the MVP, HRP, and HERC algorithms for the stocks of the *PSU banks* sector based on the historical stock prices from July 1, 2019, to June 30, 2022.

The dendrogram of the clustering of the stocks of the PSU banks sector is shown in Figure 1.51. The cluster containing the stocks BANKBARODA and CANBK is the most compact, while the one containing UNIONBANK and BANKINDIA is the least homogeneous. Figure 1.52 depicts the weight allocation done by the three portfolios to the PSU banks sector stocks. Table 1.25 shows the weight allocations for the three portfolios in tabular format.

**TABLE 1.25.** THE PORTFOLIO COMPOSITIONS OF THE PSU BANKS SECTOR
(PERIOD: JULY 1, 2019 – JUNE 30, 2022)

| Stock | MVP Portfolio | HRP Portfolio | HERC Portfolio |
|---|---|---|---|
| SBIN | 0.4890 | 0.1708 | 0.0427 |
| BANKBARODA | 0.0014 | 0.0595 | 0.0192 |
| PNB | 0.1289 | 0.1221 | 0.0401 |
| CANBK | 0.0017 | 0.0559 | 0.2334 |
| UNIONBANK | 0.0661 | 0.1303 | 0.0267 |
| INDIANB | 0.0065 | 0.0681 | 0.2418 |
| BANKINDIA | 0.0083 | 0.0824 | 0.2651 |
| MAHABANK | 0.1247 | 0.1086 | 0.0496 |
| IOB | 0.0902 | 0.0991 | 0.0604 |
| CENTRALBK | 0.0830 | 0.1032 | 0.0209 |

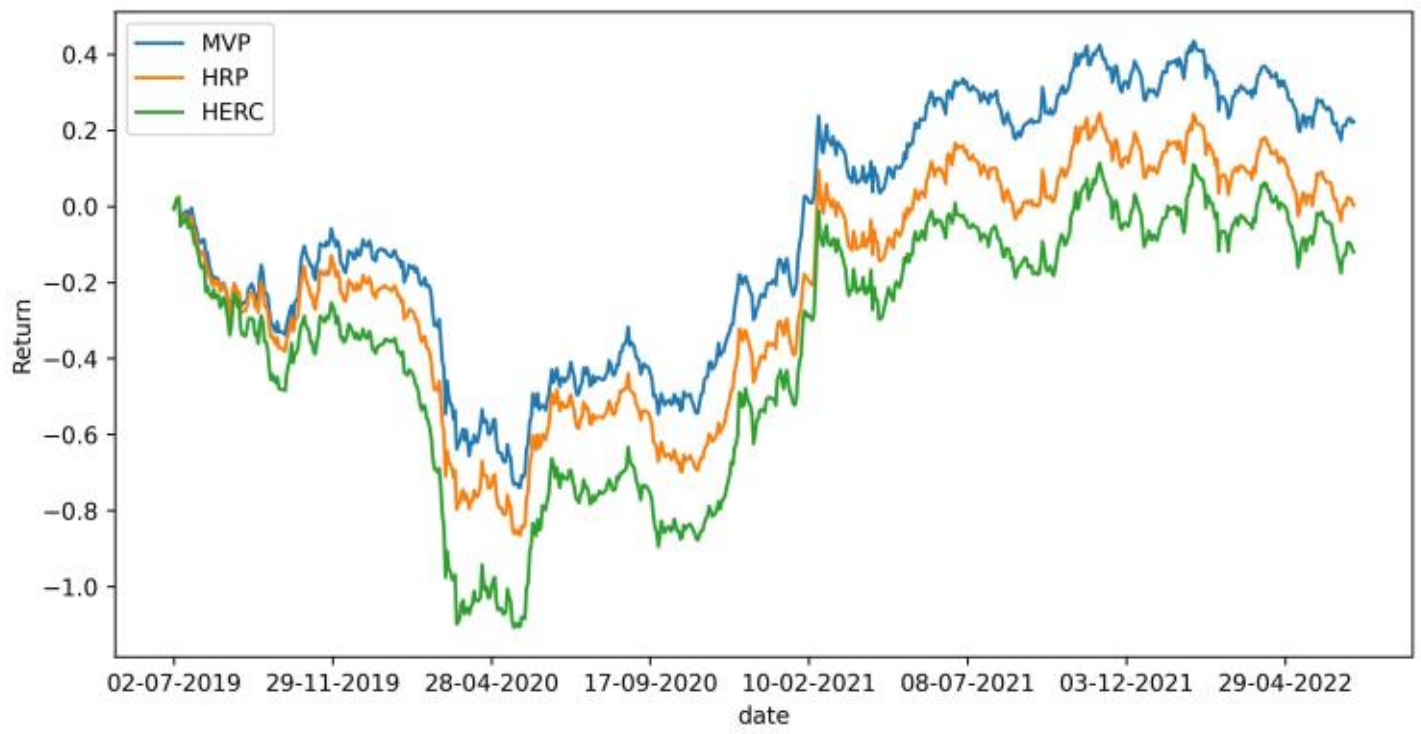

**Figure 1.53.** The cumulative returns yielded by the MVP, HRP, and HERC portfolios of the stocks from the *PSU banks* sector on the training data from July 1, 2019, to June 30, 2022.

The cumulative returns yielded by the portfolios over the training and the test periods are depicted in Figure 1.53 and Figure 1.54, respectively. Table 1.26 presents the cumulative returns, annual volatilities, and the Sharpe ratios of the three portfolios of the PSU banks sector stocks over the training and the test periods. The highest cumulative return, the lowest volatility, and the maximum Sharpe ratio over the portfolio training and test periods are indicated in red.

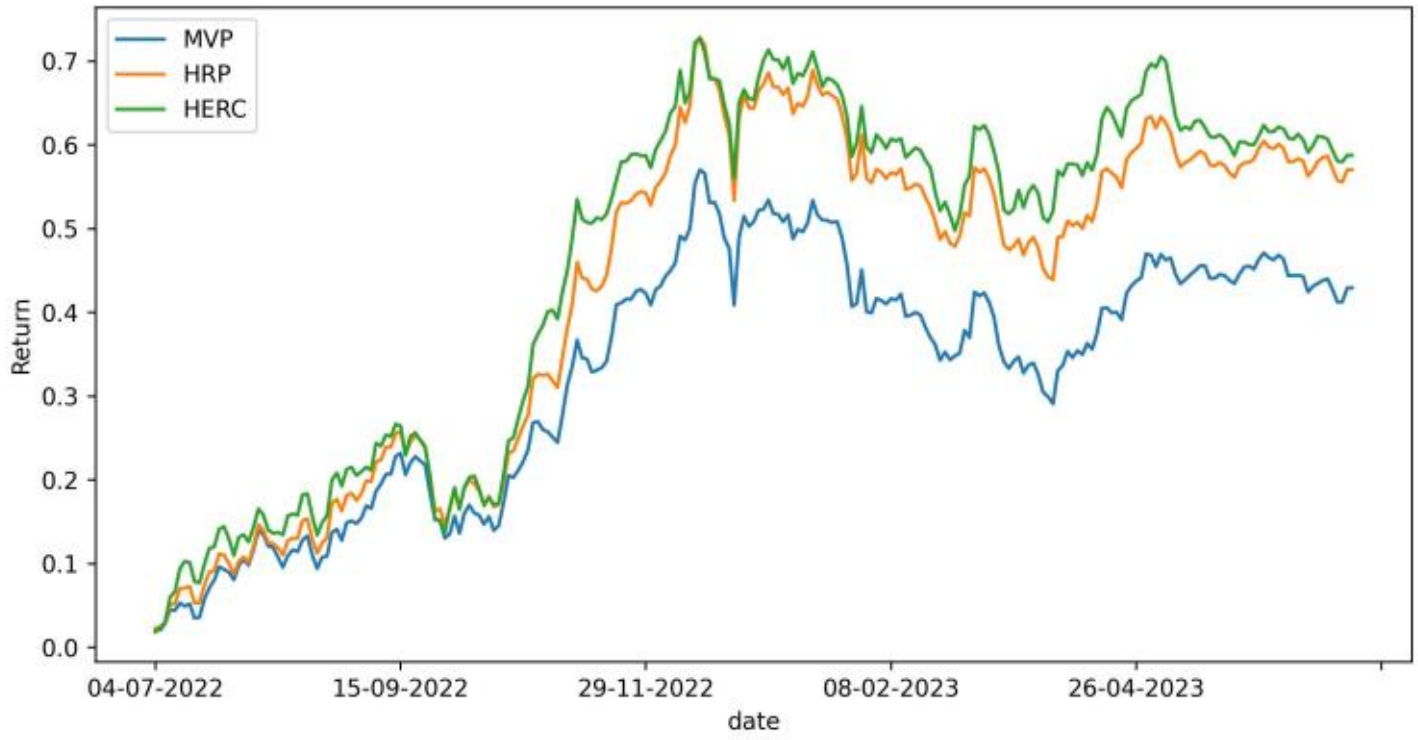

**Figure 1.54.** The cumulative returns yielded by the MVP, HRP, and HERC portfolios of the stocks from the *PSU banks* sector on the test data from July 1, 2022, to June 30, 2023.

**TABLE 1.26.** THE PERFORMANCE RESULTS OF THE PORTFOLIOS OF THE PSU BANKS SECTOR STOCKS

| Period | MVP Portfolio | | | HRP Portfolio | | | HERC Portfolio | | |
|---|---|---|---|---|---|---|---|---|---|
| | Ret (%) | Vol (%) | SR | Ret (%) | Vol (%) | SR | Ret (%) | Vol (%) | SR |
| Training | **7.55** | 35.28 | **0.2141** | 0.13 | 37.02 | 0.0035 | -0.41 | **30.44** | -0.1002 |
| Test | 44.16 | **27.34** | 1.6153 | 58.67 | 31.21 | 1.8797 | **60.45** | 31.97 | **1.8910** |

***Realty sector:*** As per the report published by the NSE on June 30, 2022, the ten stocks that have the largest free-float market capitalization in the *realty* sector and their contributions (in percent) to the overall index of the sector are as follows: (i) DLF (DLF): 25.82%, (ii) Godrej Properties (GODREJPROP): 15.95%, (iii) Macrotech Developers (LODHA): 14.71%, (iv) Phoenix Mills (PHOENIXLTD): 12.93%, (v) Oberoi Realty (OBEROIRLTY): 10.48%, (vi) Prestige Estate Projects (PRESTIGE): 6.64%, (vii) Brigade Enterprises (BRIGADE): 5.93%, (viii) Mahindra Lifespace Developers (MAHLIFE): 3.09%, (ix) Indiabulls Real Estate (IBREALEST): 2.65%, and (x) Sobha (SOBHA): 1.80% (NSE Website). Since the stock of Macrotech Developers (LODHA) was listed later than the starting date of the portfolio analysis, i.e., July 1, 2019, it could not be considered in the realty sector portfolio design. The stock of LODHA was first listed on NSE on April 19, 2021. Hence, in place of it, the stock of Sunteck Realty (SUNTECK) is considered as it has the highest free float market capitalization among the remaining stocks in the realty sector.

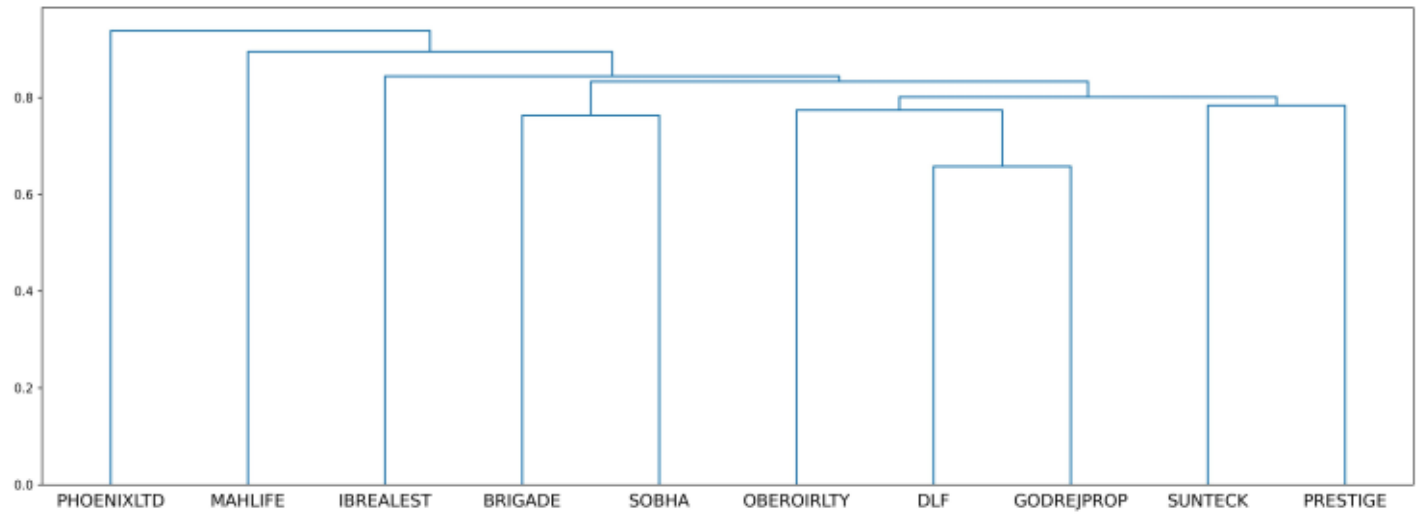

**Figure 1.55.** The dendrogram of the agglomerative clustering of the stocks from the *realty* sector is created based on the historical stock prices from July 1, 2019, to June 30, 2022.

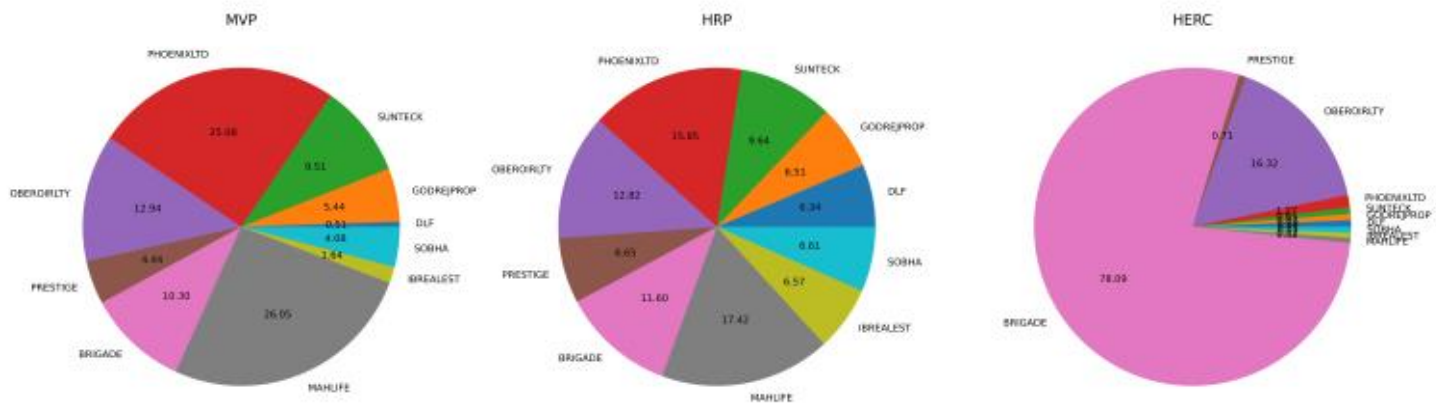

**Figure 1.56.** The allocation of weights done by the MVP, HRP, and HERC algorithms for the stocks of the *realty* sector based on the historical stock prices from July 1, 2019, to June 30, 2022.

The dendrogram of the clustering of the stocks of the realty sector is shown in Figure 1.55. The cluster containing the stocks DLF and GODREJPROP is the most compact, while the one containing SUNTECK and PRESTIGE is the least homogeneous. Figure 1.56 depicts the weight allocation done by the three portfolios to the realty sector stocks. Table 1.27 shows the weight allocations for the three portfolios in tabular format.

**TABLE 1.27.** THE PORTFOLIO COMPOSITIONS OF THE REALTY SECTOR
(PERIOD: JULY 1, 2019 – JUNE 30, 2022)

| Stock | MVP Portfolio | HRP Portfolio | HERC Portfolio |
|---|---|---|---|
| DLF | 0.0051 | 0.0634 | 0.0068 |
| GODREJPROP | 0.0544 | 0.0651 | 0.0063 |
| SUNTECK | 0.0951 | 0.0964 | 0.0065 |
| PHOENIXLTD | 0.2508 | 0.1585 | 0.0127 |
| OBEROIRLTY | 0.1294 | 0.1282 | 0.1632 |
| PRESTIGE | 0.0444 | 0.0665 | 0.0071 |
| BRIGADE | 0.1030 | 0.1160 | 0.7809 |
| MAHLIFE | 0.2605 | 0.1742 | 0.0052 |
| IBREALEST | 0.0164 | 0.0657 | 0.0052 |
| SOBHA | 0.0408 | 0.0661 | 0.0061 |

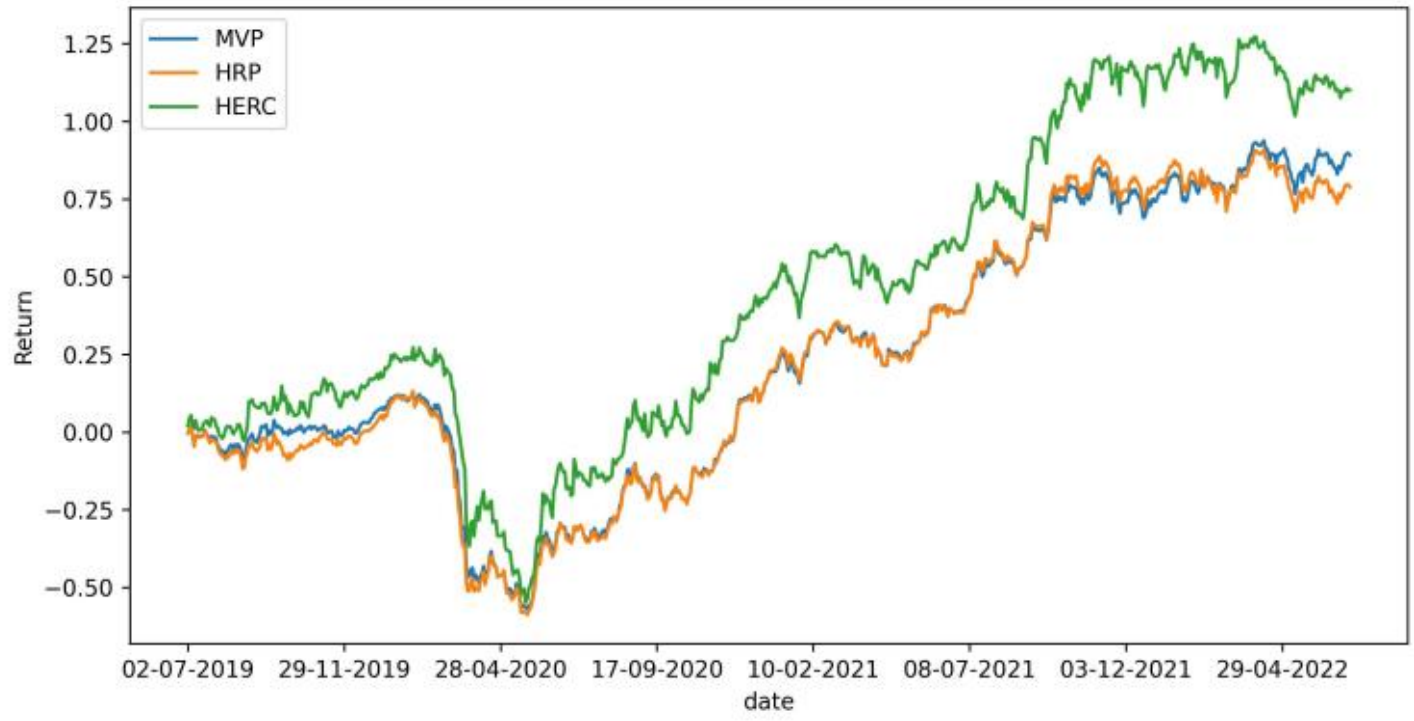

**Figure 1.57.** The cumulative returns yielded by the MVP, HRP, and HERC portfolios of the stocks from the *realty* sector on the training data from July 1, 2019, to June 30, 2022.

The cumulative returns yielded by the portfolios over the training and the test periods are depicted in Figure 1.57 and Figure 1.58, respectively. Table 1.28 presents the cumulative returns, annual volatilities, and the Sharpe ratios of the three portfolios of the realty sector stocks over the training and the test periods. The highest cumulative return, the lowest volatility, and the maximum Sharpe ratio over the portfolio training and test periods are indicated in red.

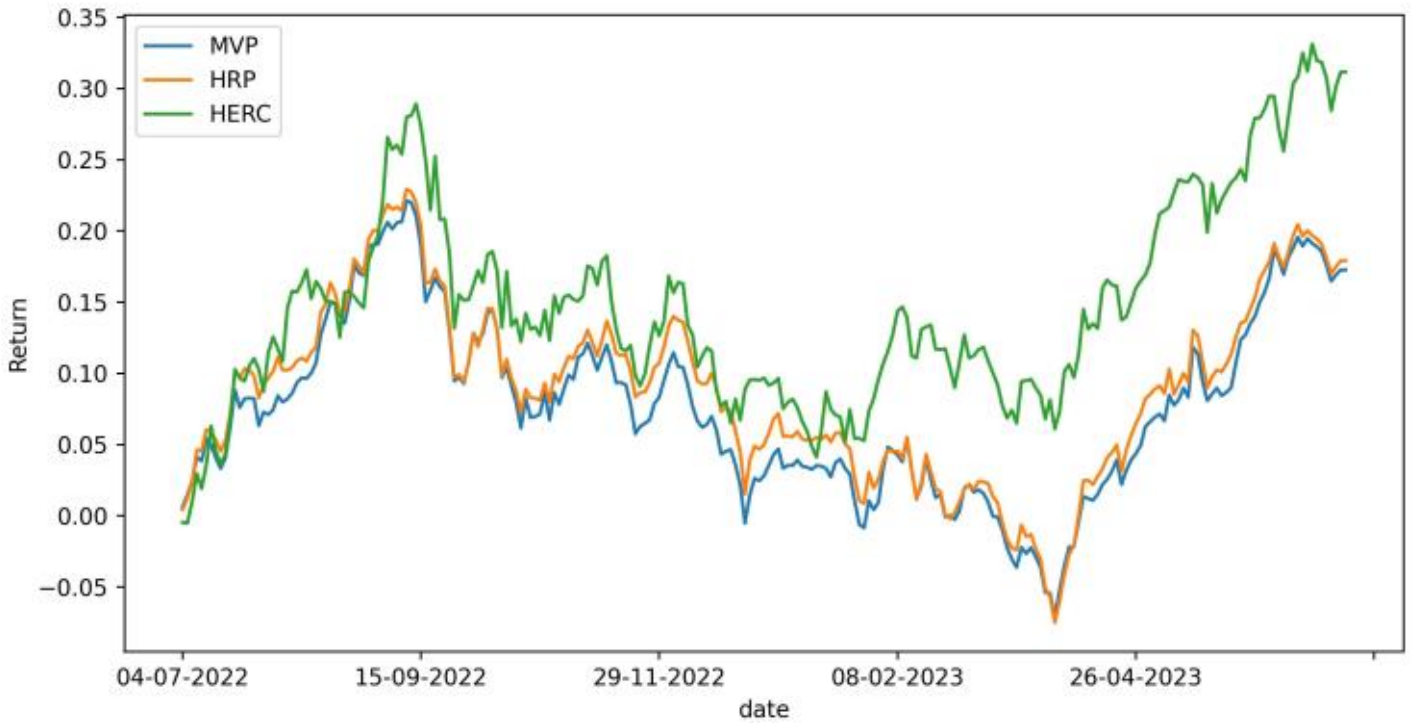

**Figure 1.58.** The cumulative returns yielded by the MVP, HRP, and HERC portfolios of the stocks from the *realty* sector on the test data from July 1, 2022, to June 30, 2023.

**TABLE 1.28.** THE PERFORMANCE RESULTS OF THE PORTFOLIOS OF THE REALTY SECTOR STOCKS

| Period | MVP Portfolio | | | HRP Portfolio | | | HERC Portfolio | | |
|---|---|---|---|---|---|---|---|---|---|
| | Ret (%) | Vol (%) | SR | Ret (%) | Vol (%) | SR | Ret (%) | Vol (%) | SR |
| Training | 30.21 | **29.35** | **1.0293** | 26.72 | 30.26 | 0.8831 | **37.29** | 40.66 | 0.9172 |
| Test | 17.76 | **18.29** | 0.9708 | 18.43 | 18.49 | 0.9966 | **32.06** | 25.59 | **1.2517** |

***NIFTY 50 stocks:*** Finally, we consider the NIFTY 50 stocks and construct three portfolios. The NIFTY 50 stocks are the market leaders across 14 sectors in the NSE and have a low-risk quotient. The NIFTY 50 index serves as the benchmark index of the Indian stock market. The index represents the weighted average of the 50 largest companies in India listed on the NSE of India (NIFTY 50 Wiki Page).

As per the NSE's report published on June 30, 2022, the stocks included in the NIFTY 50 group are the following: Adani Enterprises (ADANIENT), Adani Ports & SEZ (ADANIPORTS), Apollo Hospitals (APOLLOHOSP), Asian Paints (ASIANPAINT), Axis Bank (AXISBANK), Bajaj Auto (BAJAJ-AUTO), Bajaj Finance (BAJFINANCE), Bajaj Finserv (BAJAJFINSV), Bharat Petroleum Corporation (BPCL), Bharti Airtel (BHARTIARTL), Britannia Industries (BRITANNIA), Cipla (CIPLA), Coal India (COALINDIA), Divi's Laboratories (DIVISLAB), Dr. Reddy's Laboratories (DRREDDY), Eicher Motors (EICHERMOT), Grasim Industries (GRASIM), HCL Technologies (HCLTECH), HDFC Bank (HDFCBANK), HDFC Life (HDFCLIFE), Hero MotoCorp (HEROMOTOCO), Hindalco Industries (HINDALCO), Hindustan Unilever (HINDUNILVR), ICICI Bank (ICICIBANK), IndusInd Bank (INDUSINDBK), Infosys (INFY), ITC (ITC). JSW Steel (JSWSTEEL). Kotak Mahindra Bank (KOTAKBANK), Larsen & Toubro (LT), LTIMindtree (LTIM), Mahindra & Mahindra (M&M), Maruti Suzuki (MARUTI), Nestle India (NESTLEIND), NTPC (NTPC), Oil & Natural Gas Corporation (ONGC), Power Grid Corporation (POWERGRID), Reliance Industries (RELIANCE), SBI Life Insurance Company (SBILIFE), State Bank of India (SBIN), Sun Pharmaceutical Industries (SUNPHARMA), Tata Motors (TATAMOTORS), Tata Consultancy Services (TCS), Tata Consumer Products (TATACONSUM), Tech Mahindra (TECHM), Titan Company (TITAN), UltraTech Cement (ULTRACEMCO), UPL (UPL), and Wipro (WIPRO) (NIFTY 50 Website). The ticker names of the stocks are mentioned in parentheses. The ticker names are the unique identifiers of the stocks on the NSE.

On July 31, 2023, the contributions of different sectors to the composite NIFTY 50 index are as follows: Financial Services including Banks: 37.68%, Information Technology: 12.90%, Oil and Gas: 11.67%, Consumer Goods: 9.47%, Auto: 5.94%, Healthcare: 4.04%, Construction: 3.71%, Metals & Mining: 3.65%, Consumer Durables: 3.18%, Telecom: 2.56%, Power: 2.23%, Construction: 1.88%, Services: 0.75%, and Chemicals: 0.36% (NSE Website).

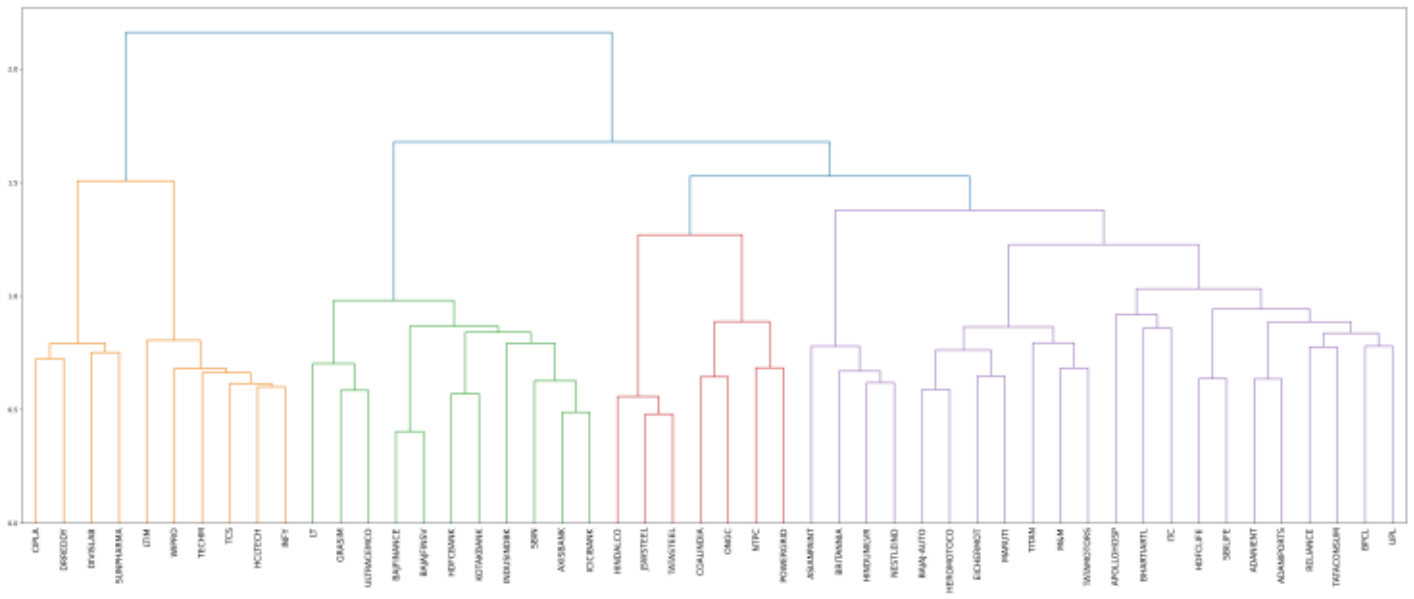

**Figure 1.59.** The dendrogram of the agglomerative clustering of the NIFTY 50 stocks is created based on the historical stock prices from July 1, 2019, to June 30, 2022.

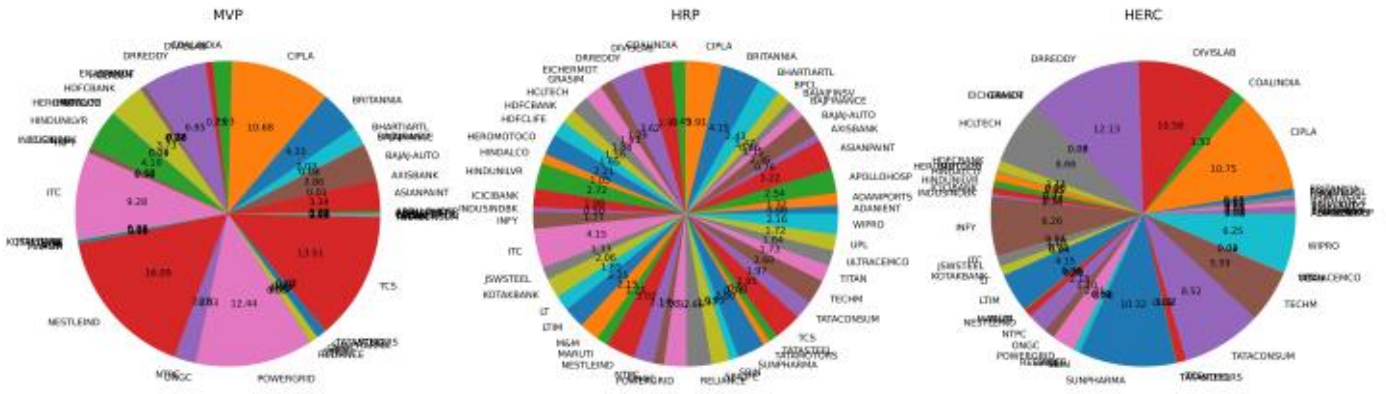

**Figure 1.60.** The allocation of weights done by the MVP, HRP, and HERC algorithms for the NIFTY 50 stocks based on the historical stock prices from July 1, 2019, to June 30, 2022.

The dendrogram of the clustering of the NIFTY 50 is shown in Figure 1.59. The cluster containing the stocks BAJFINANCE and BAJAJFINSV is the most compact, while the one containing BHARTIARTL and ITC is the least homogeneous. Figure 1.60 depicts the weight allocation done by the three portfolios to the NIFTY 50 stocks. Table 1.29 shows the weight allocations for the three portfolios in tabular format.

**TABLE 1.29.** THE PORTFOLIO COMPOSITIONS OF THE NIFTY 50 STOCKS
(PERIOD: JULY 1, 2019 – JUNE 30, 2022)

| Stock | MVP | HRP | HERC | Stock | MVP | HRP | HERC |
|---|---|---|---|---|---|---|---|
| ADANIENT | 0.0002 | 0.0088 | 0.0002 | INFY | 0.0056 | 0.0177 | 0.0626 |
| ADANIPORTS | 0.0003 | 0.0132 | 0.0003 | ITC | 0.0928 | 0.0415 | 0.0004 |
| APOLLOHOSP | 0.0022 | 0.0254 | 0.0002 | JSWSTEEL | 0.0001 | 0.0133 | 0.0110 |
| ASIANPAINT | 0.0334 | 0.0322 | 0.0004 | KOTAKBANK | 0.0007 | 0.0206 | 0.0094 |
| AXISBANK | 0.0001 | 0.0076 | 0.0057 | LT | 0.0006 | 0.0165 | 0.0004 |
| BAJAJ-AUTO | 0.0386 | 0.0236 | 0.0008 | LTIM | 0.0031 | 0.0225 | 0.0415 |
| BAJFINANCE | 0.0002 | 0.0115 | 0.0061 | M&M | 0.0003 | 0.0215 | 0.0005 |
| BAJAJFINSV | 0.0002 | 0.0086 | 0.0053 | MARUTI | 0.0003 | 0.0125 | 0.0006 |
| BPCL | 0.0003 | 0.0147 | 0.0003 | NESTLEIND | 0.1609 | 0.0307 | 0.0080 |
| BHARTIARTL | 0.0203 | 0.0243 | 0.0003 | NTPC | 0.0228 | 0.0214 | 0.0218 |
| BRITANNIA | 0.0433 | 0.0415 | 0.0067 | ONGC | 0.0003 | 0.0108 | 0.0110 |
| CIPLA | 0.1068 | 0.0391 | 0.1075 | POWERGRID | 0.1244 | 0.0237 | 0.0241 |
| COALINDIA | 0.0203 | 0.0149 | 0.0152 | RELIANCE | 0.0005 | 0.0266 | 0.0003 |
| DIVISLAB | 0.0075 | 0.0297 | 0.1058 | SBILIFE | 0.0093 | 0.0197 | 0.0004 |
| DRREDDY | 0.0685 | 0.0362 | 0.1213 | SBIN | 0.0003 | 0.0093 | 0.0070 |
| EICHERMOT | 0.0036 | 0.0126 | 0.0006 | SUNPHARMA | 0.010 | 0.0290 | 0.1032 |
| GRASIM | 0.0002 | 0.0191 | 0.0003 | TATAMOTORS | 0.0001 | 0.0076 | 0.0004 |
| HCLTECH | 0.0024 | 0.0188 | 0.0666 | TATASTEEL | 0.0002 | 0.0091 | 0.0002 |
| HDFCBANK | 0.0373 | 0.0156 | 0.0111 | TCS | 0.1351 | 0.0295 | 0.0102 |
| HDFCLIFE | 0.0004 | 0.0165 | 0.0003 | TATACONSUM | 0.0004 | 0.0197 | 0.0852 |
| HEROMOTOCO | 0.0004 | 0.0221 | 0.0006 | TECHM | 0.0004 | 0.0260 | 0.0559 |
| HINDALCO | 0.0001 | 0.0105 | 0.0095 | TITAN | 0.0005 | 0.017 | 0.0003 |
| HINDUNILVR | 0.0419 | 0.0272 | 0.0071 | ULTRACEMCO | 0.0004 | 0.0164 | 0.0004 |
| ICICIBANK | 0.0002 | 0.0198 | 0.0072 | UPL | 0.0003 | 0.0172 | 0.0002 |
| INDUSINDBK | 0.0001 | 0.0050 | 0.0030 | WIPRO | 0.0018 | 0.0216 | 0.0625 |

The cumulative returns yielded by the portfolios over the training and the test periods are depicted in Figure 1.61 and Figure 1.62, respectively. Table 1.30 presents the cumulative returns, annual volatilities, and the Sharpe ratios of the three portfolios of the NIFTY 50 stocks over the training and the test periods. The highest cumulative return, the lowest volatility, and the maximum Sharpe ratio over the portfolio training and test periods are indicated in red.

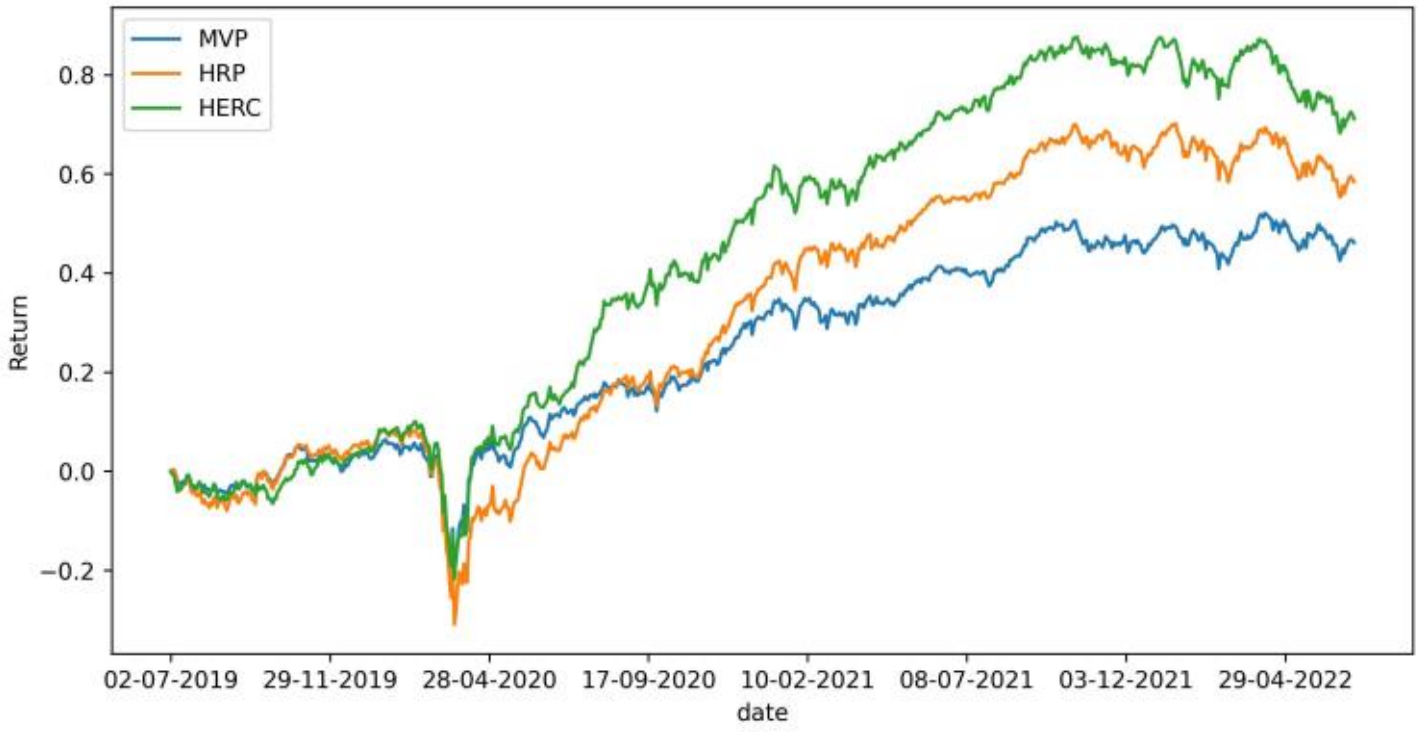

**Figure 1.61.** The cumulative returns yielded by the MVP, HRP, and HERC portfolios of the NIFTY 50 stocks on the training data from July 1, 2019, to June 30, 2022.

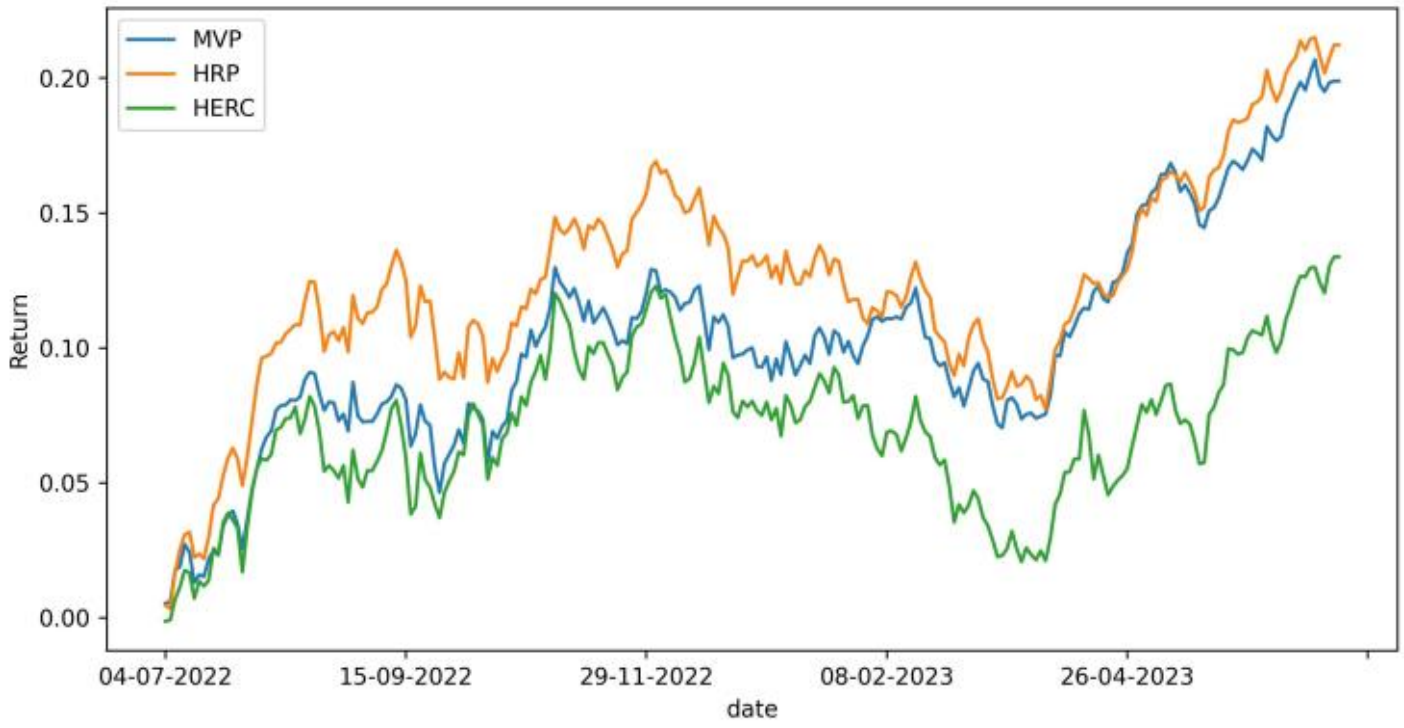

**Figure 1.62.** The cumulative returns yielded by the MVP, HRP, and HERC portfolios of the NIFTY 50 stocks on the test data from July 1, 2022, to June 30, 2023.

**TABLE 1.30.** THE PERFORMANCE RESULTS OF THE PORTFOLIOS OF THE NIFTY 50 STOCKS

| Period | MVP Portfolio | | | HRP Portfolio | | | HERC Portfolio | | |
|---|---|---|---|---|---|---|---|---|---|
| | Ret (%) | Vol (%) | SR | Ret (%) | Vol (%) | SR | Ret (%) | Vol (%) | SR |
| Training | 15.63 | **17.16** | 0.9106 | 19.80 | 20.35 | 0.9727 | **24.10** | 20.42 | **1.1802** |
| Test | 20.46 | **9.47** | **2.1607** | **21.84** | 10.28 | 2.1238 | 13.76 | 11.55 | 1.1914 |

**TABLE 1.31.** THE SUMMARY OF THE PERFORMANCES OF THE PORTFOLIOS ON THE TRAINING DATA (PERIOD: JULY 1, 2019 – JUNE 30, 2022)

| Sector | MVP | | | HRP | | | HERC | | |
|---|---|---|---|---|---|---|---|---|---|
| | Annual Return | Annual Vol | Sharpe Ratio | Annual Return | Annual Vol | Sharpe Ratio | Annual Return | Annual Vol | Sharpe Ratio |
| Auto | 26.32% | 25.35 | 1.0385 | 26.28% | 27.18 | 0.9670 | 29.38% | 41.92 | 0.7009 |
| Banking | 10.81% | 28.79% | 0.3755 | 9.91% | 31.92% | 0.3104 | 6.45% | 34.12% | 0.1890 |
| Fin. Services | 6.81% | 24.02% | 0.2834 | 10.48% | 25.88% | 0.4051 | 9.56% | 32.25% | 0.2956 |
| Cons Durables | 16.80% | 20.05% | 0.8380 | 20.13% | 20.63% | 0.9757 | 13.16% | 26.33% | 0.4998 |
| FMCG | 15.65% | 18.33% | 0.8539 | 15.83% | 18.97% | 0.8347 | 12.47% | 20.53% | 0.6071 |
| IT | 26.83% | 24.24% | 1.1023 | 29.82% | 25.42% | 1.1730 | 29.60% | 28.87% | 1.0253 |
| Media | 21.11% | 26.42% | 0.7991 | 24.85% | 27.90% | 0.8906 | 22.26% | 34.02% | 0.6543 |
| Metal | 52.03% | 33.70% | 1.5441 | 47.14% | 35.86% | 1.3145 | 46.10% | 37.29% | 1.2363 |
| Mid-Small IT & Telecom | 32.93% | 23.96% | 1.3744 | 38.43% | 24.83% | 1.5475 | 41.57% | 28.86% | 1.4404 |
| Oil & Gas | 10.02% | 23.25% | 0.4310 | 15.38% | 24.86% | 0.6189 | 21.75% | 26.67% | 0.8158 |
| Pharma | 25.38% | 20.71% | 1.2255 | 26.23% | 21.79% | 1.2039 | 26.53% | 29.33% | 0.9044 |
| Private Banks | 3.56% | 28.42% | 0.1253 | 0.27% | 31.62% | 0.0084 | -5.49% | 37.23% | -0.1474 |
| PSU Banks | 7.55% | 35.28% | 0.2141 | 0.13% | 37.02% | 0.0035 | -0.41% | 30.44% | -0.1002 |
| Realty | 30.21% | 29.35% | 1.0293 | 26.72% | 30.26% | 0.8831 | 37.29% | 40.66% | 0.9172 |
| NIFTY 50 | 15.63% | 17.16% | 0.9106 | 19.80% | 20.35% | 0.9727 | 24.10% | 20.42% | 1.1802 |
| **Overall** | **4** | **14** | **8** | **6** | **0** | **5** | **5** | **1** | **2** |

**TABLE 1.32.** THE SUMMARY OF THE PERFORMANCES OF THE PORTFOLIOS ON THE TEST DATA (PERIOD: JULY 1, 2022 – JUNE 30, 2023)

| Sector | MVP | | | HRP | | | HERC | | |
|---|---|---|---|---|---|---|---|---|---|
| | Annual Return | Annual Vol | Sharpe Ratio | Annual Return | Annual Vol | Sharpe Ratio | Annual Return | Annual Vol | Sharpe Ratio |
| Auto | 35.03% | 16.07% | 2.1795 | 32.41% | 15.39% | 2.1061 | 19.32% | 20.52% | 0.9416 |
| Banking | 27.98% | 15.88% | 1.6436 | 40.49% | 17.73% | 2.0356 | 44.10% | 18.82% | 2.1931 |
| Fin. Services | 19.79% | 13.39% | 1.4775 | 22.19% | 13.97% | 1.5889 | 30.71% | 18.72% | 1.6407 |
| Cons Durables | 16.52% | 21.18% | 0.7802 | 16.30% | 18.92% | 0.8615 | 10.41% | 18.56% | 0.5610 |
| FMCG | 32.80% | 12.75% | 2.5717 | 30.54% | 12.13% | 2.5170 | 19.98% | 14.30% | 1.3977 |
| IT | 5.78% | 19.25% | 0.3006 | 8.54% | 20.28% | 0.4214 | 11.65% | 21.00% | 0.5546 |
| Media | 17.54% | 21.66% | 0.8100 | 13.50% | 21.00% | 0.6431 | 9.13% | 22.75% | 0.4013 |
| Metal | 36.53% | 24.02% | 1.5210 | 43.03% | 23.70% | 1.8153 | 43.64% | 23.23% | 1.8783 |
| Mid-Small IT & Telecom | 36.98% | 16.96% | 2.1798 | 37.81% | 17.89% | 2.1131 | 28.32% | 20.33% | 1.3931 |
| Oil & Gas | 10.09% | 13.67% | 0.7386 | 7.27% | 13.76% | 0.5286 | -1.81% | 16.50% | -0.1097 |
| Pharma | 14.88% | 12.79% | 1.1640 | 18.20% | 13.20% | 1.3788 | 22.94% | 15.89% | 1.4437 |
| Private Banks | 17.35% | 16.94% | 1.0245 | 28.17% | 17.65% | 1.5963 | 42.85% | 20.17% | 2.1239 |
| PSU Banks | 44.16% | 27.34% | 1.6153 | 58.67% | 31.21% | 1.8797 | 60.45% | 31.97% | 1.8910 |
| Realty | 17.76% | 18.29 | 0.9708 | 18.43% | 18.49% | 0.9966 | 32.06% | 25.59% | 1.2517 |
| NIFTY 50 | 20.46% | 9.47% | 2.1607 | 21.84% | 10.28% | 2.1238 | 13.76% | 11.55% | 1.1914 |
| **Overall** | **7** | **10** | **6** | **2** | **3** | **1** | **6** | **2** | **8** |

Table 1.31 presents the summary of the results of the performances of the three portfolios on the training data. The results of the same portfolios on the test data are depicted in Table 1.32.

Some important observations are found in the results presented in Table 1.31 and Table 1.32. These observations are listed in the following.

First, on the training data, MVP portfolios have yielded the highest Sharpe ratio for 8 among the 15 sectors. The number of sectors for which HRP and HERC portfolios have yielded the highest returns are 5 and 2, respectively. Hence, the performance of MVP portfolios has been the best among the three approaches on Sharpe ratios. Based on the maximization of the annual return, HRP portfolios performed the best among the three approaches yielding the highest returns for 6 among the 15 sectors, the corresponding numbers for MVP and HERC portfolios are 4 and 5, respectively. Finally, based on minimization of the annual volatility, MVP portfolios outperformed the other two approaches yielding the lowest volatilities for 14 among the 15 sectors. While HRP portfolios could not minimize the annual returns for any sector, HERC portfolios produced the lowest volatility for 1 sector. If the Sharpe ratio is considered the most critical metric in evaluating the performance of a portfolio, MVP portfolios distinctly produce the best results on the training data.

However, the performance of the portfolios over the test data is more critical and of sole interest to investors. On the test data, HERC portfolios are found to be superior based on the Sharpe ratios, yielding the highest values of the ratio for 8 among the 15 sectors. The number of sectors for which MVP and HRP portfolios produced the highest Sharpe ratios are 6 and 1, respectively. Based on the maximization of the annual returns, MVP portfolios performed the best producing the highest returns for 7 among the 15 sectors on the test data. The number of sectors for which HRP and HERC portfolios yielded the highest returns are 2 and 6, respectively. However, MVP portfolios outperformed their counterparts based on minimization of the annual volatilities, producing the lowest volatilities for 10 among the 15 sectors, the corresponding numbers for HRP and HERC portfolios being 3 and 2, respectively. Considering the criticality of the Sharpe ratio as the metric for portfolio performance, HERC portfolios are found to have yielded the best results for the majority of sectors on the portfolio test data.

Second, on the training data, the highest Sharpe ratio of 1.5475 was produced by the HRP portfolio on the stocks of the mid-small IT and telecom sector. The MVP portfolio, on the other hand, yielded the highest Sharpe ratio of 2.5717 on the stocks of the FMCG sector. The highest annual return of 52.03% was produced by the MVP portfolio on the training data for the stocks of the metal sector. On the test data, however, the HERC portfolio on the stocks of the PSU banks yielded the highest return of 60.45%. The minimum annual volatility of 17.16% over the portfolio training data was produced by the MVP portfolio for the NIFTY 50 stocks. The MVP portfolio yielded the lowest annual volatility of 9.47% for the NIFTY 50 stocks over the test period as well. Since the performance on the test data is a true reflection of the current market situation, it may be concluded that the FMCG sector has the highest prospect of yielding the highest risk-adjusted return, the NIFTY 50 portfolio has the least volatility, and the portfolios on the PSU bank stocks have the maximum potential in yielding high returns.

Third, from the point of view of investors, the portfolios that perform the best for a given sector on the training data are the ones that are adopted for investments. For example, if the MVP portfolio yielded the highest Sharpe ratio for the auto sector over the training data, then an investor looking to maximize the risk-adjusted return will go for the MVP portfolio for the auto sector stocks for the portfolio test period. However, if the MVP portfolio for the auto sector stocks fails to achieve the highest Sharpe ratio over the test period, then the objective of the investor will not be met. Hence, it is important to identify those sectors and portfolios that are consistent in their performance over the training and the test data. It is observed that for the stocks of the auto and FMCG sectors, the MVP portfolios consistently yielded the highest Sharpe ratios for both the training and test data. The sectors that yielded the highest return for the same portfolio on both training and test data are metal (MVP), mid-small IT & telecom (HRP), pharma (HERC), and realty (HERC). The portfolios yielding the highest returns are mentioned in parentheses. Eleven sectors exhibited the lowest volatility with the same portfolio both on the training and test data. These sectors are banking, financial services, IT, mid-small IT and telecom, oil & gas, pharma, private banks, realty, and NIFTY 50 (MVP). For all these sectors, the MVP portfolios yielded the lowest volatility.

## Conclusion

This work presented in this chapter discussed three different approaches to portfolio optimization. The three approaches are mean-variance portfolio (MVP), hierarchical risk parity (HRP) portfolio, and hierarchical equal risk contribution (HERC). The portfolios are built on stock chosen from 15 important sectors listed on the NSE of India. These 15 sectors included one diversified sector consisting of the 50 stocks of NIFTY 50. For the other 14 sectors, the top 10 stocks from each sector are selected based on their free-float market capitalization. For each sector, three portfolios are built based on the historical prices of the stocks from July 1, 2019, to June 30, 2022. Finally, the portfolios are tested on the stock prices from July 1, 2022, to June 30, 2023. For evaluating the performances of the portfolios three metrics, annual return, annual volatility, and the Sharpe ratio, are used. For each sector, the portfolios that yielded the highest Sharpe ratio, the maximum annual return, and the minimum annual volatility are identified. It is observed that the overall performance of the MVP portfolio is the best among the three portfolios on the training data. However, the HERC portfolio is found to have yielded the best results on the test data. While the MVP portfolios yielded the highest Sharpe ratios for 8 sectors among the 15 sectors analyzed on the portfolio training data, the HERC portfolio produced the highest Sharpe ratios for the same number of sectors on the test data. Since the performance of the portfolios on the test data is of actual interest to the investors, the HERC portfolio appears to be the most efficient one for the Indian stock market in current times. However, from the point of view of maximization of returns and minimization of risk, the MVP portfolios outperformed their HRP and HERC counterparts on the test data. Based on the performance of the portfolios on the test data, it is also observed that the FMCG sector has currently the maximum prospects yielding the highest risk-adjusted returns, the NIFTY 50 stocks have the lowest volatility, and the PSU banks stocks are most likely to yield higher returns compared to the other sectors. From investors' perspective, portfolios yielding the best results both on the training and the test data are the consistent performers. The sectors that yielded the highest return for the same portfolio on both training and test data are metal (MVP), mid-small IT & telecom (HRP), pharma (HERC), and realty (HERC). The portfolios yielding the highest returns are mentioned in parentheses. Eleven sectors exhibited the lowest volatility with the same portfolio, both on the training and test data. These sectors are banking, financial services, IT, mid-small IT and telecom, oil & gas, pharma, private banks, realty, and NIFTY 50 (MVP). For all these sectors, the MVP portfolios yielded the lowest volatility.

The future work includes a study involving important stocks listed in the other major stock exchanges in the world to verify whether the portfolios exhibit similar behavior on those stocks.